\documentclass[pdflatex, sn-mathphys-num]{sn-jnl}

\usepackage{graphicx}%
\usepackage{multirow}%
\usepackage{amsmath,amssymb,amsfonts}%
\usepackage{amsthm}%
\usepackage[title]{appendix}%
\usepackage{xcolor}%
\usepackage{textcomp}%
\usepackage{manyfoot}%
\usepackage{booktabs}%
\usepackage{algpseudocode}%
\usepackage{listings}%
\usepackage{rotating}%
\usepackage{booktabs}%
\usepackage{longtable}%
\usepackage{pdflscape}%
\usepackage{float}%
\usepackage{needspace}%
\usepackage{afterpage}%
\usepackage{tabularx}%
\usepackage{euscript}%
\usepackage{array}%
\usepackage{hyperref}
\usepackage{url}
\usepackage{euscript} 

\theoremstyle{thmstyleone}%
\theoremstyle{thmstyletwo}%

\theoremstyle{thmstylethree}%

\begin{document}

\title[Article Title]{Spatiotemporal Learning of Brain Dynamics from fMRI Using Frequency-Specific Multi-Band Attention for Cognitive and Psychiatric Applications}


\author[1]{\fnm{Sangyoon} \sur{Bae}}\email{stellasybae@snu.ac.kr}

\author[2]{\fnm{Junbeom} \sur{Kwon}}\author[3]{\fnm{Shinjae} \sur{Yoo}}

\author*[1,4,5]{\fnm{Jiook} \sur{Cha}}\email{connectome@snu.ac.kr}

\affil[1]{\orgdiv{Interdisciplinary Program in Artificial Intelligence}, \orgname{Seoul National University}, \orgaddress{\street{Gwanak-ro 1}, \city{Seoul}, \postcode{08826}, \state{Seoul}, \country{South Korea}}}

\affil[2]{\orgdiv{Department of Psychology}, \orgname{University of Texas at Austin}, \orgaddress{\street{Dean Keeton Street}, \city{Austin}, \postcode{78712}, \state{Texas}, \country{United States}}}

\affil[3]{\orgdiv{Computational Science Initiative}, \orgname{Brookhaven National Laboratory}, \orgaddress{\street{Brookhaven Avenue}, \city{Shirley}, \postcode{11967}, \state{NewYork}, \country{United States}}}

\affil[4]{\orgdiv{Department of Psychology}, \orgname{Seoul National University}, \orgaddress{\street{Gwanak-ro 1}, \city{Seoul}, \postcode{08826}, \state{Seoul}, \country{South Korea}}}

\affil[5]{\orgdiv{Department of Brain and Cognitive Sciences}, \orgname{Seoul National University}, \orgaddress{\street{Gwanak-ro 1}, \city{Seoul}, \postcode{08826}, \state{Seoul}, \country{South Korea}}}



\abstract{Understanding how the brain's complex nonlinear dynamics give rise to cognitive function remains a central challenge in neuroscience. While brain functional dynamics exhibits scale-free and multifractal properties across temporal scales, conventional neuroimaging analytics assume linearity and stationarity, failing to capture frequency-specific neural computations. Here, we introduce Multi-Band Brain Net (MBBN), the first transformer-based framework to explicitly model frequency-specific spatiotemporal brain dynamics from fMRI. MBBN integrates biologically-grounded frequency decomposition with multi-band self-attention mechanisms, enabling discovery of previously undetectable frequency-dependent network interactions. Trained on 49,673 individuals across three large-scale cohorts (UK Biobank, ABCD, ABIDE), MBBN sets a new state-of-the-art in predicting psychiatric and cognitive outcomes (depression, ADHD, ASD), showing particular strength in classification tasks with up to 52.5\% higher AUROC and provides a novel framework for predicting cognitive intelligence scores. Frequency-resolved analyses uncover disorder-specific signatures: in ADHD, high-frequency fronto-sensorimotor connectivity is attenuated and opercular somatosensory nodes emerge as dynamic hubs; in ASD, orbitofrontal-somatosensory circuits show focal high-frequency disruption together with enhanced ultra-low-frequency coupling between the temporo-parietal junction and prefrontal cortex. By integrating scale-aware neural dynamics with deep learning, MBBN delivers more accurate and interpretable biomarkers, opening avenues for precision psychiatry and developmental neuroscience.}




\maketitle

\section{Introduction}\label{sec1}
Understanding the brain’s dynamic processes is essential for unraveling the mechanisms underlying adaptive cognition and behavior. These processes influence critical functions such as learning, decision-making, and social interactions, with mounting evidence that their dysregulation is closely linked to mental health conditions (\cite{taghia2018uncovering, he2010temporal, sendi2025brain, breakspear2017dynamic}).
While functional magnetic resonance imaging (fMRI) has been instrumental in probing brain activity, many conventional analytical methods remain constrained by assumptions of linearity, stationarity and static connectivity. Approaches ranging from mass-univariate statistical tests and seed-based correlations to static functional connectivity maps often treat the brain as a relatively fixed system, thus failing to capture the temporal fluctuations, nonlinear interactions, and adaptive reconfigurations that characterize real neural dynamics. Recent advances in artificial intelligence, particularly transformer-based architectures originally designed for sequential data, have shown promise for modeling brain activity as a continuously evolving network of interdependent processes. Yet, these approaches largely ignore the frequency-specific nature of neural oscillations (\cite{kan2022brain, bedel2023bolt, wang2023brainbert, ortega2023brainlm}).

This oversight is particularly problematic given that existing neuroimaging models, including recent deep learning approaches, treat brain signals as broadband phenomena, overlooking the fundamental frequency-specific organization of neural computation. Indeed, this organization is best understood through two fundamental properties that describe the brain's complexity: its \textbf{multifractal structure} and its \textbf{scale-free dynamics}.

Multifractality, a key structural property, points to the brain's nested, hierarchical organization where multiple scaling dimensions and power-law behaviors coexist (\cite{stanley1988multifractal, falconer2013fractal, salat2017multifractal, Guidolin2024, ciuciu2008log}). Crucially, it is within this complex structure that scale-free dynamics—a key functional property—become apparent. While a broadband signal is not inherently scale-free, decomposing it into distinct frequency components reveals power-law relationships (Equation \ref{power_law}) that signify consistent activity patterns across different timescales. These dynamics play crucial roles in brain maturation and state transitions, with temporal variability serving as a key regulatory mechanism (\cite{smit2011scale, tagliazucchi2013breakdown}). The power-law exponent ($\beta$) varies across brain regions and states, providing valuable insights into neural network heterogeneity and serving as a potential marker for neurological and psychiatric disorders (\cite{tolkunov2010power, maxim2005fractional}). Evidence from fMRI, ECoG, and EEG studies further suggests that these principles drive frequency-specific changes in connectivity and topological organization, such as small-worldness and modularity (\cite{achard2006resilient, sasai2021frequency}). Understanding these patterns is crucial for developing more effective diagnostic and therapeutic strategies.

\begin{figure}[H]
\centering
\includegraphics[width=0.8\textwidth]{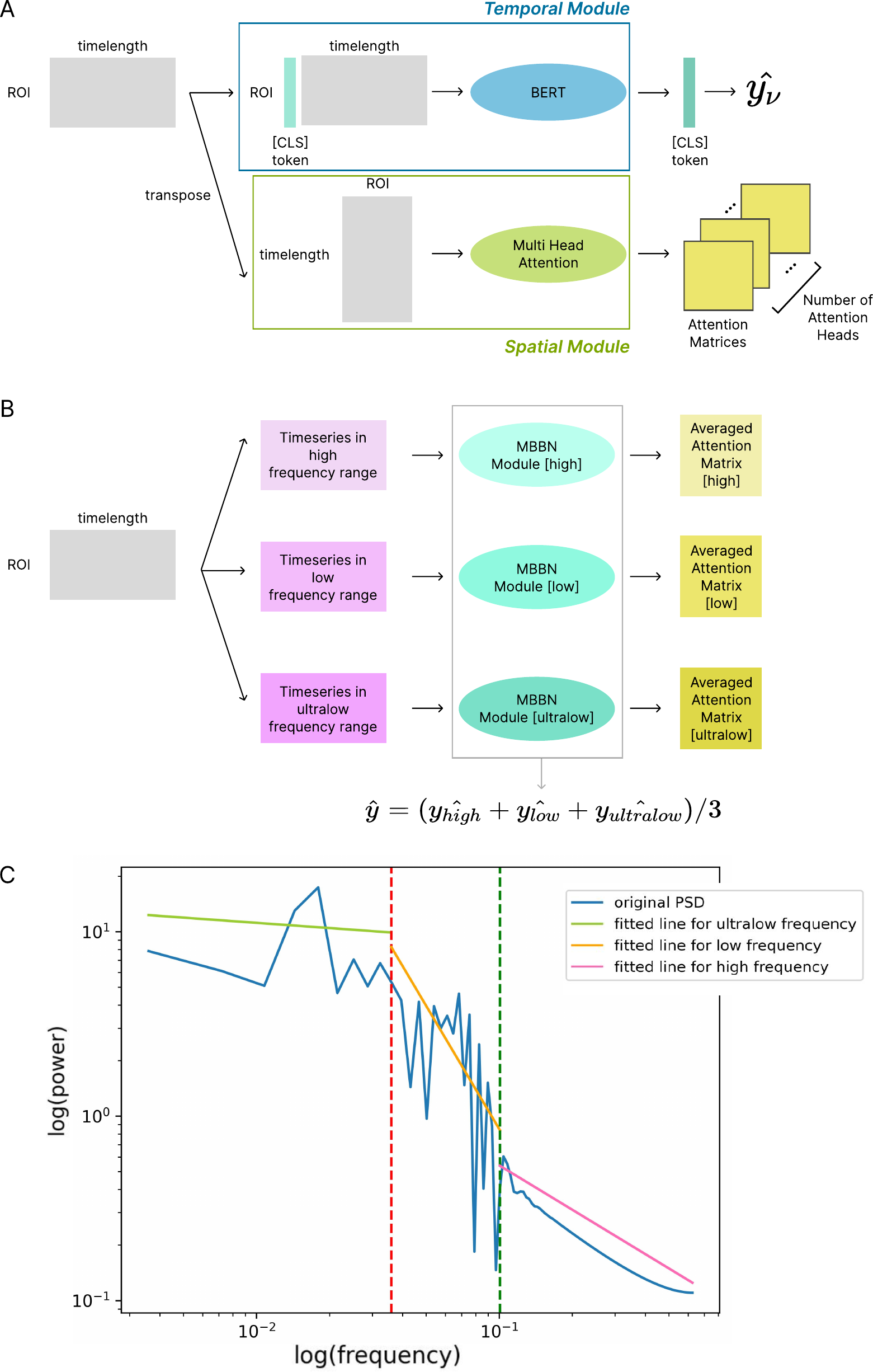}
\caption{\textbf{\textbf{Multi-Band Brain Net procedure} \\A.} The MBBN module processes each timeseries signal, divided by frequency, within specific frequency ranges. In the temporal module (denoted as BERT), BERT updates the [CLS (Classification)] token, which is then passed through a classifier to generate a prediction $\hat{y_{\nu}}$ ($\nu$ represents the frequency range). The spatial module employs a multi-head attention mechanism to compute attention matrices representing the interactions between regions of interest (ROIs). \textbf{B.} The MBBN modules are applied to timeseries signals across three frequency ranges (ultra-low, low, and high frequency). Parameter sharing is implemented in the temporal module, while spatial modules have independent parameters for each frequency range. The final output of MBBN is computed as the average of the predictions across all frequency ranges, along with attention matrices specific to each frequency range. \textbf{C.} Frequency decomposition using scale-free principles on representative UKB data. Individual-specific knee frequencies ($f_{1}$, $f_{2}$) separate three biologically distinct bands: ultralow (pink), low (yellow-green), and high (yellow) frequencies, each exhibiting characteristic power-law scaling.
The dotted red and green lines indicate the frequency boundaries: ultra-low to low ($f_1$), and low to high ($f_2$) frequencies, respectively.}\label{fig_procedure}
\end{figure}

To address these challenges, we introduce \textbf{Multi-Band Brain Net (MBBN)}, the first transformer-based framework specifically designed to model frequency-resolved spatiotemporal brain dynamics from fMRI. Unlike existing approaches that analyze broadband signals, MBBN explicitly decomposes neural activity into biologically meaningful frequency bands using scale-free principles, then applies specialized attention mechanisms to capture frequency-specific connectivity patterns incorporating network theory concepts like communicability (Figure \ref{fig_procedure}, \ref{fig_pretraining}). MBBN divides fMRI BOLD signals into distinct frequency bands, enabling the capture of band-specific functional and topological properties. This approach provides a neurobiologically informed framework for modeling the brain's multifractal and scale-free dynamics. The model features two key strengths: Firstly, its self-attention weighted connectivity dynamically models functional connectivity by capturing complex, \textbf{nonlinear interactions} and time-varying relationships between brain regions, thereby overcoming the limitations of traditional static, \textbf{correlation-based linear representations}. Secondly, a communicability-inspired pretraining loss function optimizes for biologically meaningful connectivity patterns.
By leveraging these advances, MBBN not only provides deeper insights into brain function but also identifies frequency-specific markers for neurological and psychiatric disorders, offering a powerful new tool for clinical neuroscience research and applications.

\begin{equation}\label{power_law}
log(power) \propto -\beta \cdot log(frequency)
\end{equation}

\section{Results}\label{results}
\subsection{MBBN demonstrates superior performance across diverse neuroimaging tasks}\label{results_from_scratch}

\begin{figure}[hbt!]
\centering
\includegraphics[width=0.8\textwidth]{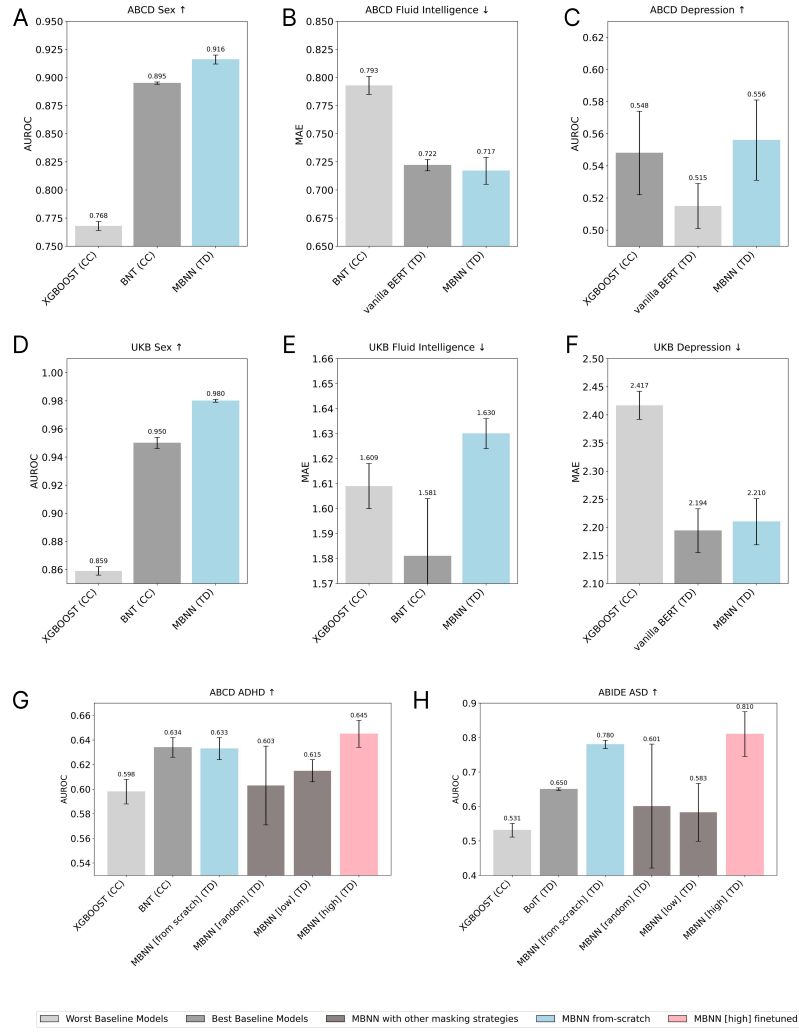}
\caption{\textbf{Comparative Analysis of fMRI Prediction Models.} Across datasets and tasks, the proposed MBBN demonstrates superior predictive performance compared to baseline models including state-of-the-art models in the field  (e.g., Brain Network Transformer (\cite{kan2022brain})), as well as 
vanilla BERT (without multifractal decomposition). [CC] indicates models based on Pearson's correlation-based connectivity; [TD] represents models accounting for temporal dynamics. The x-axis denotes the models, and the y-axis represents performance metrics, with AUROC used for classification tasks and MAE for regression tasks.
MBBN consistently outperforms baseline models across biological (A-B: sex prediction), cognitive (C-D: fluid intelligence), and clinical (E-H: depression, ADHD, ASD) prediction tasks. [CC] indicates correlation-based connectivity models; [TD] represents temporal dynamics models. AUROC is shown for classification tasks, MAE for regression tasks. Data from ABCD (A-C, G), UKB (D-F), and ABIDE (H) datasets using HCP-MMP1 atlas. A-F denotes from-scratch experiments and G-H denotes fine-tuning experiments.}
\label{fig_main_result}
\end{figure}


\begin{table}[hbt!]
\centering
\begin{tabular}{|l|c|c|c|}
\hline
\textbf{Task} & \textbf{MBBN} & \textbf{vs Best Baseline (\%)} & \textbf{vs Worst Baseline (\%)} \\
\hline
ABCD Sex & 0.916 (AUROC) & +2.346 & +19.271 \\
\hline
ABCD Fluid Intelligence & 0.717 (MAE) & +0.693 & +9.584 \\
\hline
ABCD Depression & 0.556 (AUROC) & +1.460 & +7.961 \\
\hline
UKB Sex & 0.980 (AUROC) & +3.158 & +14.086 \\
\hline
UKB Fluid Intelligence & 1.630 (MAE) &  -3.099 & -1.305 \\
\hline
UKB Depression & 2.21 (MAE) &  -1.231 & +0.924 \\
\hline
ABCD ADHD & 0.645 (AUROC) & +1.735 & +7.860 \\
\hline
ABIDE ASD & 0.810 (AUROC) & +24.615 & +52.542 \\
\hline
\end{tabular}
\caption{\textbf{Performance gain of MBBN between best and worst baseline model.} We illustrated finetuned MBBN for comparison of ABCD ADHD classification and ABIDE ASD classification.}
\end{table}


To evaluate its performance, MBBN was trained from scratch on a diverse set of downstream tasks spanning biological, cognitive, and clinical domains, utilizing both the HCP-MMP1 and Schaefer atlases with the UKB and ABCD datasets. While MBBN demonstrated unparalleled performance in classification tasks, its advantage in regression tasks (e.g., fluid intelligence and depression score prediction) was less pronounced, suggesting that the model's architecture is particularly optimized for distinguishing categorical phenotypes rather than predicting continuous scores. The primary results presented in Figure \ref{fig_main_result} A-F focus on experiments conducted with the HCP-MMP1 atlas; corresponding analyses using the Schaefer atlas are detailed in the Appendix \ref{Appendix_Results}.

For biological phenotype prediction, MBBN achieved substantial improvements in sex classification with gains up to 14.09\% (UKB) and 19.27\% (ABCD) using HCP-MMP1 atlas. Cognitive assessment showed marked improvements, with fluid intelligence prediction demonstrating up to 9.58\% enhancement on ABCD. Also clinical prediction tasks revealed MBBN's therapeutic potential, achieving 0.92\% improvement in depression prediction (UKB) and exceptional performance in psychiatric disorders (detailed in Supplementary Materials).

\subsection{Communicability-based pretraining enables superior clinical prediction performance}\label{result_finetune}

The exceptional performance of MBBN, particularly in clinical prediction tasks, stems from our novel communicability-based pretraining strategy that masks nodes with high communicability within the brain network. This strategy was predicated on the hypothesis that selectively reconstructing these highly influential nodes during pretraining would better represent the complex communication patterns within the brain. To validate this hypothesis and evaluate the downstream impact of this pretraining strategy, MBBN was subsequently fine-tuned on two clinically relevant classification tasks: ADHD classification using the ABCD dataset, and ASD classification using the ABIDE dataset. These datasets and tasks were chosen for their established clinical significance and high prevalence rates.

Fine-tuning experiments (Figure \ref{fig_main_result} G-H) revealed a clear advantage for the \textit{MBBN-high} model, which employed a pretraining strategy that selectively masked nodes with high communicability. This approach resulted in significant performance improvements compared to baselines: a 7.86\% gain in ADHD classification AUROC and a 52.54\% gain in ASD classification AUROC. In contrast, models using random node masking (\textit{MBBN-random}) or masking nodes with low communicability (\textit{MBBN-low}) showed inferior performance, even compared to the MBBN model trained from scratch (\textit{MBBN-from scratch}). These findings underscore the critical role of communicability-based pretraining in effectively leveraging the underlying network structure of the data to enhance predictive performance. Detailed results and statistical details are available in Appendix \ref{Appendix_Results}. These results demonstrate that communicability-based loss better captures a complex information flow of the brain than conventional masking loss, and MBBN's frequency-resolved approach captures clinically relevant neural signatures that are missed by conventional broadband analysis methods.

\subsection{Frequency-specific connectivity patterns reveal distinct psychiatric biomarkers}\label{results_interpretability}

\begin{figure}[hbt!]
\centering
\includegraphics[width=0.8\textwidth]{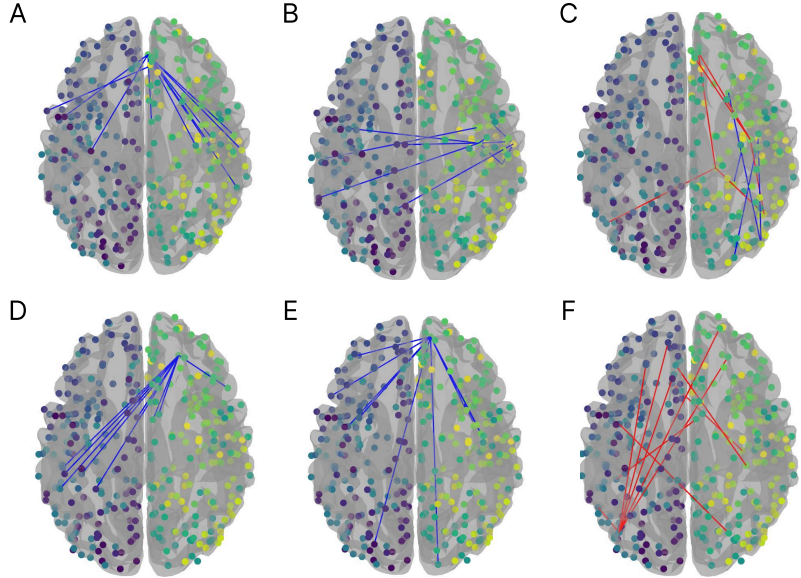}
\caption{\textbf{Frequency-specific connectivity disruptions in ADHD and ASD.} Statistically significant connections (FDR-corrected $p < 0.05$, top 10 effect sizes) distinguishing ADHD (A-C) and ASD (D-F) from healthy controls across high, low, and ultralow frequency bands. Red: increased connectivity in disorder; Blue: decreased connectivity. Detailed results in Supplementary Materials.}
\label{fig_interpretation}
\end{figure}
 
To identify frequency-specific biomarkers underlying psychiatric disorders, we applied Grad-CAM-based attribution analysis to extract self-attention-weighted connectivity patterns that distinguished patient groups from healthy controls. This approach revealed distinct frequency-dependent neural signatures for ADHD and ASD that were previously undetectable using conventional broadband connectivity analysis. Details are described in Section \ref{methods_9_interpretability}.

In the ABCD test dataset (HCP-MMP1 atlas), frequency-dependent differences in functional connectivity patterns were identified between the ADHD and HC groups. ADHD showed reduced connectivity in high and low frequencies, while exhibiting increased connectivity in the ultralow-frequency range (Figure \ref{fig_interpretation}). These findings reveal frequency-specific neural connectivity alterations in ADHD.

ADHD exhibits frequency-dependent connectivity alterations with high-frequency disruptions predominating. Comparing ADHD patients to healthy controls in the ABCD dataset, we identified distinct patterns of connectivity changes across frequency bands (Figure \ref{fig_interpretation}  A-C). Statistical analysis revealed that high-frequency connections showed the most extensive alterations, with 8.43\% of all significant connections ($p < 0.05$) occurring in this band and a 75th percentile Cohen's d of 0.332, indicating moderate effect sizes. In contrast, low-frequency and ultralow-frequency bands showed more limited but still significant changes, accounting for 3.46\% (Cohen's d 75th percentile = 0.220) and 3.56\% (Cohen's d 75th percentile = 0.117) of significant connections, respectively.

Anatomically, high-frequency disruptions centered on Area 10v (ventral part of Brodmann Area 10), indicating impaired executive control networks. Low-frequency alterations consistently involved Area OP2-3-VS (opercular parts of Areas 2 and 3 and adjacent Ventral Somatosensory / Sylvian region) and Auditory 4 Complex, suggesting disrupted sensorimotor-auditory integration critical for attention regulation. Ultralow-frequency changes in Medial Belt Complex and ParaHippocampal regions, while showing smaller effect sizes, point to subtle alterations in memory and emotional processing circuits.



ASD demonstrated a contrasting frequency profile with ultralow and low frequencies most severely affected (Figure \ref{fig_interpretation} D-F). T-test comparisons between ASD patients and healthy controls in the ABIDE dataset revealed a different pattern from ADHD. Ultralow-frequency connections showed the most extensive disruptions, representing 4.54\% of all significant connections with large effect sizes (75th percentile Cohen's d = 0.717). Low-frequency alterations were similarly prominent, accounting for 4.04\% of significant connections with substantial effect sizes (Cohen's d 75th percentile = 0.714). Notably, high-frequency changes were minimal, comprising only 1.81\% of significant connections, though with moderate effect sizes (Cohen's d 75th percentile = 0.575).

High-frequency changes, while limited in extent, localized specifically to Area 11l (lateral part of Brodmann Area 11), consistent with known social-cognitive deficits. Low-frequency disruptions prominently affected Area 10pp (frontopolar cortex), suggesting impaired higher-order executive functions. Most significantly, the extensive ultralow-frequency alterations involved Area TPOJ3 (temporo-parieto-occipital junction) and multiple frontal regions, indicating fundamental disruptions in social cognition and language networks—core features of ASD pathophysiology.



In both the ABCD and ABIDE datasets, analysis of false positive and false negative subjects (based on predicted values) showed no statistically significant connections in the self-attention-based connectivity analysis. These frequency-resolved patterns reveal disorder-specific neural signatures: ADHD primarily affects high-frequency sensorimotor networks with compensatory ultralow-frequency changes, while ASD shows pervasive low and ultralow-frequency disruptions in social-cognitive circuits. Importantly, false positive and false negative cases showed no significant connectivity patterns, confirming the specificity of these frequency-dependent biomarkers for accurate psychiatric classification.

\section{Methods} \label{methods}
\subsection{Preprocessing} \label{methods_0_preprocesing}
The ABCD, UKB, and ABIDE datasets underwent dataset-specific preprocessing to account for acquisition differences and multi-site harmonization requirements. In ABCD, a conservative bandpass filter (0.009-0.08 Hz) was applied following consortium protocols for adolescent populations (\cite{hagler2019image}). UKB employed a broader filter (0.008-0.1 Hz) optimized for adult populations to minimize heart rate variability effects (\cite{marek2022reproducible}). ABIDE used finite impulse response filtering (0.01-0.1 Hz) specifically designed for multi-site ASD data integration (\cite{ingalhalikar2021functional}).

For the present study, additional preprocessing was performed on the resting-state BOLD data. The following preprocessing procedures were applied to each participant’s resting-state BOLD data: First, skull-stripping, slice-timing correction, and spatial normalization to MNI space were conducted using fMRIPrep (\cite{esteban2019fmriprep}). Subsequently, head motion artifacts and signals from white matter were removed using component-based noise correction (CompCor) (\cite{behzadi2007component}). The mean time-series signals from predefined regions of interest (ROIs) were extracted using the Nilearn package, with HCP-MMP1 asymmetric (\cite{glasser2016multi}) and Schaefer 400 (\cite{schaefer2018local}) atlases employed for ROI definitions. Participants whose data contained voxels with NaN values, typically arising from the regression of unreliable white-matter signals, were excluded from the analysis. Additionally, to minimize spurious signal fluctuations due to scanner instability, the first 20 volumes of each time-series were discarded.

For deep neural network training, uniform sequence lengths across all participants were required. To standardize the sequence length, it was set to the smallest multiple of 8 that was greater than or equal to the shortest sequence length in the dataset, optimizing computational efficiency for GPU-based matrix operations. The final sequence lengths for the datasets were as follows: UK Biobank (UKB) 464, Adolescent Brain Cognitive Development (ABCD) 348, and Autism Brain Imaging Data Exchange (ABIDE) 280.

\subsection{Dataset} \label{methods_1_dataset}

Demographic and diagnosis distribution details for the three datasets (UKB, ABCD, ABIDE) are summarized in Table \ref{dataset_metadata}. The UK Biobank (UKB) dataset, which includes 41,283 participants, was used for pretraining due to its large sample size and broad demographic representation. This extensive dataset enabled the model to capture generalizable brain dynamics patterns. The Adolescent Brain Cognitive Development (ABCD) dataset, consisting of 4,527 participants (including individuals with ADHD and healthy controls), and the Autism Brain Imaging Data Exchange (ABIDE) dataset, which includes 141 participants (comprising individuals with ASD and healthy controls), were employed for fine-tuning to capture disorder-specific patterns.

ADHD diagnosis was based on CBCL attention and ADHD scores exceeding a T-score threshold of 65, while healthy controls were defined as individuals with no diagnosed mental health disorders (\cite{cordova2022attention}). This classification ensured a distinct separation between clinical and non-clinical groups, facilitating model training and evaluation.

\begin{table}[!ht]
    \centering
    \begin{tabular}{cccc}
    \hline
        Features & UKB & ABCD & ABIDE \\ \hline
        Gender (M/F) & 40,699 (21,556/19,143) & 8,833 (4,635/4,198) & 141 (114/27) \\ 
        Age (year) & 54.9$\pm$7.5 & - & 14.3$\pm$3.6\\
        Age (month) & - & 119.2$\pm$7.5 & - \\ 
        Depression (PHQ-9 score) & 11.5$\pm$3.5 & - & - \\ 
        Depression (disorder/controls) & - & 3,113 (2,652/461) & - \\
        ADHD (disorder/controls) & - & 4,289 (2,439/1,850) & - \\
        ASD (disorder/controls) & - & - & 141 (77/64) \\ \hline
    \end{tabular}
    \caption{\textbf{Features of the UKB, ABCD, and ABIDE subjects.} Disorder denotes ADHD in the ABCD dataset and ASD in the ABIDE dataset.}\label{dataset_metadata}
\end{table}

\subsection{Experimental Settings} \label{methods_2_exp_settings}
To address the imbalance in clinical variables within the dataset, stratified sampling was employed to ensure balanced distributions of target variables (ADHD, depression, ASD) across the training, validation, and test sets. A similar approach was used for the biological phenotype (sex), which showed minimal imbalance. To enhance model robustness and reduce overfitting, training was conducted using three random seeds, and the average performance across these runs was reported.

Gradient clipping with a norm of one and gradient accumulation were applied to stabilize training and improve predictive accuracy. To accelerate the training process, Automatic Mixed Precision (AMP) was utilized. The BERT model architecture comprised 8 hidden layers and 8 or 12 attention heads, depending on the number of regions of interest (ROIs) in the atlas used. The number of attention heads was chosen based on the constraint that it should be a divisor of the total number of ROIs. Hyperparameter optimization included testing learning rates ([1e-5, 1e-2]), weight decay ([1e-3, 1e-2]), learning rate policy (step, SGDR), and optimizers (Adam, AdamW, RMSprop).

For computational efficiency, pretraining was performed on a single A100 GPU, while training from scratch and fine-tuning were conducted using a single NVIDIA RTX 3090 GPU. The number of epochs was set to 100 for training from scratch. Efficient pretraining was achieved by determining the optimal number of epochs using weightwatcher (\cite{martin2021predicting}), which identifies well-trained layers based on statistical learning theory. Optimal pretraining epochs were set to 1,000 for the HCP-MMP1 atlas and 400 for the Schaefer 400 atlas. Additional methodological details are provided in Supplementary Materials \ref{Appendix_Pretraining_MBBN_weightwatcher}.

\subsection{Frequency-resolved decomposition based on scale-free principles}

Brain activity exhibits multifractal properties with distinct scaling behaviors across frequency ranges, reflecting the hierarchical organization of neural networks from local circuits to global integration pathways. Conventional broadband connectivity analysis obscures these frequency-specific dynamics, potentially missing critical neurophysiological signatures of psychiatric disorders. To capture these frequency-dependent neural processes, we developed a systematic two-step fitting procedure to identify physiologically meaningful frequency boundaries (knee frequencies) that demarcate distinct scaling regimes in the power spectral density of resting-state fMRI signals.
Following frequency decomposition using the identified knee frequencies, we confirmed that each of the three segments exhibited distinct power-law scaling behavior, as evidenced by significantly different $\beta$ values across frequency bands ($p < 0.001$).

\subsubsection{Step 1: Ultralow frequency boundary identification using Lorentzian fitting}

We first applied the Lorentzian equation (Equation \ref{Lorentzian equation}) to the entire power spectral density to identify $f_{1}$, the primary knee frequency that separates ultralow frequency activity from higher-frequency neural processes:

\begin{equation}\label{Lorentzian equation}
Power(f) = \frac{A \cdot f_{1}^2}{f^2 + f_{1}^2}
\end{equation}

where $A$ represents the amplitude parameter and $f_1$ denotes the characteristic frequency at which the power spectrum transitions from flat (low-frequency plateau) to steep decay (high-frequency roll-off). This Lorentzian model effectively captures the fundamental 1/f characteristics of neural signals while identifying the critical transition point. The fitting was performed using SciPy's curve\_fit function, which is well-suited for this two-parameter optimization problem. Physiologically, $f_1$ corresponds to the boundary between slow cortical potentials and faster network dynamics, typically associated with default mode network fluctuations and global integration processes.

\subsubsection{Step 2: Low-high frequency boundary identification using spline multifractal modeling}

For frequencies above $f_1$, we applied a sophisticated spline multifractal equation to identify $f_2$, the second knee frequency that distinguishes low-frequency network connectivity from high-frequency local processing:

To ensure smooth transitions between different scaling regimes and prevent fitting artifacts, we implemented a cubic spline weight function:

\begin{equation}\label{spline_multifractal_equation_weight}
w(f) = \begin{cases}
0 & \text{if } \log(f) < \log(f_2) - s \\
1 & \text{if } \log(f) > \log(f_2) + s \\
\frac{1}{2}(1 - \cos(\frac{\pi(\log(f) - \log(f_2) + s)}{2s})) & \text{otherwise}
\end{cases}
\end{equation}

This weight function $w(f)$ provides a smooth, differentiable transition between two distinct power-law regimes. When $f \ll f_2$, $w(f) \approx 0$, ensuring the function follows low-frequency scaling behavior characterized by $\beta_{low}$. Conversely, when $f \gg f_2$, $w(f) \approx 1$, shifting to high-frequency scaling defined by $\beta_{high}$. The smoothness parameter $s$ controls the sharpness of this transition, preventing abrupt discontinuities that could introduce fitting artifacts.

The complete spline multifractal function then becomes:

\begin{equation}\label{spline_multifractal_equation}
Power(f) = A \cdot f^{\beta_{low}(1-w(f)) + \beta_{high}w(f)}
\end{equation}

where:
\begin{itemize}
    \item $A$ is the amplitude scaling factor
    \item $\beta_{low}$ is the low-frequency scaling exponent (typically steeper, reflecting long-range correlations)
    \item $\beta_{high}$ is the high-frequency scaling exponent (typically flatter, reflecting local processing)
    \item $f_2$ is the second knee frequency (transition point between network and local dynamics)
    \item $s$ is the smoothness parameter controlling transition width
\end{itemize}

This five-parameter model was fitted using the iminuit package, specifically designed for high-dimensional parameter estimation with robust convergence properties. Physiologically, $f_2$ represents the transition from large-scale network connectivity to localized neural processing, typically corresponding to the boundary between task-negative and task-positive network activity.

\subsubsection{Frequency band segmentation and filtering implementation}

Based on the identified knee frequencies $f_1$ and $f_2$, we segmented each participant's time-series data into three physiologically distinct frequency bands:

\begin{enumerate}
    \item \textbf{Ultra-low frequency band ($< f_1$)}: Captures slow cortical potentials, default mode network activity, and global integration processes
    \item \textbf{Low frequency band ($f_1$ to $f_2$)}: Represents intermediate-scale network connectivity and inter-regional communication
    \item \textbf{High frequency band ($> f_2$)}: Reflects local circuit dynamics, sensorimotor processing, and rapid cognitive operations
\end{enumerate}

The segmentation was implemented using finite impulse response (FIR) filters from the \textit{nitime} package for ultra-low and low-frequency bands, and Boxcar filters for high-frequency extraction. The choice of filter types was optimized through systematic hyperparameter tuning on validation sets, evaluating their impact on downstream classification performance. FIR filters were preferred for lower frequencies due to their superior phase preservation properties, while Boxcar filters provided computational efficiency for high-frequency components without compromising signal quality.

\subsubsection{Validation of frequency-specific scaling properties}

\begin{table}[!ht]
    \centering
    \begin{tabular}{ccc}
    \hline
        Dataset & $f_{1}$ & $f_{2}$ \\ \hline
        ABCD & 0.0456 $\pm$ 0.0106 & 0.0784 $\pm$ 0.0151 \\ 
        UKB & 0.0624 $\pm$ 0.0113 & 0.0961 $\pm$ 0.0157 \\
        ABIDE & 0.0182 $\pm$ 0.0121 & 0.0299 $\pm$ 0.0256 \\ \hline
    \end{tabular}
    \caption{Different $f_{1}$ and $f_{2}$ among datasets}\label{tab_different_f1_f2}
\end{table}

\begin{table}[!ht]
    \centering
    \begin{tabular}{cccc}
    \hline
        Dataset & $\beta_{ultralow}$  & $\beta_{low}$  & $\beta_{high}$ \\ \hline
        ABCD & 0.2442$\pm$0.2991  &  1.7845$\pm$0.3836 & 1.1595$\pm$0.4260 \\ 
        UKB &  0.0209$\pm$0.2185 &  2.8959$\pm$0.4562 & 1.1232$\pm$0.4107  \\
        ABIDE & 0.5805$\pm$0.3925  &  2.5939$\pm$0.3923  & 0.9563$\pm$0.2228 \\  \hline
    \end{tabular}
    \caption{Different $\beta$ value among frequency ranges and datasets}\label{tab_different_beta}
\end{table}

\begin{table}[!ht]
    \centering
    \begin{tabular}{lcc}
        \hline
        \multirow{2}{*}{Dataset} & Lorentzian Equation & Spline Multifractal Equation \\
        \cmidrule{2-3}
        & Mean $R^2$ & Mean $R^2$ \\
        \hline
        ABCD & 0.7423 $\pm$ 0.0765 & 0.6822 $\pm$ 0.2375  \\
        UKB & 0.6605 $\pm$ 0.0786 & 0.8292 $\pm$ 0.0854  \\
        ABIDE & 0.7435 $\pm$ 0.1018 & 0.8309 $\pm$ 0.1019  \\
        \hline
    \end{tabular}
    \caption{Goodness of Fit ($R^2$) Statistics for Lorentzian and Spline Multifractal Equations Across Datasets}\label{table:goodness_of_fit}
    \begin{tablenotes}
      \small
      \item Note: Values are presented as mean$\pm$standard deviation. $R^2$ values indicate the proportion of variance explained by each model.
    \end{tablenotes}
\end{table}

To validate that our frequency decomposition captured distinct neurophysiological processes, we computed scaling exponents ($\beta$ values) for each frequency band using power-law fitting:

Statistical analysis revealed significant differences in $\beta$ values across all frequency bands ($p < 0.001$), confirming that each band exhibits unique scale-free dynamics. Ultralow frequencies showed the flattest scaling (smallest $\beta$), consistent with long-range temporal correlations. High frequencies exhibited intermediate scaling, reflecting network-level integration. Low frequencies displayed the steepest scaling (largest $\beta$), characteristic of local, rapidly decorrelated processes. These findings support the hypothesis that frequency-specific decomposition captures fundamentally different aspects of brain organization, from global integration to local specialization.

\subsection{Model Design} \label{methods_5_model_architecture}
\subsubsection{Model Architecture} 

MBBN integrates temporal and spatial processing through a dual-module architecture that consists of the temporal module and the spatial module. These modules work synergistically to encode both the temporal and spatial characteristics of frequency-divided timeseries data, as shown in Figure \ref{fig_procedure} A. To ensure an effective balance between generalization and specificity, a parameter-sharing strategy is employed, as outlined below.

The temporal module (depicted as a blue oval in Figure \ref{fig_procedure} A) is designed to capture the temporal dynamics of the data. For each frequency-divided timeseries (ultra-low, low, and high), the data are processed independently using a shared BERT encoder, reflecting the temporal similarities across frequency bands. The BERT encoder outputs a sequence of hidden states for each timeseries, with the final hidden state corresponding to the [CLS] token passed to a classifier. This process produces logits ($\hat{y_{ultralow}}$, $\hat{y_{low}}$, and $\hat{y_{high}}$), which represent the model's predictions for each frequency range. The hidden dimension size of the MBBN temporal module is set to match the number of ROIs. Due to the constraints imposed by the BERT backbone in MBBN, the number of attention heads must be a divisor of the hidden dimension size. Consequently, when using the Schaefer atlas, the number of attention heads is set to 8, whereas for the HCP MMP1 atlas, it is set to 12.

The spatial module (depicted as a yellow-green oval in Figure \ref{fig_procedure} A) focuses on the spatial connectivity among regions of interest (ROIs). Unlike the temporal module, the spatial module uses a frequency-specific approach to capture the distinct spatial patterns of each frequency band. Specifically, a multi-head attention mechanism is applied to the transposed timeseries data, generating a set of attention matrices for each frequency band. These attention matrices—reflecting the distinct focus of different heads—are averaged to produce a single attention matrix per band. These averaged matrices represent the spatial connectivity patterns, which are subsequently used to compute the spatial loss (Equation \ref{Spatial loss term}). In practice, the MBBN spatial module receives the transposed input of the temporal module, so the number of attention heads must divide the sequence length. For example, we set 12 heads for ABCD (sequence length is 348) and 8 heads for both UKB (464) and ABIDE (280).

Although the temporal and spatial modules do not directly exchange hidden states, they remain closely coupled through the composite loss function. The total loss combines cross-entropy (or task-specific) loss with the spatial loss, so backpropagation updates parameters across the entire network. As a result, while the BERT-based temporal module primarily absorbs gradient updates from the mask and task objectives, and the attention-based spatial module is mostly driven by spatial loss, their optimization processes are fundamentally interdependent. Changes in one module can indirectly shift the parameter landscape of the other, ensuring a shared optimization of temporal and spatial patterns.

This architectural design supports module specialization—BERT focuses on temporal dependencies, while the attention module models spatial relationships—without direct interference. Yet both modules ultimately contribute to the network’s overall optimization, facilitating a more integrated representation of brain dynamics. Additionally, by treating these modules as partly separable, the framework allows for independent evaluation and targeted enhancements in each module, preserving their complementary roles in the final model.

As shown in Figure \ref{fig_procedure} B, the temporal and spatial modules operate on each frequency range, generating corresponding logits and attention matrices. To obtain the final prediction, the logits from the three frequency ranges are averaged and compared with the true label ($y_{true}$) using a task-specific loss function ($Loss_{task}$). In parallel, a spatial loss is computed to enforce distinct spatial connectivity patterns across frequency bands.
The composite loss function enforces both task performance and frequency-specific spatial patterns:

\begin{equation}\label{Total loss term}
Loss_{total} = Loss_{task}(y_{true}, \frac{\hat{y_{ultralow}} + \hat{y_{low}} + \hat{y_{high}}}{3}) + \lambda \cdot Loss_{spatial}
\end{equation}

where task-specific loss ensures predictive accuracy while spatial loss encourages distinct connectivity patterns across frequency bands.

In Equation \ref{Total loss term}, $Loss_{task}$ represents the cross-entropy loss for classification tasks or L1 loss for regression tasks. The $Loss_{spatial}$ term, as described in Equation \ref{Spatial loss term}, emphasizes the distinct spatial connectivity patterns inherent to each frequency band. $\lambda$ was determined by hyperparameter tuning. The chosen $\lambda$ value and the filter type were shown in Table \ref{tab:hyperparameters}.

 \begin{table}[!ht]
  \centering
  \begin{tabular}{lcc}
    \hline
    \textbf{Dataset} & \textbf{Filter Type} & \textbf{Spatial Factor $\lambda$} \\
    \hline
    \multicolumn{3}{l}{\textit{Schaefer atlas}} \\
    ABCD & Boxcar & 10 \\
    UKB & Boxcar & 1 \\
    ABIDE & Boxcar & 100 \\
    \hline
    \multicolumn{3}{l}{\textit{HCPMMP1 atlas}} \\
    ABCD & FIR & 10 \\
    UKB & Boxcar & 1 \\
    ABIDE & Boxcar & 100 \\
    \hline
  \end{tabular}
  \caption{Optimal filter types and $\lambda$ determined through hyperparameter tuning}
  \label{tab:hyperparameters}
\end{table}

\subsubsection{Parameter Sharing}
To enhance model performance, we adopted a multi-view learning approach, which is well-suited for frequency-divided timeseries data. We treat the ultra-low, low, and high frequency time series as three ‘views’ of the same underlying data, where different representations of the same underlying phenomenon are considered. Multi-view data enables the model to leverage shared information across views, allowing it to capture complementary aspects of the data.

For the temporal module, we utilized a parameter-sharing strategy. Given that temporal dynamics exhibited minimal differences across frequency bands (as indicated by similar autocorrelation structures), a single shared BERT encoder was used to process timeseries data from all frequency bands. This approach allows the model to exploit common temporal patterns across frequency bands, leading to a robust and generalized representation of temporal dynamics.

In contrast, the spatial module required a frequency-specific approach. Previous studies (\cite{shim2013frequency}) and our analysis have shown that spatial connectivity patterns differ significantly across frequency bands. Therefore, we implemented frequency-specific parameters for the spatial module. Self-attention-weighted connectivity was calculated independently for each frequency range, with each frequency band using its own set of parameters to capture the unique spatial relationships among ROIs. This design ensures that the model preserves the distinct connectivity characteristics of each frequency band while maintaining the interpretability of spatial attention.

This parameter-sharing strategy strikes a balance between generalization and specificity. Shared temporal parameters capture global patterns common across frequency bands, while frequency-specific spatial parameters emphasize the unique spatial relationships within each frequency band. By integrating this approach with task-specific and spatial loss functions (Equation \ref{Total loss term}), the model effectively captures the multi-level organization of frequency-divided timeseries data, facilitating the identification of relevant biomarkers across multiple scales.

\subsection{Temporal BERT Module} \label{methods_4_temporal_BERT}
\subsubsection{Data Encoding}
To capture temporal dependencies across regions of interest (ROIs), the frequency-divided time-series data were sequentially input into the BERT model. Each signal at a single time point was treated as a distinct token, with the token length corresponding to the total number of ROIs. This encoding strategy effectively represented both the temporal and spatial dynamics of the fMRI data, ensuring that these features were preserved for downstream analysis.

\subsubsection{Using Original Form of BERT}
FMRI BOLD signal data consists of long sequences, typically ranging from 280 to 464 tokens, necessitating a model that can handle such extended sequences. BERT, originally developed for natural language processing (NLP) tasks, is particularly well-suited for this purpose due to its ability to capture complex contextual relationships within sequences (\cite{devlin2018bert}). This makes BERT an ideal model for modeling the intricate temporal dependencies inherent in fMRI data.

In contrast, models like GPT are optimized for text generation tasks and do not emphasize inter-sequence contextual relationships, which are critical for our analysis. Unlike GPT’s causal masking—which is tailored to text generation and unidirectional context—BERT’s bidirectional attention allows it to capture long-range temporal dependencies from both past and future time points simultaneously, making it more suitable for modeling complex fMRI time series. Although some previous studies have used convolutional filters as a preprocessing step before entering signals into BERT (\cite{baevski2020wav2vec}), we chose to bypass this step. Although convolutional preprocessing can be computationally efficient, it may sacrifice the preservation of fine-grained temporal dynamics within the fMRI data, which could impact model performance (see Supplementary Materials \ref{Appendix_Ablations}). By directly inputting the time-series data into the original BERT structure, we aimed to retain the full complexity of the temporal dynamics, allowing the model to effectively capture both short-term and long-term dependencies.

\subsection{Spatial Attention Module} \label{methods_5_spatial_attention}
\subsubsection{Self-Attention-Weighted Connectivity}
Traditional methods of functional connectivity based on statistical correlations assume linear relationships and static connectivity (\cite{mohanty2020rethinking}). However, these approaches do not account for the nonlinear and time-varying nature of brain connectivity. To address this limitation, we employed a dynamic method that captures the complex, evolving interactions within resting-state functional networks. This method provides a more accurate representation of the diverse and temporal relationships among brain regions.

Given that spatial connectivity reflects the relationships between brain regions, we hypothesized that a self-attention mechanism, as used in transformer models (\cite{vaswani2017attention}), would be a suitable alternative. Self-attention mechanisms are designed to capture relationships between tokens (or regions, in our case), and the attention matrices they generate are diverse, reflecting the multifaceted nature of these relationships. While self-attention with multiple parallel attention heads is referred to as multi-head attention, our implementation specifically employed the multi-head attention mechanism to capture diverse representational information simultaneously. Multi-head attention mechanisms allow neural networks to focus on different aspects of input data simultaneously, similar to how different neural circuits in the brain can process the same sensory information through specialized pathways. In this approach, the input is projected into multiple representation spaces where separate attention computations occur in parallel, enabling the model to capture diverse relationships within the data. This parallel processing architecture has proven remarkably effective for complex tasks like language understanding, where different heads can specialize in detecting various linguistic patterns such as syntactic relationships, semantic connections, or long-range dependencies.
In our transformer model architecture, multiple attention heads are employed, with each head generating its own $N_{ROI} \times N_{ROI}$ attention matrix, where $N_{ROI}$ denotes number of ROIs. These attention matrices differ across heads as each head learns to capture different aspects of the relationships between ROIs. For example, one head might focus more strongly on frontal-temporal relationships while another might emphasize occipital-parietal connections. The final attention-based connectivity is then computed by averaging these head-specific attention matrices. 
To more effectively capture the complex dynamic relationships between ROIs compared to traditional correlation-based methods, we defined self-attention-weighted connectivity as the mean of the attention matrices (\textit{Attention matrix} $\in \mathbb{R}^{(N_{ROI} \times N_{ROI})}$), where each matrix represents an estimate of spatial connectivity between ROIs.

\subsubsection{Spatial Loss}
The three frequency-divided components, separated by frequencies ($f_1$, $f_2$), exhibit more distinct differences in connectivity patterns (\cite{shim2013frequency}). To highlight these differences, we introduced a spatial loss term designed to encourage different spatial connectivity patterns for each frequency range. Specifically, the loss term maximizes the sum of L1 losses between attention matrices derived from different frequency ranges (Equation \ref{Spatial loss term}). The use of the negative logarithm in the loss function magnifies smaller differences between frequency bands, ensuring that even subtle distinctions are emphasized during training. This approach enables the model to learn unique spatial connectivity patterns for each frequency band, reflecting the potentially distinct contributions of these bands to brain activity. Moreover, this spatial loss term works in conjunction with other loss functions in the model to strike a balance between generalization and specificity.

\begin{equation}\label{spatial-loss-high-low}
Loss_{high-low} = Loss_{L1}(attmat_{high}, attmat_{low})
\end{equation}
\begin{equation}\label{spatial-loss-high-ultralow}
Loss_{high-ultralow} = Loss_{L1}(attmat_{high}, attmat_{ultralow})
\end{equation}
\begin{equation}\label{spatial-loss-ultralow-low}
Loss_{ultralow-low} = Loss_{L1}(attmat_{ultralow}, attmat_{low})
\end{equation}
\begin{equation}\label{Spatial loss term}
Loss_{spatial} = -log(Loss_{high-low}+Loss_{high-ultralow}+Loss_{ultralow-low})
\end{equation}

The negative logarithm amplifies subtle connectivity differences, ensuring the model learns distinct spatial patterns even when frequency bands show similar overall connectivity structures.
We confirmed that this approach indeed derives neurobiologically plausible network properties (see Appendix \ref{Appendix_NWproperties}).

\subsection{Communicability-based Pretraining Strategy} \label{methods_7_pretraining}
\subsubsection{Masking Strategy and Pretraining Loss}
The goal of pretraining is to enable the model to learn the underlying structure of brain signal data in a self-supervised manner. To achieve this, we incorporated the concept of communicability into the pretraining process. A detailed definition of communicability used in this study is provided in Supplementary Materials \ref{Appendix_Pretraining_Communicability}.

High-communicability nodes act as critical hubs in brain networks, integrating information across distributed regions. By masking these influential nodes during pretraining, we force the model to learn their functional roles through contextual inference, similar to masked language modeling but tailored to brain network topology.

Our pretraining strategy draws inspiration from masking techniques commonly used in language models to capture contextual relationships and data patterns (\cite{devlin2018bert}). We hypothesized that masking nodes with high communicability would encourage the model to infer the activity of these influential nodes based on the surrounding network dynamics. This approach highlights the role of high-communicability hubs in shaping the overall structure of resting-state networks.

To validate this hypothesis, we compared three masking strategies:

\begin{itemize}
    \item Masking nodes with high communicability.
    \item Masking nodes with low communicability.
    \item Randomly masking regions of interest (ROIs).
\end{itemize}

Detailed model performance results (Supplementary Materials \ref{Appendix_Results}) show that masking high-communicability nodes led to the most effective learning of the data structure, resulting in superior performance in downstream tasks.

\begin{figure}[hbt!]
\centering
\includegraphics[width=0.8\textwidth]{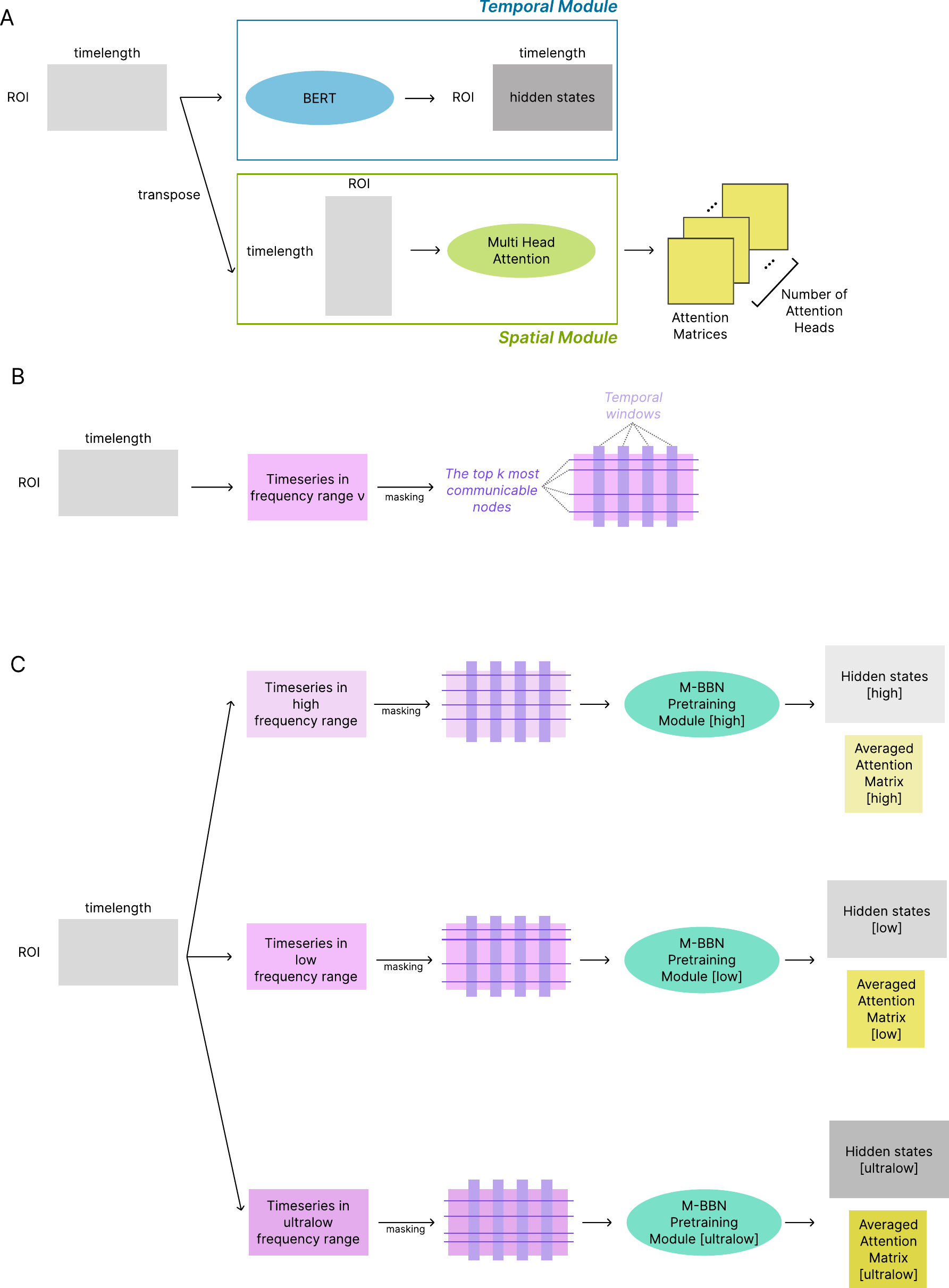}
\caption{\textbf{MBBN Pretraining Workflow.} 
\textbf{A.} Each frequency-divided timeseries is processed using the MBBN pretraining module, where the temporal module extracts hidden states using a BERT encoder, and the spatial module computes attention matrices via a multi-head attention mechanism. \textbf{B.} Masking strategies: spatial masking nullifies the time series data of the top $k$ nodes with the highest communicability scores, while temporal masking zeros out ROI signals within predefined time windows. \textbf{C.} Pretraining is conducted separately for high, low, and ultralow frequency bands. Parameter sharing is applied in the temporal module but not in the spatial module. Outputs include hidden states and averaged attention matrices for each frequency range.}
\label{fig_pretraining}
\end{figure}

Pretraining involved two complementary masking strategies: temporal masking and spatial masking (Figure \ref{fig_pretraining}). Spatial masking involved masking the timeseries data of the top $k$ nodes with the highest communicability scores, where $k$ is less than the total number of ROIs. Temporal masking, on the other hand, initialized all ROI signals to zero within predefined time windows. Both strategies were applied to each frequency band (ultralow, low, and high), denoted as $\nu$.

The masking loss ($Loss_{mask}$) was calculated as the sum of L1 losses between the original BOLD signals ($BOLD_{\nu}$) and their masked counterparts ($masked BOLD_{\nu}$):

\begin{equation}\label{mask loss term}
Loss_{mask} = \sum\limits_{\nu}Loss_{L1}(BOLD_{\nu}, masked BOLD_{\nu})
\end{equation}

The pretraining loss was the combination of the masking loss and the spatial loss (Equation \ref{Spatial loss term}), which enables the model to simultaneously learn node relationships and recover masked regions. $\lambda$ employed in the pretraining loss function was maintained at the same value as that used in the total loss calculation (Equation \ref{Total loss term}) during from-scratch training.

\begin{equation}\label{pretraining loss term}
Loss_{pretrain} = Loss_{mask} + \lambda \cdot Loss_{spatial}
\end{equation}

\subsubsection{Temporal Masking Optimization}
In temporal masking, rather than predicting a single time point, the model is tasked with predicting entire time windows. This strategy makes the task more challenging, encouraging the model to learn temporal dependencies more effectively. Masking was implemented by setting ROI signals to zero within predefined time windows.

To optimize temporal masking, we empirically tested various configurations by adjusting two parameters:

\begin{itemize}
    \item $N$: the length of the masked time window.
    \item $k$: the distance between consecutive masked windows.
\end{itemize}

The best performance in downstream tasks was achieved with $N=20$ and $k=1$, where $N$ ranged from 5 to 45 in multiples of 5, and $k$ was an integer between 1 and 5.

\subsubsection{Implementation Details and Dataset}
Pretraining was conducted using the UK Biobank dataset, with a sequence length of 464. During finetuning, padding was applied to align shorter sequences ($\tau$) to this fixed length, with padding added equally to both ends of the sequence ($(464 - \tau) / 2$). The outputs from the pretraining process, which include hidden states and attention matrices for each frequency band, were used to initialize the MBBN model for downstream tasks. Figure \ref{fig_pretraining} illustrates the pretraining workflow, including the temporal and spatial masking procedures. Temporal masking is represented by lavender rectangles, while spatial masking is depicted with purple lines.

\subsection{Interpretability} \label{methods_9_interpretability}

MBBN computes attention matrices between Regions of Interest (ROIs) across different frequency ranges. These matrices serve as estimates of functional connectivity, making it crucial to analyze individual connections to understand their contributions to phenotype classification. This analysis offers insights into the neural mechanisms underlying the phenotype of interest.

Traditional attribution methods are not directly applicable to this model, as the relationship between attention matrices and target values is indirect. To address this challenge, we developed a GradCAM (\cite{selvaraju2017grad})-inspired approach to identify frequency-specific connections driving psychiatric classification. This method highlights connections where gradient magnitude (importance) aligns with attention strength (activation), revealing clinically relevant biomarkers.

\begin{equation}\label{equation_heatmap}
Heatmap_{k} = \frac{\partial {Loss}_{spatial}}{\partial f_k(input_k)} \odot f_{k}(input_k)
\end{equation}

GradCAM typically highlights important regions by computing the gradient of the loss for the target class and weighting it by the output feature maps. Similarly, we compute heatmaps by multiplying the activation values (attention matrices) with their respective gradients, as described in Equation \ref{equation_heatmap}. This method highlights specific functional connectivity patterns that are relevant to mental disorders. In Equation \ref{equation_heatmap}, $f$ refers to the spatial attention module, ${Loss}_{spatial}$ corresponds to the loss term from Equation \ref{Spatial loss term}, $k$ represents the frequency range (high, low, ultralow), and $input_k$ denotes the fMRI BOLD data for frequency range $k$ ($Heatmap_k \in \mathbb{R}^{(N_{ROI} \times N_{ROI})}$).

To interpret the model's predictions, we selected the best-performing model based on the validation dataset and applied it to the test dataset. Heatmaps were generated for true positive subjects. Analysis for false positive and false negative subjects was conducted to validate the method.
For inter-group comparisons, we conducted t-tests to identify statistically significant connections ($p < 0.05$). For multiple comparison correction, we applied False Discovery Rate (FDR) (Benjamin-Hochberg method) to control for Type I errors. Then we calculated the effect size using Cohen's d (\cite{cohen2013statistical}). The sign of the t-statistic indicated which group had a higher attribution score for a given connection. For example, in the classification of ADHD using the ABCD dataset, a positive t-statistic suggested that a connection contributed more significantly to the Healthy Control (HC) group than to the ADHD group.

Statistically significant connections ($p < 0.05$) and regions were analyzed, but the number of connections was too large to handle effectively. To address this, we selected only a few top connections based on the absolute values of Cohen's d for further evaluation of their impact. An exhaustive list of all significant connections can be found in Supplementary Materials \ref{Appendix_Interpretability}.

\subsection{Performance Comparison with Baseline Models} \label{methods_10_other_models}
To evaluate the effectiveness of MBBN in learning spatiotemporal dynamics, we compared its performance with baseline models, including functional connectivity-based models (XGBoost (\cite{chen2015xgboost}), BNT (\cite{kan2022brain})) and temporal dynamics models (BolT (\cite{bedel2023bolt}), vanilla BERT (\cite{devlin2018bert})). These models were selected based on their relevance to brain dynamics interpretation, either through static functional connectivity or sequential processing of temporal data. While functional connectivity-based models primarily rely on static representations, temporal dynamics models capture sequential patterns but do not explicitly integrate spatial relationships. MBBN addresses this gap by combining both spatiotemporal characteristics using its frequency-dividing encoder and spatial attention module. Given that MBBN is an extension of vanilla BERT with additional architectural enhancements, a direct comparison between the two provides a robust evaluation of these modifications. While MBBN demonstrated superior performance across several downstream tasks, including the ABCD ADHD classification task, with significant improvements in accuracy and AUC scores, not all tasks showed a statistically significant advantage. Detailed performance comparison is described in Supplementary Materials \ref{Appendix_Results}.

To ensure a fair evaluation, we conducted statistical tests (paired t-tests) to compare the performance of MBBN with baseline models on each task. The p-values for these comparisons indicated that, although MBBN did not outperform all baseline models in every metric, the differences were not statistically significant ($p > 0.05$) in cases where baseline models showed slightly better performance. These results suggest that MBBN encodes information at a level comparable to or better than existing models while providing the additional advantage of capturing spatiotemporal dynamics, which baseline models are unable to incorporate.

\section{Discussion}\label{discussion}

This study presents MBBN, a novel deep learning framework that integrates biologically grounded principles with frequency-aware self-attention to model spatiotemporal brain dynamics. By leveraging scale-free and multifractal properties of neural signals, MBBN captures essential patterns of brain organization that are often overlooked by conventional transformer-based models.
Additionally, MBBN employs communicability-based masking and frequency-aware self-attention to capture biologically meaningful connectivity patterns across multiple frequency ranges, enabling accurate and interpretable predictions of cognitive and clinical outcomes.

Unlike previous approaches that rely on broadband or static representations (\cite{kan2022brain, bedel2023bolt}), MBBN applies a frequency decomposition strategy to encode neural oscillations as discrete features, enabling a richer characterization of dynamic connectivity patterns. This frequency-resolved approach is particularly critical given that neural communication is mediated by distinct oscillatory dynamics across frequency bands, with high-frequency oscillations supporting local processing (\cite{kucewicz2024high}) and low-frequency rhythms facilitating long-range integration (\cite{costa2024distinct}). Traditional broadband or narrowband analyses often obscure these nuanced patterns by averaging across spectrally diverse signals, potentially masking functionally specific effects. The MBBN framework overcomes these limitations by capturing such frequency-specific properties, allowing for a more precise and physiologically grounded understanding of brain network dynamics. Furthermore, our communicability-based (\cite{estrada2008communicability, estrada2012physics}) masking method ensures that topological structures within brain networks are embedded into learned representations, improving both robustness and interpretability.

Importantly, MBBN-derived connectivity patterns complement rather than replicate traditional functional connectivity. While moderate anticorrelation existed at the broadband level, frequency-specific analyses revealed unique network dynamics (Supplementary Table \ref{tab:correlation_analysis}), confirming that MBBN captures neurophysiologically distinct information critical for clinical interpretation.


The biological plausibility of MBBN is further supported by its incorporation of the scale-free property, characterized by a power-law exponent ($\beta$), which differentiates brain states and regions based on their temporal correlation profiles (\cite{he2010temporal, bullmore2001colored, fox2007spontaneous, he2011scale, mandelbrot1968fractional, eke2002fractal}). Variations in $\beta$ values are associated with functional specialization and pathological alterations (\cite{tolkunov2010power, maxim2005fractional, radulescu2012power, cha2016clinically}), underscoring the importance of integrating these intrinsic properties into computational models. By incorporating communicability-based masking, MBBN systematically accounts for graph-structured brain data, addressing the limitations of random masking strategies (\cite{tu2023rare, liu2024mask}) and enhancing understanding of network-level interactions (\cite{estrada2008communicability, harush2017dynamic}).

When applied to graph-structured functional brain data, random masking fails to consider node importance and the non-Euclidean nature of the data (\cite{tu2023rare, liu2024mask}), leading to uncertainties in reconstructing masked information. MBBN overcomes these limitations through a communicability-based masking strategy, which quantifies information flow across all possible network paths and emphasizes the role of high-communicability hubs in shaping network dynamics (\cite{estrada2008communicability, harush2017dynamic}). By selectively masking these critical hubs and reconstructing their activity through interactions with surrounding nodes, MBBN learns structural dependencies intrinsic to neural systems, enhancing both interpretability and robustness. This approach establishes MBBN as a foundational framework for advancing neuroimaging analysis.

Across diverse cohorts—including UK Biobank, ABCD, and ABIDE—MBBN consistently demonstrated strong predictive performance, capturing neural organizational principles across age groups. Unlike many existing models that are validated only on narrow demographics, MBBN successfully generalized across developmental stages, revealing its potential as a lifespan-inclusive neuroimaging tool.

Our frequency-specific analyses uncovered distinct connectivity disruptions in ADHD and ASD. In ADHD, we observed attenuated high-frequency connectivity in fronto-sensorimotor networks, which may underlie core symptoms such as inattention and executive dysfunction (\cite{burgess2013rostral, roth2004executive}). At lower frequencies, altered sensory and language processing circuits were evident (\cite{baker2018connectomic5, peterson2023neuroanatomy, ito2009somatosensory, conant2014speech}), and in the ultralow-frequency range, both hypo- and hyperconnectivity patterns suggested inefficient compensatory mechanisms, particularly involving the parahippocampal and cingulate regions (\cite{laflamme2021parahippocampal, bohbot1998spatial, skodzik2017long, steinberg2015developmental, bunford2015adhd}).

A striking finding was the frequency-dependent shift in Area OP2-3-VS (Opercular parts of Areas 2 and 3 and adjacent Ventral Somatosensory / Sylvian region) connectivity in ADHD. This region consistently emerged as a central node across different frequency bands, suggesting it may serve as a dynamic hub that potentially coordinates neural communication based on oscillatory states. Rather than subserving a fixed function, Area OP2-3-VS appears to assume context-dependent roles, possibly regulating cross-frequency information flow or facilitating inter-network switching—mechanisms that could be critical for coherent cognitive function. In ADHD, disruptions in these putative regulatory processes may contribute to frequency-specific connectivity alterations and the disorder's heterogeneous symptomatology, underscoring the potential importance of spectral dynamics in characterizing functional hubs.

In ASD, high-frequency hubs involving orbitofrontal and somatosensory circuits were prominent, consistent with known deficits in emotional and social processing (\cite{rolls2020orbitofrontal, case2015systematic, cibralic2019systematic, velikonja2019patterns}). At low frequencies, the posterior frontopolar cortex emerged as a key integrative hub; disruptions here may impair the coordination of memory, decision-making, and multisensory integration essential for goal-directed behavior (\cite{baker2018connectomic2, law2023frontopolar, ferrucci2025reward, polonyiova2024roots}). Interestingly, ultralow-frequency analyses revealed heightened connectivity between the temporo-parietal junction and prefrontal regions, possibly reflecting compensatory cognitive strategies or atypical resource allocation during social cognition tasks (\cite{de2014anatomo, seghatol2020hyperconnectivity, supekar2013brain, chita2016social}).

Collectively, these results illustrate that MBBN not only outperforms prior models in predictive capacity but also provides interpretable biomarkers aligned with known neurodevelopmental patterns. By aligning transformer-based modeling with core neuroscientific principles—particularly the relevance of neural oscillations—MBBN offers a scalable, biologically plausible framework for understanding individualized brain dynamics.

While MBBN offers notable methodological advancements, some limitations warrant consideration. First, our frequency decomposition employs theory-based spectral bands. Although grounded in established neurophysiological knowledge, this approach may not capture finer-grained oscillatory phenomena that deviate from these canonical bands. Future work could explore more adaptive approaches to move beyond theory-based bands, for example, by employing more granular spectral tiling (e.g., via wavelet packet decomposition \cite{coifman1992entropy}) or by developing a fully learnable filter-bank layer within the MBBN architecture. This might allow for end-to-end optimization of frequency bands for specific downstream tasks. Second, the communicability-based masking, while effective in promoting network-aware learning, is computationally demanding. To enhance scalability for ever-larger datasets, future iterations could investigate computationally efficient approximations of communicability, such as those based on polynomial expansions or random walks (\cite{estrada2012physics}). Third, the assessment of test-retest reliability for the derived frequency-specific connectivity patterns would be crucial. Establishing such reliability is not only fundamental to evaluating the robustness of the MBBN model itself but also essential for validating the temporal stability of these patterns. This is particularly important for applications involving the tracking of neurodevelopmental trajectories or disease progression, where consistent and dependable measures are paramount (\cite{noble2021guide}).

Future extensions of MBBN could incorporate multimodal data (e.g., structural MRI, diffusion MRI) and expand into task-based paradigms to examine how frequency-specific connectivity evolves under cognitive demands. Additionally, exploring cross-frequency coupling may offer deeper insights into hierarchical brain dynamics beyond individual frequency bands. MBBN establishes a new standard for interpretable, frequency-resolved brain modeling. By integrating neurophysiological principles with transformer architectures, it offers both predictive power and mechanistic insights into brain function, laying the foundation for precision neuroscience applications.

\backmatter

\section*{Data and Code Availability}

The code is available on Github (https://github.com/Transconnectome/MBBN). The UKB, ABCD, and ABIDE datasets are publicly available to researchers.

\section*{Author Contributions}

Sangyoon Bae conceptualized the research, implemented the core algorithms, and conducted all experiments. Junbeom Kwon contributed to code optimization and manuscript preparation.

\section*{Declarations}

Nothing to declare.

\section*{Funding}

This work was supported by the National Research Foundation of Korea(NRF) grant funded by the Korea government(MSIT) (No. 2021R1C1C1006503, RS-2023-00266787, RS-2023-00265406, RS-2024-00421268), by Creative-Pioneering Researchers Program through Seoul National University(No. 200-20240057), by Semi-Supervised Learning Research Grant by SAMSUNG(No.A0426-20220118), by Identify the network of brain preparation steps for concentration Research Grant by LooxidLabs(No.339-20230001), by Institute of Information \& communications Technology Planning \& Evaluation (IITP) grant funded by the Korea government(MSIT) [NO.RS-2021-II211343, Artificial Intelligence Graduate School Program (Seoul National University)] by the MSIT(Ministry of Science, ICT), Korea, under the Global Research Support Program in the Digital Field program(RS-2024-00421268) supervised by the IITP(Institute for Information \& Communications Technology Planning \& Evaluation), by the National Supercomputing Center with supercomputing resources including technical support(KSC-2023-CRE-0568) and
by the Ministry of Education of the Republic of Korea and the National Research Foundation of Korea (NRF-2021S1A3A2A02090597), and by Artificial intelligence industrial convergence cluster development project funded by the Ministry of Science and ICT(MSIT, Korea) \& Gwangju Metropolitan City.


\begin{appendices}
\section{Pretraining} \label{Appendix_Pretraining}
\subsection{Communicability} \label{Appendix_Pretraining_Communicability}
Communicability encompasses not only the shortest paths but also other types of walks between nodes $p$ and $q$ within a network. A network can be represented as a graph $G = (V, E)$, where $V$ represents nodes and $E$ represents the edges connecting the nodes. Let us denote the number of nodes as $n$ and the number of edges as $m$. The adjacency matrix of the graph is $A(G) = A$, where $A_{ij}$ takes a value of 1 if nodes $i$ and $j$ are connected and 0 otherwise. The communicability between nodes $p$ and $q$ in the network, denoted as $G_{pq}$, incorporates eigenvalues and eigenvectors of the adjacency matrix. Let $\lambda_{1} \geq \lambda_{2} \geq \cdots \geq \lambda_{n}$ be the eigenvalues in non-increasing order, and let $\phi_{j}(p)$ represent the $p$-th element of the $j$-th orthonormal eigenvector corresponding to eigenvalue $\lambda_{j}$. Then, communicability is calculated as follows:

\begin{equation}\label{communicability}
G_{pq} = \sum_{j=1}^{\infty} \phi_{j}(p)\phi_{j}(q)e^{\lambda_{j}}
\end{equation}

Communicability provides a measure of how efficiently information flows between nodes by considering not just direct connections but also indirect pathways weighted by their importance. This property makes communicability particularly suitable for identifying high-influence nodes in brain networks, which are then used in our pretraining masking strategy.

\subsection{Effective Pretraining: Finding the Optimal Number of Epochs}
\label{Appendix_Pretraining_MBBN_weightwatcher}

\begin{figure}[hbt!]
\setcounter{figure}{0}
\renewcommand{\figurename}{Supplementary Figure}
\centering
\includegraphics[width=0.7\textwidth]{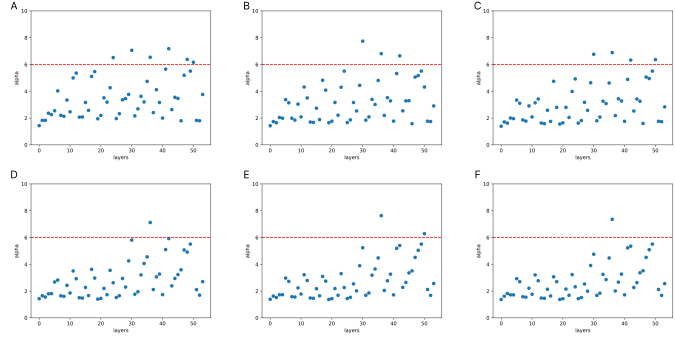}
\caption{\textbf{Alpha Value Changes by Pretraining Epochs (HCPMMP1 Atlas, Epochs: 100-600)}\\ }\label{fig_setting_pretraining_epoch_HCPMMP1_high_comm_100_to_600}
\end{figure}

Determining the optimal number of epochs for pretraining is crucial for creating an effective pretraining strategy. We utilized weightwatcher (\cite{martin2021predicting}), a diagnostic tool based on traditional statistical learning theory and statistical mechanics, to identify well-trained models. Weightwatcher analyzes the norms of the weight matrix $W$ and parameters of power-law (PL) fit on the eigenvalues of $W$ to assess training quality without requiring data.

According to the Heavy-Tailed Self-Regularization (HT-SR) Theory, lower PL exponent ($\alpha$) values indicate stronger implicit self-regularization, leading to better generalization. Consistently low $\alpha$ values across layers allow effective correlation flow, enabling the gradual transformation of data correlations through each layer. As shown in previous work, layers with $\alpha$ values below 6 are deemed well-trained, while those exceeding 6 are considered poorly trained.

In the following figures, the x-axis represents the layers, and the y-axis represents the $\alpha$ values. The red dotted line marks the threshold of $\alpha = 6$.
The number of layers with $\alpha$ values above 6 converges to a single layer after epoch 500 (Supplementary Figure \ref{fig_setting_pretraining_epoch_HCPMMP1_high_comm_100_to_600}). We hypothesize that the epoch at which the pattern of layers with $\alpha$ values exceeding 6 stabilizes marks the optimal pretraining epoch. In our monitoring process, we utilized WeightWatcher to analyze the pretrained model at intervals of 100 epochs. We identified the critical epoch $n$ at which the number of layer with $\alpha > 6$ converges. Subsequently, we conducted a series of fine-tuning experiments at 10-epoch intervals between epochs $n-100$ and $n$ to empirically determine the optimal epoch for model performance. In HCPMMP1 atlas, $n$ was 600.

Selecting these epochs ensures that most layers in the model achieve sufficient self-regularization, improving overall generalization performance during fine-tuning.

\section{Interpretation of Network Properties}
\label{Appendix_NWproperties}
\begin{table}[ht]
\centering
\setcounter{table}{0}
\renewcommand{\tablename}{Supplementary Table}
\begin{tabular}{l|cc|ccc|ccc}
\hline
\multirow{2}{*}{\textbf{Measure}} & \multicolumn{2}{c|}{\textbf{Modularity}} & \multicolumn{6}{c}{\textbf{Small-worldness Parameters}} \\
\cmidrule{2-9}
 & \textbf{FC} & \textbf{Attention} & \multicolumn{3}{c|}{\textbf{FC}} & \multicolumn{3}{c}{\textbf{Attention}} \\
 &  &  & $\gamma$ & $\lambda$ & $\sigma$ & $\gamma$ & $\lambda$ & $\sigma$ \\
\hline
High frequency & 0.0278 & 0.0576 & 4.7683 & 1.4239 & 3.3487 & 9.3960 & 0.9220 & 10.1913 \\
Low frequency & 0.0251 & 0.0701 & 4.8803 & 1.4140 & 3.4514 & 5.6272 & 0.9953 & 5.6539 \\
Ultralow frequency & 0.0350 & 0.0991 & 4.8600 & 1.4794 & 3.2850 & 8.3929 & 0.9229 & 9.0941 \\
Raw (Non-dividing) & 0.0630 & $<0.0001$ & 7.7804 & 1.6143 & 4.8197 & 5.9527 & 0.9254 & 6.4329 \\
\hline
\end{tabular}
\caption{Comparison of Network Properties between FC and Attention Matrix}
\label{tab:network_properties}
\end{table}

Our analysis of network properties provides strong evidence that frequency-specific attention matrices capture meaningful connectivity patterns that differ from traditional functional connectivity (FC).

\begin{itemize}
    \item \textbf{Modularity} quantifies the strength of division of a network into modules or communities. It measures how well a network can be decomposed into distinct subgroups, with higher values indicating more distinct community structures where connections within communities are denser than connections between communities. Formally, it is defined as the fraction of edges that fall within given communities minus the expected fraction if edges were distributed at random while preserving node degrees.
    
    \item \textbf{Small-worldness parameters} characterize the topological properties of a network that exhibits both high clustering and short path lengths:
    
    \begin{itemize}
        \item \textbf{Gamma ($\gamma$)}: The clustering coefficient ratio comparing the observed network to a random network with the same number of nodes and edges. Higher values indicate greater local connectivity and clustering, reflecting more efficient local information processing.
        
        \item \textbf{Lambda ($\lambda$)}: The characteristic path length ratio comparing the observed network to a random network. Values closer to 1 indicate efficient global integration and information transfer across the network.
        
        \item \textbf{Sigma ($\sigma$)}: The small-worldness index calculated as $\gamma/\lambda$. Values significantly greater than 1 indicate that the network exhibits small-world properties, characterized by high local clustering and relatively short path lengths. Small-world networks are considered optimal for information processing, combining efficient local processing with global integration.
    \end{itemize}
\end{itemize}

These network metrics provide quantitative measures for assessing the topological organization of brain networks and are widely used in neuroimaging studies to characterize functional and structural connectivity patterns. Several key observations can be made from Supplementary Table~\ref{tab:network_properties}:

\begin{enumerate}
    \item \textbf{Frequency-dependent modularity}: The attention matrices exhibit distinct modularity and small-worldness patterns across frequency bands (high, low, and ultralow frequencies). A clear trend of increasing modularity with decreasing frequency (0.0576 → 0.0701 → 0.0991) is observed. This pattern aligns with findings from \cite{kajimura2023frequency}, which demonstrated that resting-state BOLD signals display frequency-specific network architecture, where networks that are finely subdivided in lower frequency bands become integrated into fewer networks in higher frequency bands rather than reconfigured. Such evidence supports the observation that modularity values are higher in lower frequency bands, consistent with our results.
    
    \item \textbf{Contrasting patterns with FC}: While FC exhibits its highest modularity in the raw (unfiltered) condition (0.0630), attention-based connectivity shows essentially no modularity in the raw condition ($<0.0001$). This striking difference suggests that our attention mechanism captures frequency-specific network organization that is lost when analyzing the full frequency spectrum together.
    
    \item \textbf{Small-worldness characteristics}: The small-worldness parameters ($\gamma$, $\lambda$, and $\sigma$) reveal distinct network architectures between FC and attention-based connectivity. Notably, attention matrices consistently show higher clustering ($\gamma$) and small-worldness ($\sigma$) values across frequency bands compared to FC, while maintaining more optimal path length properties ($\lambda$ closer to 1). This indicates that attention-based connectivity may better preserve the small-world architecture that is characteristic of brain networks.
    
    \item \textbf{Frequency-specific information}: Raw data without frequency decomposition fails to exhibit meaningful modularity in the attention matrix, whereas frequency-decomposed data reveals distinct modular structures. These findings suggest effective coupling between the spatial module and frequency-dividing strategy, enabling the extraction of neurobiologically relevant patterns. This observation underscores the importance of frequency-specific analysis when employing attention-based connectivity measures to characterize brain network organization.
\end{enumerate}

These findings support our claim that frequency-specific attention matrices represent a valid and interpretable new form of connectivity measure, capturing network properties that complement traditional FC measures.

\section{Correlation Analysis Between FC and Attention-Based Connectivity}

To further investigate the relationship between functional connectivity (FC) and our proposed attention-based connectivity measure, we conducted correlation analyses across different frequency bands. Spearman correlation measures the monotonic relationship between two variables by evaluating the rank order association, making it robust to outliers and nonlinear relationships. Kendall tau is a similar non-parametric measure that assesses the ordinal association between two variables by analyzing the number of concordant and discordant pairs, providing a measure of agreement in rankings. The results are presented in Table~\ref{tab:correlation_analysis}.

\begin{table}[ht]
\centering
\renewcommand{\tablename}{Supplementary Table}
\begin{tabular}{l|cc}
\hline
\textbf{Frequency Band} & \textbf{Spearman Correlation} & \textbf{Kendall Tau} \\
\hline
Raw (Non-dividing) & -0.6081 & -0.4367 \\
High frequency & -0.0952 & -0.0700 \\
Low frequency & -0.1459 & -0.1072 \\
Ultralow frequency & -0.1973 & -0.1451 \\
\hline
\end{tabular}
\caption{Correlation Analysis Between FC and Attention Matrix}
\label{tab:correlation_analysis}
\end{table}

The correlation analysis reveals several important insights:

\begin{enumerate}
    \item \textbf{Complementary information}: The strong negative correlation between FC and attention-based connectivity in the raw condition (Spearman $\rho = -0.6081$, Kendall $\tau = -0.4367$) indicates that these two measures capture fundamentally different aspects of brain connectivity. This negative correlation suggests that our attention-based connectivity is not merely replicating traditional FC but providing complementary information.
    
    \item \textbf{Frequency-specific patterns}: When examining frequency-specific bands, the correlations become considerably weaker (ranging from approximately -0.10 to -0.20), suggesting that each frequency band captures unique connectivity patterns.
    
    \item \textbf{Methodological validity}: The consistent negative correlations across all frequency bands, albeit weaker in the frequency-specific analyses, indicate that our attention mechanism systematically extracts connectivity information that differs from traditional FC. This suggests that attention-based connectivity represents a valid new form of connectivity measure with its interpretable meaning.
\end{enumerate}

These correlation findings, combined with the network property analysis, provide strong evidence that our frequency-specific attention-based connectivity measure captures meaningful and interpretable connectivity patterns that complement traditional FC analysis. The method does not simply maximize differences between frequencies but extracts biologically relevant connectivity information specific to each frequency band.

\section{Ablations} \label{Appendix_Ablations}
\subsection{The performance of the model when provided with timeseries data of a single frequency range.}
\begin{table}[ht]
\label{tab_single_frequency}
\renewcommand{\tablename}{Supplementary Table}
\begin{tabular*}{\textwidth}{@{\extracolsep\fill}cccccc}
\toprule
& \multicolumn{2}{@{}c@{}}{ABCD sex classification} & \multicolumn{2}{@{}c@{}}{UKB sex classification} \\
\cmidrule{2-3}\cmidrule{4-5}
Data & AUROC ↑ & Accuracy ↑ & AUROC ↑ & Accuracy ↑ \\
\midrule
Z-scored original signal & 0.901$\pm$0.003 & 0.814$\pm$0.0031 & 0.942$\pm$0.014 & 0.876$\pm$0.006 \\
High Frequency  & 0.888$\pm$0.007 & 0.797$\pm$0.011 & 0.933$\pm$0.007 & 0.857$\pm$0.003 \\
Low frequency   & 0.884$\pm$0.005 & 0.784$\pm$0.011 & 0.936$\pm$0.004 & 0.850$\pm$0.015 \\
Ultra-low Frequency  & 0.870$\pm$0.010 & 0.779$\pm$0.013 & 0.935$\pm$0.003 & 0.862$\pm$0.007 \\
Divided Frequencies  & \textbf{0.916$\pm$0.004} & \textbf{0.844$\pm$0.006} & \textbf{0.980$\pm$0.001} & \textbf{0.920$\pm$0.001} \\
\bottomrule
\end{tabular*}
\caption{Results of BERT models trained with single frequency range in the timeseries data extracted by HCP-MMP1 atlas}
\end{table}

The core innovation of our model lies in encoding signals through frequency division and merging them within the modeling process. To identify the source of performance improvement, we conducted ablation studies focusing on single-frequency range data. 

First, we examined whether frequency-divided timeseries data alone could capture the dominant signals essential for phenotype prediction. If successful, single-frequency data would outperform other inputs. However, as shown in Table~\ref{tab_single_frequency}, single-frequency data consistently underperformed compared to the z-scored original signal. This finding suggests that the model does not rely solely on dominant signals from specific ranges but rather leverages contributions from all three frequency ranges.

Second, we analyzed the impact of merging signals from multiple frequency ranges. Results indicate that integrating signals within the model, through parameter sharing across multi-views, significantly enhances performance compared to using frequency-separated data alone. This demonstrates the importance of combining signals at the model level rather than isolating them at the data preprocessing stage.

\subsection{Treating fMRI signal as a sequence of discrete words.}
\begin{table}[ht]
\label{tab_ablation_conv}
\renewcommand{\tablename}{Supplementary Table}
\begin{tabular*}{\textwidth}{@{\extracolsep\fill}cccccc}
\toprule
& \multicolumn{2}{@{}c@{}}{ABCD sex classification} & \multicolumn{2}{@{}c@{}}{UKB sex classification} \\
\cmidrule{2-3}\cmidrule{4-5}
Model & AUROC ↑ & Accuracy ↑ & AUROC ↑ & Accuracy ↑ \\
\midrule
MBBN with conv & 0.825$\pm$0.009 & 0.726$\pm$0.009 & 0.947$\pm$0.003 & 0.870$\pm$0.002 \\
MBBN without conv  & \textbf{0.916$\pm$0.004} & \textbf{0.844$\pm$0.006} & \textbf{0.980$\pm$0.001} & \textbf{0.920$\pm$0.001} \\
\bottomrule
\end{tabular*}
\caption{Results of BERT models with a convolutional layer before BERT model} 
\end{table}

Determining whether fMRI data should be treated as continuous variables, akin to speech, or discrete sequences, similar to natural language, is a critical design choice. Given the high sample rate of fMRI signals, we hypothesized that they resemble written sentences more closely than speech signals.

To evaluate this, we introduced convolutional layers before feeding the input into the model, inspired by speech recognition frameworks such as Wav2Vec \cite{baevski2020wav2vec}. Table~\ref{tab_ablation_conv} reveals that treating fMRI signals as discrete word sequences yields superior performance compared to processing them as continuous variables. This result highlights the effectiveness of viewing fMRI signals as structured sequences analogous to natural language.

\subsection{Frequency-dividing strategies.}
\begin{table}[ht]
\label{tab_frequency_dividing_strategy}
\renewcommand{\tablename}{Supplementary Table}
\begin{tabular*}{\textwidth}{@{\extracolsep\fill}cccccc}
\toprule
& \multicolumn{2}{@{}c@{}}{ABCD sex classification} & \multicolumn{2}{@{}c@{}}{UKB sex classification} \\
\cmidrule{2-3}\cmidrule{4-5}
Dividing Strategy & AUROC ↑ & Accuracy ↑ & AUROC ↑ & Accuracy ↑ \\
\midrule
No division & 0.901$\pm$0.003 & 0.814$\pm$0.003 & 0.942$\pm$0.014 & 0.876$\pm$0.006 \\
Random 2 Freq & 0.903$\pm$0.004 & 0.818$\pm$0.004 & 0.969$\pm$0.003 & 0.911$\pm$0.003 \\
Lorentizan-based 2 Freq  & 0.903$\pm$0.006 & 0.805$\pm$0.010 & 0.973$\pm$0.007 & 0.918$\pm$0.005 \\
Random 3 Freq & 0.900$\pm$0.006 & 0.803$\pm$0.008 & 0.972$\pm$0.002 & 0.918$\pm$0.003 \\
Ours  & \textbf{0.916$\pm$0.004} & \textbf{0.844$\pm$0.006} & \textbf{0.980$\pm$0.001} & \textbf{0.920$\pm$0.001} \\
\bottomrule
\end{tabular*}
\caption{Results of BERT variants trained with different frequency-dividing strategies in the timeseries data extracted by HCP-MMP1 atlas}
\end{table}

To optimize the frequency-dividing strategy, we compared methods using Lorentzian and spline multifractal equations with alternative approaches. Results are presented in Table~\ref{tab_frequency_dividing_strategy}.

First, we tested random division by selecting two random cutoff frequencies from a normal distribution. Compared to this baseline (Random 3 Freq), our method (Ours) improved AUROC by 1.11\%, demonstrating the effectiveness of Lorentzian and spline multifractal equations.

Second, we evaluated division into two frequency ranges instead of three. Using Lorentzian-based boundaries (Lorentzian-based 2 Freq) led to reasonable performance but was outperformed by our three-range strategy, with a 1.44\% improvement in AUROC. This emphasizes the importance of high-frequency signals in predicting phenotypes.

In conclusion, dividing frequencies using Lorentzian and spline multifractal equations into three ranges yields superior performance, highlighting the biological significance of high-frequency components.

\subsection{Spatial module and temporal module.}
\begin{table}[ht]
\renewcommand{\tablename}{Supplementary Table}
\begin{tabular*}{\textwidth}{@{\extracolsep\fill}cccccc}
\toprule
& \multicolumn{2}{@{}c@{}}{ABCD sex classification} & \multicolumn{2}{@{}c@{}}{UKB sex classification} \\
Module & AUROC ↑& Accuracy ↑ & AUROC ↑ & Accuracy ↑ \\
\midrule
Spatial & 0.553$\pm$0.007 & 0.540$\pm$0.018 & 0.856$\pm$0.004 & 0.778$\pm$0.004 \\
Temporal & 0.909$\pm$0.007 & 0.831$\pm$0.011 & 0.964$\pm$0.001 & 0.915$\pm$0.001 \\
Spatiotemporal (MBBN) & \textbf{0.916$\pm$0.004} & \textbf{0.844$\pm$0.006} & \textbf{0.980$\pm$0.001} & \textbf{0.921$\pm$0.006} \\
\bottomrule
\end{tabular*}
\caption{Results of spatial and temporal modules of MBBN in the timeseries data extracted by HCP-MMP1 atlas}
\label{tab_spatial_temporal_module}
\end{table}

Table \ref{tab_spatial_temporal_module} illustrates the performance of individual spatial and temporal modules compared to the full MBBN spatiotemporal architecture on ABCD/UKB sex classification tasks. The results demonstrate that spatial and temporal modules interact synergistically through joint parameter updates during training. When used independently, both modules show diminished performance compared to the integrated spatiotemporal approach (MBBN). The spatial module achieves moderate performance, while the temporal module performs considerably worse on the ABCD dataset.
These findings suggest that the neural dynamics captured in fMRI data require both spatial and temporal processing to maximize predictive power. The superior performance of the integrated MBBN architecture implies that allowing these modules to interact through shared loss functions and simultaneous parameter updates creates representations that better capture the complex spatiotemporal patterns in brain activity. This emphasizes the importance of modeling both spatial and temporal dimensions of neural data rather than treating them as independent components.

\subsection{Number of parameters and FLOPs}\label{Appendix_Ablation_num_of_parameters}
\begin{table}[ht]
\renewcommand{\tablename}{Supplementary Table}
\begin{tabular*}{\textwidth}{@{\extracolsep\fill}cccccc}
\toprule
Module & AUROC ↑ & Accuracy ↑ & FLOPs & Number of Params \\
\midrule
Vanilla BERT (FLOPs-matched) & 0.886$\pm$0.000 & 0.807$\pm$0.008 & 70,489,026,464 & 29,894,149 \\
Vanilla BERT (Params-matched) & 0.884$\pm$0.006 & 0.778$\pm$0.027 & 23,497,736,096 & 10,050,757 \\
MBBN & 0.916$\pm$0.004 & 0.844$\pm$0.006 & 70,528,086,432 & 11,140,695 \\
\bottomrule
\end{tabular*}
\caption{Performance of vanilla BERT with parameter count or FLOPs matched to MBBN levels. Experiments were conducted on the ABCD dataset, and task was sex classification.}
\label{tab_ablation_FLOPs_num_parameters}
\end{table}

Table \ref{tab_ablation_FLOPs_num_parameters} demonstrates a fair comparison between MBBN and vanilla BERT models with matched computational complexity. Two variants of vanilla BERT were created: one matching MBBN's FLOPs (70.5B vs 70.5B) and another matching MBBN's parameter count (10.1M vs 11.1M). Despite having similar computational resources or parameter counts, MBBN significantly outperforms both Vanilla BERT variants in AUROC and accuracy.
These results strongly suggest that MBBN's superior performance cannot be attributed simply to increased model capacity or computational complexity. Rather, the architectural innovations in MBBN appear to be responsible for its enhanced effectiveness compared to vanilla BERT baselines.

\section{Comparison with other models} \label{Appendix_comparison_with_other_models}
\subsection{Performance comparison} \label{Appendix_Results}
\begin{table}[hbt!]
\renewcommand{\tablename}{Supplementary Table}
\begin{tabular*}{\textwidth}{@{\extracolsep\fill}cccccc}
\toprule
& \multicolumn{2}{@{}c@{}}{ABCD sex} & \multicolumn{2}{@{}c@{}}{UKB sex}  \\
\cmidrule{2-3}\cmidrule{4-5}
Models (data form) & AUROC ↑ & Accuracy ↑ & AUROC ↑ & Accuracy ↑ \\
\midrule
XGBOOST (CC) & 0.785$\pm$0.005 & 0.709$\pm$0.003 & 0.859$\pm$003 & 0.768$\pm$0.005 \\
BNT (CC) & 0.872$\pm$0.005 & 0.795$\pm$0.007 & 0.950$\pm$0.004 & 0.864$\pm$0.017  \\
BolT (TD) & 0.886$\pm$0.002& 0.795$\pm$0.002 & 0.950$\pm$0.000 & 0.878$\pm$0.001  \\
vanilla BERT (TD) & 0.864$\pm$0.004 & 0.777$\pm$0.006 & 0.942$\pm$0.014 & 0.876$\pm$0.006 \\
MBBN (TD) & \textbf{0.913$\pm$0.002} & \textbf{0.823$\pm$0.009} & \textbf{0.980$\pm$0.001} & \textbf{0.921$\pm$0.006} \\
\bottomrule
\end{tabular*}
\caption{Results of models trained from scratch predicting biological phenotypes. Atlas is HCP MMP 1. CC denotes correlation-based connectivity and TD denotes temporal dynamics.}
\label{tab_from_scratch_biological_HCPMMP1}
\end{table}

\begin{table}[hbt!]
\renewcommand{\tablename}{Supplementary Table}
\begin{tabular*}{\textwidth}{@{\extracolsep\fill}cccccc}
\toprule
& \multicolumn{2}{@{}c@{}}{ABCD sex} & \multicolumn{2}{@{}c@{}}{UKB sex}  \\
\cmidrule{2-3}\cmidrule{4-5}
Models & AUROC ↑ & Accuracy ↑ & AUROC ↑ & Accuracy ↑ \\
\midrule
XGBOOST (CC)  & 0.718$\pm$0.004 & 0.659$\pm$0.003 & 0.905$\pm$0.004 & 0.884$\pm$0.002 \\
BNT (CC)  & 0.920$\pm$0.006 & 0.839$\pm$0.012 & 0.982$\pm$0.001 & 0.934$\pm$0.092  \\
BolT (TD)  & 0.911$\pm$0.002& 0.826$\pm$0.003 & 0.982$\pm$0.000 & 0.932$\pm$0.001  \\
vanilla BERT (TD)  & 0.901$\pm$0.007 & 0.818$\pm$0.017 & 0.980$\pm$0.001 & 0.925$\pm$0.005 \\
MBBN (TD)  & \textbf{0.918$\pm$0.002} &\textbf{0.831$\pm$0.009} & \textbf{0.987$\pm$0.001} & \textbf{0.934$\pm$0.008} \\
\bottomrule
\end{tabular*}
\caption{Results of models trained from scratch predicting biological phenotypes. Atlas is Schaefer 400, 17 Networks. CC denotes correlation-based connectivity and TD denotes temporal dynamics.}
\label{tab_from_scratch_biological_Schaefer400}
\end{table}

\begin{table}[hbt!]
\renewcommand{\tablename}{Supplementary Table}
\begin{tabular*}{\textwidth}{@{\extracolsep\fill}cccccc}
\toprule
& \multicolumn{2}{@{}c@{}}{UKB depression} & \multicolumn{2}{@{}c@{}}{ABCD depression} \\
\cmidrule{2-3}\cmidrule{4-5}
Models (data form) & MAE ↓ & MSE ↓ & AUROC ↑ & Accuracy ↑ \\
\midrule
XGBOOST (CC)  & 2.417$\pm$0.025 & 12.342$\pm$0.151 & 0.548$\pm$0.026 & 0.829$\pm$0.000 \\
BNT (CC)  & 2.276$\pm$0.015 & 13.989$\pm$0.682 & 0.546$\pm$0.015 & 0.847$\pm$0.006 \\
BolT (TD)  & 2.217$\pm$0.022& 12.729$\pm$0.290 & 0.588$\pm$0.006 & 0.847$\pm$0.001\\
vanilla BERT (TD)  & 2.194$\pm$0.039 & 12.751$\pm$0.780 & 0.546$\pm$0.015 & 0.818$\pm$0.051  \\
MBBN (TD)  & \textbf{2.210 $\pm$0.041} &\textbf{13.205$\pm$0.956} & \textbf{0.590$\pm$0.013} & \textbf{0.750$\pm$0.008} \\
\bottomrule
\end{tabular*}
\caption{Results of models trained from scratch predicting clinical outcomes, atlas is HCP-MMP1. CC denotes correlation-based connectivity and TD denotes temporal dynamics.}
\label{tab_from_scratch_clinical_HCP_MMP1}
\end{table}

\begin{table}[hbt!]
\renewcommand{\tablename}{Supplementary Table}
\begin{tabular*}{\textwidth}{@{\extracolsep\fill}cccccc}
\toprule
& \multicolumn{2}{c}{UKB depression} & \multicolumn{2}{c}{ABCD depression} \\
\cmidrule{2-3}\cmidrule{4-5}
Models (data form) & MAE ↓ & MSE ↓ & AUROC ↑ & Accuracy ↑ \\
\midrule
XGBOOST (CC) & 2.408$\pm$0.002 & 11.942$\pm$0.052 & 0.581$\pm$0.012 & 0.833$\pm$0.003 \\
BNT (CC) & 2.329$\pm$0.012 & 15.179$\pm$2.730 & 0.579$\pm$0.018 & 0.842$\pm$0.010 \\
BolT (TD) & 2.201$\pm$0.004& 13.044$\pm$0.301 & 0.585$\pm$0.011 & 0.844$\pm$0.005\\
vanilla BERT (TD) & 2.230$\pm$0.006 & 13.359$\pm$0.479 & 0.586$\pm$0.040 & 0.846$\pm$0.004 \\
MBBN (TD) & \textbf{2.249 $\pm$0.005} &\textbf{14.486$\pm$0.282} & \textbf{0.641$\pm$0.018} & \textbf{0.777$\pm$0.013}\\
\bottomrule
\end{tabular*}
\caption{Results of models trained from scratch predicting clinical outcomes, atlas is Schaefer 400, 17 Networks. CC denotes correlation-based connectivity, and TD denotes temporal dynamics.}
\label{tab_from_scratch_clinical_Schaefer_400}
\end{table}

\begin{table}[hbt!]
\renewcommand{\tablename}{Supplementary Table}
\begin{tabular*}{\textwidth}{@{\extracolsep\fill}cccccc}
\toprule
& \multicolumn{2}{@{}c@{}}{UKB fluid intelligence} & \multicolumn{2}{@{}c@{}}{ABCD fluid intelligence} \\
\cmidrule{2-3}\cmidrule{4-5}
Models (data form) & MAE ↓ & MSE ↓ & MAE ↓ & MSE ↓ \\
\midrule
XGBOOST (CC)  & 1.609$\pm$0.009 & 4.078$\pm$0.086 &  0.720$\pm$0.003 & 0.827$\pm$0.007 \\
BNT (CC)  & 1.581$\pm$0.023 & 3.993$\pm$0.516 & 0.763$\pm$0.020 & 1.061$\pm$0.093 \\
BolT (TD)  & 1.589$\pm$0.002& 3.973$\pm$0.006 & 0.728$\pm$0.004 & 0.845$\pm$0.010\\
vanilla BERT (TD)  & 1.597$\pm$0.015 & 4.081$\pm$0.094 & 0.699$\pm$0.016 & 0.788$\pm$0.036  \\
MBBN (TD)  & \textbf{1.630 $\pm$0.006} &\textbf{4.225$\pm$0.125} & \textbf{0.694$\pm$0.018} & \textbf{0.774$\pm$0.041} \\
\bottomrule
\end{tabular*}
\caption{Results of models trained from scratch predicting cognitive outcomes, atlas is HCPMMP1. CC denotes correlation-based connectivity and TD denotes temporal dynamics.}
\label{tab_from_scratch_cognitive_HCP_MMP1}
\end{table}

\begin{table}[hbt!]
\renewcommand{\tablename}{Supplementary Table}
\begin{tabular*}{\textwidth}{@{\extracolsep\fill}cccccc}
\toprule
& \multicolumn{2}{c}{UKB fluid intelligence} & \multicolumn{2}{c}{ABCD fluid intelligence} \\
\cmidrule{2-3}\cmidrule{4-5}
Models (data form) & MAE ↓ & MSE ↓ & MAE ↓ & MSE ↓ \\
\midrule
XGBOOST (CC) & 1.600$\pm$0.002 & 4.003$\pm$0.025 & 0.779$\pm$0.008 & 0.970$\pm$0.011 \\
BNT (CC) & 1.552$\pm$0.019 & 4.459$\pm$2.127 & 0.781$\pm$0.020 & 0.845$\pm$0.086 \\
BolT (TD) & 1.576$\pm$0.006& 3.919$\pm$0.013 & 0.718$\pm$0.005 & 0.822$\pm$0.016\\
vanilla BERT (TD) & 1.585$\pm$0.020 & 3.985$\pm$0.105 & 0.721$\pm$0.018 & 0.823$\pm$0.039  \\
MBBN (TD) & \textbf{1.628$\pm$0.051} &\textbf{4.205$\pm$0.282} & \textbf{0.719$\pm$0.013} & \textbf{0.828$\pm$0.028} \\
\bottomrule
\end{tabular*}
\caption{Results of models trained from scratch predicting cognitive outcomes, atlas is Schaefer 400, 17 Networks. CC denotes correlation-based connectivity, and TD denotes temporal dynamics.}
\label{tab_from_scratch_cognitive_Schaefer_400}
\end{table}

\begin{table}[hbt!]
\renewcommand{\tablename}{Supplementary Table}
\begin{tabular*}{\textwidth}{@{\extracolsep\fill}ccccc}
\toprule
& \multicolumn{2}{@{}c@{}}{ABCD ADHD classification} & \multicolumn{2}{@{}c@{}}{ABIDE ASD classification}\\
\cmidrule{2-3} \cmidrule{4-5}
Models (data form) & AUROC ↑ & Accuracy ↑ & AUROC ↑ & Accuracy ↑\\
\midrule
XGBOOST (CC)  & 0.598$\pm$0.010 & 0.570$\pm$0.005 & 0.573$\pm$0.053 & 0.546$\pm$0.046 \\
BNT (CC)  & 0.634$\pm$0.008 & 0.606$\pm$0.014 & 0.691$\pm$0.168 & 0.617$\pm$0.062 \\
BolT (TD)  & 0.603$\pm$0.005 & 0.592$\pm$0.010 & 0.632$\pm$0.005 & 0.566$\pm$0.003 \\
Vanilla BERT (TD)  & 0.609$\pm$0.020 & 0.592$\pm$0.015 & 0.644$\pm$0.089 & 0.625$\pm$0.051 \\
MBBN [from scratch] (TD)  & 0.633$\pm$0.009 & 0.593$\pm$0.015 & 0.780$\pm$0.012 & 0.657$\pm$0.038 \\
MBBN [random] (TD) & 0.603$\pm$0.032 & 0.589$\pm$0.026 & 0.601$\pm$0.180 & 0.534$\pm$0.082 \\
MBBN [low] (TD) & 0.615$\pm$0.009 & 0.602$\pm$0.011 & 0.583$\pm$0.084 & 0.547$\pm$0.098 \\
MBBN [high] (TD) & \textbf{0.645$\pm$0.011} &\textbf{0.601$\pm$0.015} & \textbf{0.810$\pm$0.065} & \textbf{0.717$\pm$0.019} \\
\bottomrule
\end{tabular*}
\caption{Finetuning results of pre-trained BERT models with different pretraining loss (HCP-MMP1 atlas). CC denotes correlation-based connectivity, TD denotes temporal dynamics, and MBBN represents our proposed model. The [random], [low], and [high] variants indicate different pretraining strategies based on masking node communicability.}
\label{tab_from_pretrain_HCP_MMP1}
\end{table}

\begin{table}[hbt!]
\renewcommand{\tablename}{Supplementary Table}
\begin{tabular*}{\textwidth}{@{\extracolsep\fill}ccccc}
\toprule
& \multicolumn{2}{@{}c@{}}{ABCD ADHD classification} & \multicolumn{2}{@{}c@{}}{ABIDE ASD classification}\\
\cmidrule{2-3} \cmidrule{4-5}
Models (data form) & AUROC ↑ & Accuracy ↑ & AUROC ↑ & Accuracy ↑\\
\midrule
XGBOOST (CC)  & 0.594$\pm$0.008 & 0.564$\pm$0.006 & 0.660$\pm$0.023 & 0.629$\pm$0.030 \\
BNT (CC)  & 0.620$\pm$0.018 & 0.599$\pm$0.018 & 0.715$\pm$0.045 & 0.652$\pm$0.077 \\
BolT (TD)  & 0.599$\pm$0.010 & 0.589$\pm$0.011 & 0.608$\pm$0.028 & 0.578$\pm$0.027 \\
Vanilla BERT (TD)  & 0.604$\pm$0.017 & 0.600$\pm$0.008 & 0.664$\pm$0.148 & 0.667$\pm$0.059 \\
MBBN [from scratch] (TD)  & 0.614$\pm$0.020 & 0.610$\pm$0.011 & 0.800$\pm$0.067 & 0.683$\pm$0.062 \\
MBBN [random] (TD) & 0.596$\pm$0.014 & 0.593$\pm$0.032 & 0.696$\pm$0.052 & 0.563$\pm$0.088\\
MBBN [low] (TD) & 0.584$\pm$0.033 & 0.574$\pm$0.034 & 0.500$\pm$0.051 & 0.479$\pm$0.059 \\
MBBN [high] (TD) & \textbf{0.618$\pm$0.019} &\textbf{0.580$\pm$0.035} & \textbf{0.822$\pm$0.019} & \textbf{0.768$\pm$0.051} \\
\bottomrule
\end{tabular*}
\caption{Finetuning results of pre-trained BERT models with different pretraining loss (Schaefer 400 atlas). CC denotes correlation-based connectivity, TD denotes temporal dynamics, and MBBN represents our proposed model. The [random], [low], and [high] variants indicate different pretraining strategies based on masking node communicability.}
\label{tab_from_pretrain_Schaefer}
\end{table}

The tables above (Supplementary Table \ref{tab_from_scratch_biological_HCPMMP1}, \ref{tab_from_scratch_biological_Schaefer400}, \ref{tab_from_scratch_clinical_HCP_MMP1}, \ref{tab_from_scratch_clinical_Schaefer_400}, \ref{tab_from_scratch_cognitive_HCP_MMP1}, \ref{tab_from_scratch_cognitive_Schaefer_400}) present a comparison of various models, atlases, and metrics for different downstream tasks trained from scratch. In these tables, (CC) refers to correlation-based connectivity, and (TD) represents temporal dynamics. Specifically, (CC) denotes Pearson correlation computed on temporal dynamics. Across most tasks, MBBN (highlighted in color) generally outperforms or shows comparable performance to other baseline models.

For instance, in the ABCD depression classification task, the dataset’s imbalance in the depression-HC ratio naturally leads to high regular accuracy. However, MBBN demonstrates robustness against this imbalance by achieving more balanced performance metrics compared to other models (Table \ref{tab_from_scratch_clinical_HCP_MMP1}, \ref{tab_from_scratch_clinical_Schaefer_400}).

Tables \ref{tab_from_pretrain_HCP_MMP1} and \ref{tab_from_pretrain_Schaefer} present a comparison across models and metrics for various fine-tuning tasks. Here, MBBN is compared under different pretraining strategies:
\begin{itemize}
    \item MBBN [from scratch]: Trained without pretraining.
    \item MBBN [low]: Pretrained with masking loss that targets nodes with low communicability.
    \item MBBN [random]: Pretrained with random masking loss.
    \item MBBN [high]: Pretrained with masking loss that targets nodes with high communicability.
\end{itemize}

MBBN [high], highlighted in color, consistently outperforms or matches the performance of other baseline models and pretraining strategies across various tasks. This result emphasizes the effectiveness of the high-communicability masking strategy in enhancing model generalization and fine-tuning performance.

\subsection{FLOPs and number of parameters}\label{Appendix_comparison_FLOPs_parameter_size}

The parameter count difference between our proposed MBBN model and the baseline models (Table \ref{tab_FLOPs_num_parameters}) stems from our architectural design, which processes data through three separate streams. This design choice was intentional to capture complex patterns through specialized processing pathways. While this results in a higher parameter count, our ablation studies (Appendix \ref{Appendix_Ablation_num_of_parameters}) demonstrate that the performance improvements arise primarily from the model's architecture rather than simply from increased capacity.

\begin{table}[hbt!]
\renewcommand{\tablename}{Supplementary Table}
\begin{tabular*}{\textwidth}{@{\extracolsep\fill}ccc}
\toprule
Models (data form) & Number of parameters & FLOPs \\
\midrule
XGBOOST (CC)  & 6004 & 400 \\
BNT (CC)  & 11,678,810 & 13,084,233,536  \\
BolT (TD)  & 20,697,661 & 50,032,700,640 \\
Vanilla BERT (TD)  & 22,280,137 & 69,848,980,256 \\
MBBN  (TD)  & 23,370,075 & 209,581,818,912 \\
\bottomrule
\end{tabular*}
\caption{The table shows the number of parameters and FLOPs of baseline models used in this study and MBBN (our model).}
\label{tab_FLOPs_num_parameters}
\end{table}

\section{Interpretability}\label{Appendix_Interpretability}
In the following tables, two regions involved in connectivity (based on HCP-MMP1) are presented, along with whether the connectivity shows higher scores in the disorder group or the healthy control (HC) group. Additionally, corrected p-values and Cohen's d scores were rounded to three decimal places. Due to space limitations, only the top 100 connectivity pairs are displayed.

\renewcommand{\tablename}{Supplementary Table}
\begin{landscape}
\setlength{\LTcapwidth}{1.0\textwidth}
\begin{longtable}{@{}p{6cm}p{6cm}ccc@{}}
\caption{Top connectivity features in high-frequency range of ABCD ADHD prediction. \label{tab:connectivity_high_frequency_ABCD_ADHD}}\\
\toprule
\textbf{Region 1} & \textbf{Region 2} & \textbf{Corrected p-value} & \textbf{Disorder/HC} & \textbf{Cohen's d} \\ \midrule
\endfirsthead

\multicolumn{5}{c}%
{{\tablename\ \thetable{} -- continued from previous page}} \\
\toprule
\textbf{Region 1} & \textbf{Region 2} & \textbf{Disorder/HC} & \textbf{Corrected p-value} & \textbf{Cohen's d} \\ \midrule
\endhead

\midrule \multicolumn{5}{r}{{Continued on next page}} \\ \midrule
\endfoot

\bottomrule
\endlastfoot

Area OP4-PV R & Area 10v R & HC & $<0.001$ & 0.564 \\
Primary Motor Cortex R & Area 10v R & HC & $<0.001$ & 0.561 \\
Primary Motor Cortex L & Area 10v R & HC & $<0.001$ & 0.56 \\
Medial Belt Complex R & Area 10v R & HC & $<0.001$ & 0.559 \\
Area OP1-SII R & Area 10v R & HC & $<0.001$ & 0.559 \\
Area PF Opercular R & Area 10v R & HC & $<0.001$ & 0.558 \\
Area 43 L & Area 10v R & HC & $<0.001$ & 0.557 \\
Area a24 R & Ventral Area 6 L & HC & $<0.001$ & 0.554 \\
Superior Temporal Visual Area R & Area 10v R & HC & $<0.001$ & 0.554 \\
Area Posterior 24 prime R & Area 10v R & HC & $<0.001$ & 0.554 \\
Area 1 R & Area 10v R & HC & $<0.001$ & 0.553 \\
PreSubiculum R & Area 10v R & HC & $<0.001$ & 0.551 \\
Anterior IntraParietal Area R & Area 10v R & HC & $<0.001$ & 0.55 \\
Area PFt R & Area 10v R & HC & $<0.001$ & 0.549 \\
Area posterior 24 L & Ventral Area 6 L & HC & $<0.001$ & 0.548 \\
Ventral Area 6 L & Area 10v R & HC & $<0.001$ & 0.547 \\
Area 2 R & Area 10v R & HC & $<0.001$ & 0.546 \\
Anterior Ventral Insular Area L & Area 10v R & HC & $<0.001$ & 0.545 \\
Area PFcm R & Area 10v R & HC & $<0.001$ & 0.544 \\
Frontal Opercular Area 3 L & Area 10v R & HC & $<0.001$ & 0.543 \\
Ventral IntraParietal Complex L & Area 10v R & HC & $<0.001$ & 0.541 \\
Area PFm Complex R & Area 10v R & HC & $<0.001$ & 0.536 \\
VentroMedial Visual Area 1 R & Area 10v R & HC & $<0.001$ & 0.536 \\
Area 7PC R & Area 10v R & HC & $<0.001$ & 0.536 \\
Area V3B R & Area 10v R & HC & $<0.001$ & 0.535 \\
Area posterior 24 L & Area 10v R & HC & $<0.001$ & 0.535 \\
RetroInsular Cortex R & Area 10v R & HC & $<0.001$ & 0.535 \\
Area TE1 anterior L & Ventral Area 6 L & HC & $<0.001$ & 0.534 \\
Ventral Area 24d R & Area 10v R & HC & $<0.001$ & 0.534 \\
Area 25 R & Ventral Area 6 L & HC & $<0.001$ & 0.534 \\
Area p32 prime R & Area 10v R & HC & $<0.001$ & 0.534 \\
Insular Granular Complex R & Area 10v R & HC & $<0.001$ & 0.534 \\
Area 43 R & Area 10v R & HC & $<0.001$ & 0.531 \\
Area 1 L & Area 10v R & HC & $<0.001$ & 0.53 \\
Area TA2 L & Area 10v R & HC & $<0.001$ & 0.53 \\
Primary Sensory Cortex R & Area 10v R & HC & $<0.001$ & 0.53 \\
Area PGi R & Area 10v R & HC & $<0.001$ & 0.53 \\
Para-Insular Area L & Area 10v R & HC & $<0.001$ & 0.529 \\
Frontal Eye Fields L & Area 10v R & HC & $<0.001$ & 0.529 \\
Area 46 R & Area 10v R & HC & $<0.001$ & 0.528 \\
Area OP2-3-VS R & Area 10v R & HC & $<0.001$ & 0.527 \\
Area 55b R & Area 10v R & HC & $<0.001$ & 0.527 \\
Area dorsal 32 R & Ventral Area 6 L & HC & $<0.001$ & 0.527 \\
VentroMedial Visual Area 2 R & Area 10v R & HC & $<0.001$ & 0.526 \\
Dorsal Transitional Visual Area R & Area 10v R & HC & $<0.001$ & 0.526 \\
Area dorsal 32 L & Area 11l R & HC & $<0.001$ & 0.525 \\
Frontal Opercular Area 2 R & Area 10v R & HC & $<0.001$ & 0.525 \\
Area OP2-3-VS L & Area 10v R & HC & $<0.001$ & 0.524 \\
Area 10r R & Ventral Area 6 L & HC & $<0.001$ & 0.524 \\
Area 10v L & Ventral Area 6 L & HC & $<0.001$ & 0.524 \\
Frontal Opercular Area 4 L & Ventral Area 6 L & HC & $<0.001$ & 0.524 \\
Rostral Area 6 L & Area 10v R & HC & $<0.001$ & 0.523 \\
Frontal Opercular Area 1 R & Area 10v R & HC & $<0.001$ & 0.523 \\
Area PF Complex R & Area 10v R & HC & $<0.001$ & 0.522 \\
Area STSv anterior L & Ventral Area 6 L & HC & $<0.001$ & 0.521 \\
Area s32 L & Ventral Area 6 L & HC & $<0.001$ & 0.52 \\
Area TemporoParietoOccipital Junction 1 L & Area 10v R & HC & $<0.001$ & 0.52 \\
Frontal Opercular Area 4 L & Area 10v R & HC & $<0.001$ & 0.52 \\
Area 13l R & Area 10v R & HC & $<0.001$ & 0.52 \\
Area 3a R & Area 10v R & HC & $<0.001$ & 0.519 \\
Area PGi R & Primary Motor Cortex R & HC & $<0.001$ & 0.518 \\
Dorsal Area 24d R & Area 10v R & HC & $<0.001$ & 0.518 \\
Area dorsal 32 L & Ventral Area 6 L & HC & $<0.001$ & 0.518 \\
Area Posterior 24 prime L & Area 10v R & HC & $<0.001$ & 0.518 \\
Medial Belt Complex R & Primary Motor Cortex R & HC & $<0.001$ & 0.518 \\
Area posterior 24 R & Ventral Area 6 L & HC & $<0.001$ & 0.517 \\
Area PF Opercular L & Area 10v R & HC & $<0.001$ & 0.517 \\
Area IntraParietal 2 R & Area 10v R & HC & $<0.001$ & 0.517 \\
Area 47m R & Area 10v R & HC & $<0.001$ & 0.517 \\
Area 25 L & Ventral Area 6 L & HC & $<0.001$ & 0.517 \\
RetroSplenial Complex R & Area 10v R & HC & $<0.001$ & 0.516 \\
Parieto-Occipital Sulcus Area 1 L & Area 10v R & HC & $<0.001$ & 0.516 \\
Sixth Visual Area R & Area 10v R & HC & $<0.001$ & 0.516 \\
Area Frontal Opercular 5 L & Area 11l R & HC & $<0.001$ & 0.516 \\
Orbital Frontal Complex L & Ventral Area 6 L & HC & $<0.001$ & 0.515 \\
ParaHippocampal Area 3 R & Area 10v R & HC & $<0.001$ & 0.515 \\
Area p32 R & Ventral Area 6 L & HC & $<0.001$ & 0.514 \\
Area 10d L & Ventral Area 6 L & HC & $<0.001$ & 0.514 \\
Area a24 L & Ventral Area 6 L & HC & $<0.001$ & 0.514 \\
Area 43 L & Ventral Area 6 L & HC & $<0.001$ & 0.514 \\
Frontal Opercular Area 3 R & Area 10v R & HC & $<0.001$ & 0.513 \\
Supplementary and Cingulate Eye Field R & Area 10v R & HC & $<0.001$ & 0.513 \\
Dorsal area 6 L & Area 10v R & HC & $<0.001$ & 0.513 \\
Area PFt L & Area 10v R & HC & $<0.001$ & 0.513 \\
Area V6A R & Area 10v R & HC & $<0.001$ & 0.513 \\
Insular Granular Complex L & Area 10v R & HC & $<0.001$ & 0.513 \\
Area PFcm L & Area 10v R & HC & $<0.001$ & 0.513 \\
Anterior 24 prime R & Area 10v R & HC & $<0.001$ & 0.513 \\
Anterior Ventral Insular Area L & Ventral Area 6 L & HC & $<0.001$ & 0.513 \\
ParaHippocampal Area 1 R & Area 10v R & HC & $<0.001$ & 0.512 \\
PeriSylvian Language Area L & Area 10v R & HC & $<0.001$ & 0.511 \\
RetroSplenial Complex R & Area Posterior Insular 1 L & HC & $<0.001$ & 0.511 \\
Area 9-46d R & Area 10v R & HC & $<0.001$ & 0.511 \\
Area posterior 24 L & Area 11l R & HC & $<0.001$ & 0.511 \\
VentroMedial Visual Area 3 R & Area 10v R & HC & $<0.001$ & 0.511 \\
Area 47s L & Area 11l R & HC & $<0.001$ & 0.511 \\
Parieto-Occipital Sulcus Area 2 R & Area 10v R & HC & $<0.001$ & 0.51 \\
Area TG dorsal L & Area 10v R & HC & $<0.001$ & 0.51 \\
Area 2 L & Area 10v R & HC & $<0.001$ & 0.51 \\
Supplementary and Cingulate Eye Field L & Area 10v R & HC & $<0.001$ & 0.51 \\

\end{longtable}
\end{landscape}

\renewcommand{\tablename}{Supplementary Table}
\begin{landscape}
\setlength{\LTcapwidth}{1.0\textwidth}
\begin{longtable}{@{}p{6cm}p{6cm}ccc@{}}
\caption{Top connectivity features in low frequency range of ABCD ADHD prediction\label{tab:connectivity_low_frequency_ABCD_ADHD}}\\
\toprule
\textbf{Region 1} & \textbf{Region 2} & \textbf{Disorder/HC} & \textbf{Corrected p-value}  & \textbf{Cohen's d} \\ \midrule
\endfirsthead

\multicolumn{5}{c}%
{{\tablename\ \thetable{} -- continued from previous page}} \\
\toprule
\textbf{Region 1} & \textbf{Region 2} & \textbf{Disorder/HC} & \textbf{Corrected p-value}  & \textbf{Cohen's d} \\ \midrule
\endhead

\midrule \multicolumn{5}{r}{{Continued on next page}} \\ \midrule
\endfoot

\bottomrule
\endlastfoot

PeriSylvian Language Area L & Area OP2-3-VS R & HC & $<0.001$ & 0.441 \\
Dorsal Area 24d L & Area OP2-3-VS R & HC & $<0.001$ & 0.441 \\
Area 31pd L & Auditory 4 Complex R & HC & $<0.001$ & 0.44 \\
Area STSv posterior R & Auditory 4 Complex R & HC & $<0.001$ & 0.44 \\
Area STSv posterior R & Area OP2-3-VS R & HC & $<0.001$ & 0.439 \\
Dorsal area 6 R & Area OP2-3-VS R & HC & $<0.001$ & 0.435 \\
Area TE1 Middle L & Frontal Opercular Area 2 R & HC & $<0.001$ & 0.435 \\
Area TE1 anterior R & Auditory 4 Complex R & HC & $<0.001$ & 0.434 \\
Area STSd anterior L & Area OP2-3-VS R & HC & $<0.001$ & 0.433 \\
Auditory 5 Complex R & Area OP2-3-VS R & HC & $<0.001$ & 0.431 \\
Superior Temporal Visual Area L & Insular Granular Complex L & HC & $<0.001$ & 0.43 \\
RetroInsular Cortex L & Area OP2-3-VS R & HC & $<0.001$ & 0.429 \\
Area TE1 anterior R & Area OP2-3-VS R & HC & $<0.001$ & 0.429 \\
Area STSd posterior L & Area OP2-3-VS R & HC & $<0.001$ & 0.428 \\
Area 23d L & Frontal Opercular Area 2 R & HC & $<0.001$ & 0.424 \\
Area TG dorsal L & Insular Granular Complex L & HC & $<0.001$ & 0.423 \\
Sixth Visual Area R & Area OP2-3-VS R & HC & $<0.001$ & 0.423 \\
PeriSylvian Language Area L & Insular Granular Complex L & HC & $<0.001$ & 0.423 \\
PeriSylvian Language Area L & Frontal Opercular Area 2 R & HC & $<0.001$ & 0.422 \\
Entorhinal Cortex R & Area OP2-3-VS R & HC & $<0.001$ & 0.422 \\
PeriSylvian Language Area L & Auditory 4 Complex R & HC & $<0.001$ & 0.422 \\
Area STSv anterior L & Area OP2-3-VS R & HC & $<0.001$ & 0.421 \\
Area s32 L & Frontal Opercular Area 2 R & HC & $<0.001$ & 0.421 \\
Area s32 L & Area OP2-3-VS R & HC & $<0.001$ & 0.421 \\
Area TE1 Middle L & Area OP2-3-VS R & HC & $<0.001$ & 0.42 \\
Area 8Av L & Insular Granular Complex L & HC & $<0.001$ & 0.42 \\
Area STSd anterior R & Area OP2-3-VS R & HC & $<0.001$ & 0.418 \\
Area PHT R & Area OP2-3-VS R & HC & $<0.001$ & 0.417 \\
VentroMedial Visual Area 3 L & Insular Granular Complex L & HC & $<0.001$ & 0.417 \\
Area V6A L & Area OP2-3-VS R & HC & $<0.001$ & 0.417 \\
Area 47l (47 lateral) R & Area OP2-3-VS R & HC & $<0.001$ & 0.416 \\
Dorsal area 6 L & Area OP2-3-VS R & HC & $<0.001$ & 0.415 \\
Area 8Ad L & Insular Granular Complex L & HC & $<0.001$ & 0.414 \\
Area 8C L & Insular Granular Complex L & HC & $<0.001$ & 0.414 \\
Area TE1 anterior L & Auditory 4 Complex R & HC & $<0.001$ & 0.413 \\
Area V6A L & Frontal Opercular Area 2 R & HC & $<0.001$ & 0.413 \\
Area TG dorsal L & Auditory 4 Complex R & HC & $<0.001$ & 0.413 \\
Auditory 5 Complex L & Area OP2-3-VS R & HC & $<0.001$ & 0.412 \\
Area 10d R & Frontal Opercular Area 2 R & HC & $<0.001$ & 0.412 \\
Area 47s R & Frontal Opercular Area 2 R & HC & $<0.001$ & 0.411 \\
Area TE1 Middle L & Auditory 4 Complex R & HC & $<0.001$ & 0.411 \\
Area TE1 posterior L & Insular Granular Complex L & HC & $<0.001$ & 0.411 \\
Hippocampus R & Auditory 4 Complex R & HC & $<0.001$ & 0.411 \\
Area TE1 Middle L & Insular Granular Complex L & HC & $<0.001$ & 0.41 \\
Area STSd posterior L & Auditory 4 Complex R & HC & $<0.001$ & 0.41 \\
Primary Auditory Cortex R & Auditory 4 Complex R & HC & $<0.001$ & 0.41 \\
Dorsal Area 24d L & Auditory 4 Complex R & HC & $<0.001$ & 0.41 \\
Area dorsal 32 L & Area OP2-3-VS R & HC & $<0.001$ & 0.41 \\
Area TE1 posterior R & Area OP2-3-VS R & HC & $<0.001$ & 0.409 \\
Dorsal area 6 R & Insular Granular Complex L & HC & $<0.001$ & 0.409 \\
Primary Motor Cortex R & Auditory 4 Complex R & HC & $<0.001$ & 0.409 \\
Area TemporoParietoOccipital Junction 1 R & Frontal Opercular Area 2 R & HC & $<0.001$ & 0.409 \\
Area TE1 anterior R & Frontal Opercular Area 2 R & HC & $<0.001$ & 0.409 \\
Area TE1 Middle R & Area OP2-3-VS R & HC & $<0.001$ & 0.408 \\
Dorsal area 6 R & Frontal Opercular Area 2 R & HC & $<0.001$ & 0.408 \\
Lateral Belt Complex R & Auditory 4 Complex R & HC & $<0.001$ & 0.408 \\
Area 31pd L & Insular Granular Complex L & HC & $<0.001$ & 0.408 \\
Superior 6-8 Transitional Area L & Insular Granular Complex L & HC & $<0.001$ & 0.408 \\
Lateral Belt Complex R & Area OP2-3-VS R & HC & $<0.001$ & 0.407 \\
Area TG dorsal R & Auditory 4 Complex R & HC & $<0.001$ & 0.407 \\
Area 31pd L & Area OP2-3-VS R & HC & $<0.001$ & 0.407 \\
Area dorsal 32 L & Frontal Opercular Area 2 R & HC & $<0.001$ & 0.407 \\
Area 10d R & Area OP2-3-VS R & HC & $<0.001$ & 0.407 \\
Area STSd anterior L & Insular Granular Complex L & HC & $<0.001$ & 0.406 \\
Area 8B Lateral L & Insular Granular Complex L & HC & $<0.001$ & 0.405 \\
Area 31pd R & Auditory 4 Complex R & HC & $<0.001$ & 0.405 \\
Area 8Ad L & Area OP2-3-VS R & HC & $<0.001$ & 0.405 \\
Pirform Cortex L & Area OP2-3-VS R & HC & $<0.001$ & 0.405 \\
Frontal Opercular Area 1 R & Area OP2-3-VS R & HC & $<0.001$ & 0.405 \\
Area TemporoParietoOccipital Junction 1 R & Auditory 4 Complex R & HC & $<0.001$ & 0.405 \\
Dorsal area 6 R & Medial Belt Complex R & HC & $<0.001$ & 0.405 \\
Area 8B Lateral L & Area OP2-3-VS R & HC & $<0.001$ & 0.404 \\
Area PGs L & Area OP2-3-VS R & HC & $<0.001$ & 0.404 \\
Area TemporoParietoOccipital Junction 1 R & Area OP2-3-VS R & HC & $<0.001$ & 0.404 \\
Area 6mp R & Auditory 4 Complex R & HC & $<0.001$ & 0.403 \\
Area IFJa R & Auditory 4 Complex R & HC & $<0.001$ & 0.403 \\
Area 45 L & Insular Granular Complex L & HC & $<0.001$ & 0.403 \\
Area 31pd R & Insular Granular Complex L & HC & $<0.001$ & 0.403 \\
Ventral IntraParietal Complex R & Area OP2-3-VS R & HC & $<0.001$ & 0.403 \\
Auditory 5 Complex L & Insular Granular Complex L & HC & $<0.001$ & 0.403 \\
Auditory 4 Complex L & Area OP2-3-VS R & HC & $<0.001$ & 0.403 \\
Area PFcm R & Auditory 4 Complex R & HC & $<0.001$ & 0.403 \\
Area 6 anterior L & Area OP2-3-VS R & HC & $<0.001$ & 0.403 \\
Area dorsal 32 L & Insular Granular Complex L & HC & $<0.001$ & 0.403 \\
Area 10d L & Area OP2-3-VS R & HC & $<0.001$ & 0.403 \\
Area 9 Posterior L & Insular Granular Complex L & HC & $<0.001$ & 0.402 \\
Middle Temporal Area R & Area OP2-3-VS R & HC & $<0.001$ & 0.402 \\
Area 8Ad L & Frontal Opercular Area 2 R & HC & $<0.001$ & 0.402 \\
Area TE1 Middle L & Posterior Insular Area 2 R & HC & $<0.001$ & 0.401 \\
Area STSd anterior R & Auditory 4 Complex R & HC & $<0.001$ & 0.401 \\
Area TG dorsal R & Frontal Opercular Area 2 R & HC & $<0.001$ & 0.401 \\
Area TE1 posterior R & Frontal Opercular Area 2 R & HC & $<0.001$ & 0.401 \\
Area 7m R & Auditory 4 Complex R & HC & $<0.001$ & 0.401 \\
Entorhinal Cortex R & Auditory 4 Complex R & HC & $<0.001$ & 0.401 \\
Area 47l (47 lateral) L & Area OP2-3-VS R & HC & $<0.001$ & 0.4 \\
PreCuneus Visual Area L & Insular Granular Complex L & HC & $<0.001$ & 0.4 \\
Area STGa R & Area OP2-3-VS R & HC & $<0.001$ & 0.4 \\
Superior Temporal Visual Area L & Area OP2-3-VS R & HC & $<0.001$ & 0.4 \\
Area TG dorsal L & Area OP2-3-VS R & HC & $<0.001$ & 0.4 \\
Medial Belt Complex R & Area OP2-3-VS R & HC & $<0.001$ & 0.399 \\

\end{longtable}
\end{landscape}

\renewcommand{\tablename}{Supplementary Table}
\begin{landscape}
\setlength{\LTcapwidth}{1.0\textwidth}
\begin{longtable}{@{}p{6cm}p{6cm}ccc@{}}
\caption{Top connectivity features in ultralow frequency range of ABCD ADHD prediction\label{tab:connectivity_ultralow_frequency_ABCD_ADHD}}\\
\toprule
\textbf{Region 1} & \textbf{Region 2} & \textbf{Disorder/HC} & \textbf{Corrected p-value}  & \textbf{Cohen's d} \\ \midrule
\endfirsthead

\multicolumn{5}{c}%
{{\tablename\ \thetable{} -- continued from previous page}} \\
\toprule
\textbf{Region 1} & \textbf{Region 2} & \textbf{Corrected p-value} & \textbf{Disorder/HC} & \textbf{Cohen's d} \\ \midrule
\endhead

\midrule \multicolumn{5}{r}{{Continued on next page}} \\ \midrule
\endfoot

\bottomrule
\endlastfoot

Medial Belt Complex R & Area PGp R & HC & $<0.001$ & 0.373 \\
Area dorsal 32 R & ParaHippocampal Area 1 R & ADHD & $<0.001$ & 0.372 \\
Area FST R & ParaHippocampal Area 1 R & ADHD & $<0.001$ & 0.368 \\
Area dorsal 32 R & Area OP2-3-VS R & ADHD & $<0.001$ & 0.367 \\
Medial Belt Complex R & Seventh Visual Area R & HC & $<0.001$ & 0.363 \\
Area PGi L & ParaHippocampal Area 1 R & ADHD & $<0.001$ & 0.361 \\
ParaHippocampal Area 3 R & Area TF R & HC & 0.001 & 0.346 \\
Area 47s R & Area PGp R & HC & 0.001 & 0.344 \\
Area 8BM R & Area OP2-3-VS R & ADHD & 0.001 & 0.343 \\
Area FST R & Area OP2-3-VS R & ADHD & 0.001 & 0.343 \\
Area 11l L & Area PGp R & HC & 0.001 & 0.343 \\
Area 8Av R & ParaHippocampal Area 1 R & ADHD & 0.001 & 0.343 \\
Area 11l L & Anterior IntraParietal Area L & HC & 0.001 & 0.342 \\
Area 47l (47 lateral) L & Anterior IntraParietal Area L & HC & 0.001 & 0.341 \\
Area PGi L & Area OP2-3-VS R & ADHD & 0.001 & 0.339 \\
Area 8BM R & ParaHippocampal Area 1 R & ADHD & 0.001 & 0.339 \\
Area FST R & Area Posterior Insular 1 L & ADHD & 0.001 & 0.338 \\
Area TemporoParietoOccipital Junction 2 L & ParaHippocampal Area 1 R & ADHD & 0.001 & 0.337 \\
Area 47s R & Anterior IntraParietal Area L & HC & 0.001 & 0.336 \\
Area OP4-PV R & Area PGp R & HC & 0.001 & 0.336 \\
Pirform Cortex R & Area PGp R & HC & 0.001 & 0.336 \\
Dorsal Transitional Visual Area R & ParaHippocampal Area 1 R & ADHD & 0.001 & 0.335 \\
Area dorsal 23 a+b R & ParaHippocampal Area 1 R & ADHD & 0.001 & 0.334 \\
Fourth Visual Area L & Area OP2-3-VS R & ADHD & 0.001 & 0.334 \\
Area STSv posterior R & Area 47s R & ADHD & 0.001 & 0.333 \\
Area Posterior Insular 1 L & Medial Superior Temporal Area L & HC & 0.001 & 0.33 \\
Area 47s R & Seventh Visual Area R & HC & 0.001 & 0.33 \\
Area STSv posterior R & Polar 10p R & ADHD & 0.001 & 0.33 \\
Area 11l L & Seventh Visual Area R & HC & 0.001 & 0.33 \\
Area OP4-PV R & Area TF R & HC & 0.001 & 0.329 \\
Medial Belt Complex R & Sixth Visual Area L & HC & 0.001 & 0.328 \\
Area STSv posterior R & Anterior Agranular Insula Complex R & ADHD & 0.002 & 0.328 \\
Area STSv posterior R & ParaHippocampal Area 1 R & ADHD & 0.002 & 0.328 \\
Area 5L R & ParaHippocampal Area 1 R & ADHD & 0.002 & 0.328 \\
Area IFSa R & Area OP2-3-VS R & ADHD & 0.002 & 0.327 \\
Medial Belt Complex R & Medial Superior Temporal Area L & HC & 0.002 & 0.326 \\
Area STGa R & Area TF R & HC & 0.002 & 0.325 \\
Area 11l L & Area 7PC L & HC & 0.002 & 0.325 \\
Area V3A L & Area OP2-3-VS R & ADHD & 0.002 & 0.324 \\
Area posterior 9-46v R & Area OP2-3-VS R & ADHD & 0.002 & 0.324 \\
Area anterior 9-46v R & Area OP2-3-VS R & ADHD & 0.002 & 0.324 \\
Fourth Visual Area R & Area 10v L & ADHD & 0.002 & 0.323 \\
Medial Belt Complex R & Anterior IntraParietal Area L & HC & 0.002 & 0.323 \\
Area STSv posterior R & Orbital Frontal Complex L & ADHD & 0.002 & 0.323 \\
Area PFm Complex L & Area OP2-3-VS R & ADHD & 0.002 & 0.322 \\
Area STSv posterior R & Area OP2-3-VS R & ADHD & 0.002 & 0.322 \\
Medial Area 7A R & ParaHippocampal Area 1 R & ADHD & 0.002 & 0.321 \\
Medial Belt Complex R & Area V3A L & HC & 0.002 & 0.32 \\
Area OP4-PV R & Medial Superior Temporal Area L & HC & 0.002 & 0.32 \\
Area STSv posterior R & Area 10v L & ADHD & 0.002 & 0.32 \\
Area 11l L & Fourth Visual Area L & HC & 0.002 & 0.32 \\
Area STSd posterior L & ParaHippocampal Area 1 R & ADHD & 0.002 & 0.319 \\
Area dorsal 32 R & Area Posterior 24 prime R & ADHD & 0.002 & 0.319 \\
Area dorsal 32 R & Hippocampus R & ADHD & 0.002 & 0.319 \\
Area Posterior Insular 1 L & Anterior IntraParietal Area L & HC & 0.002 & 0.318 \\
Area PFm Complex L & ParaHippocampal Area 1 R & ADHD & 0.002 & 0.318 \\
Third Visual Area R & Area OP2-3-VS R & ADHD & 0.002 & 0.318 \\
Area FST L & ParaHippocampal Area 1 R & ADHD & 0.002 & 0.317 \\
Area FST R & Pirform Cortex L & ADHD & 0.002 & 0.317 \\
Area FST R & Area p32 L & ADHD & 0.002 & 0.317 \\
Anterior IntraParietal Area L & ParaHippocampal Area 1 R & ADHD & 0.002 & 0.316 \\
Area 11l L & Ventral IntraParietal Complex L & HC & 0.002 & 0.316 \\
Area STSv posterior R & Area 47l (47 lateral) R & ADHD & 0.002 & 0.316 \\
Area STSd posterior L & Area OP2-3-VS R & ADHD & 0.002 & 0.316 \\
Area ventral 23 a+b R & ParaHippocampal Area 1 R & ADHD & 0.002 & 0.315 \\
Area STSv posterior R & Pirform Cortex L & ADHD & 0.002 & 0.315 \\
Medial Belt Complex R & Second Visual Area L & HC & 0.002 & 0.315 \\
Area Posterior Insular 1 L & Area PGp R & HC & 0.002 & 0.315 \\
Area 47l (47 lateral) L & Medial Superior Temporal Area L & HC & 0.002 & 0.315 \\
Area 8Av R & Area OP2-3-VS R & ADHD & 0.002 & 0.314 \\
Pirform Cortex R & Seventh Visual Area R & HC & 0.002 & 0.314 \\
Sixth Visual Area L & ParaHippocampal Area 1 R & ADHD & 0.002 & 0.314 \\
Area 43 L & Area OP2-3-VS R & ADHD & 0.002 & 0.313 \\
Area STSv posterior R & Area p32 L & ADHD & 0.002 & 0.313 \\
Area TG dorsal R & Area PGp R & HC & 0.002 & 0.313 \\
Medial Superior Temporal Area L & ParaHippocampal Area 1 R & ADHD & 0.003 & 0.312 \\
Area dorsal 32 R & Anterior Agranular Insula Complex R & ADHD & 0.002 & 0.312 \\
Area p32 R & ParaHippocampal Area 1 R & ADHD & 0.003 & 0.312 \\
Area PGi L & Hippocampus R & ADHD & 0.003 & 0.312 \\
Area STGa L & Anterior IntraParietal Area L & HC & 0.002 & 0.312 \\
Medial Area 7A R & Area OP2-3-VS R & ADHD & 0.003 & 0.311 \\
Dorsal area 6 R & Medial Superior Temporal Area L & HC & 0.003 & 0.311 \\
Area TE1 Middle R & ParaHippocampal Area 1 R & ADHD & 0.003 & 0.311 \\
Area p32 L & Medial Superior Temporal Area L & HC & 0.003 & 0.31 \\
IntraParietal Sulcus Area 1 R & ParaHippocampal Area 1 R & ADHD & 0.003 & 0.31 \\
Fusiform Face Complex R & ParaHippocampal Area 1 R & ADHD & 0.003 & 0.309 \\
Area 11l L & Medial Superior Temporal Area L & HC & 0.003 & 0.309 \\
Area 44 R & ParaHippocampal Area 1 R & ADHD & 0.003 & 0.309 \\
Area 23c R & ParaHippocampal Area 1 R & ADHD & 0.003 & 0.309 \\
Medial Belt Complex R & Area V6A R & HC & 0.003 & 0.309 \\
Parieto-Occipital Sulcus Area 1 R & Anterior IntraParietal Area L & HC & 0.003 & 0.308 \\
Medial Belt Complex R & Ventral IntraParietal Complex L & HC & 0.003 & 0.308 \\
Dorsal area 6 R & Seventh Visual Area R & HC & 0.003 & 0.308 \\
Area 47l (47 lateral) L & Area PGp R & HC & 0.003 & 0.308 \\
Area STSv posterior R & Area anterior 10p L & ADHD & 0.003 & 0.307 \\
Ventral Area 6 L & Medial Superior Temporal Area L & HC & 0.003 & 0.306 \\
Area TemporoParietoOccipital Junction 1 R & ParaHippocampal Area 1 R & ADHD & 0.003 & 0.306 \\
Pirform Cortex L & Area PGp R & HC & 0.003 & 0.306 \\
Area STSv posterior R & Area 47s L & ADHD & 0.003 & 0.305 \\
Parieto-Occipital Sulcus Area 2 L & ParaHippocampal Area 1 R & ADHD & 0.003 & 0.305 \\

\end{longtable}
\end{landscape}

\renewcommand{\tablename}{Supplementary Table}  
\begin{landscape}
\setlength{\LTcapwidth}{1.0\textwidth}
\begin{longtable}{@{}p{6cm}p{6cm}ccc@{}}
\caption{Top connectivity features in high frequency range of ABIDE ASD prediction\label{tab:connectivity_high_frequency_ABIDE_ASD}}\\
\toprule
\textbf{Region 1} & \textbf{Region 2} & \textbf{Disorder/HC} & \textbf{Corrected p-value}  & \textbf{Cohen's d} \\ \midrule
\endfirsthead

\multicolumn{5}{c}%
{{\tablename\ \thetable{} -- continued from previous page}} \\
\toprule
\textbf{Region 1} & \textbf{Region 2} & \textbf{Corrected p-value} & \textbf{Disorder/HC} & \textbf{Cohen's d} \\ \midrule
\endhead

\midrule \multicolumn{5}{r}{{Continued on next page}} \\ \midrule
\endfoot

\bottomrule
\endlastfoot

Area 2 L & Area 11l R & HC & 0.002 & 2.232 \\
Area PFcm L & Area 11l R & HC & 0.003 & 2.191 \\
RetroInsular Cortex L & Area 11l R & HC & 0.003 & 2.162 \\
Ventral Area 24d L & Area 11l R & HC & 0.003 & 2.162 \\
Area IntraParietal 2 L & Area 11l R & HC & 0.003 & 2.149 \\
Area Posterior 24 prime R & Area 11l R & HC & 0.003 & 2.147 \\
Area Posterior 24 prime L & Area 11l R & HC & 0.003 & 2.147 \\
Ventral Area 24d R & Area 11l R & HC & 0.003 & 2.145 \\
Area 1 L & Area 11l R & HC & 0.003 & 2.141 \\
Area 44 R & Area 11l R & HC & 0.003 & 2.14 \\
Primary Auditory Cortex L & Area 11l R & HC & 0.003 & 2.127 \\
Area dorsal 32 R & Area 11l R & HC & 0.004 & 2.11 \\
Area Frontal Opercular 5 L & Area 11l R & HC & 0.004 & 2.106 \\
Area anterior 32 prime L & Area 11l R & HC & 0.004 & 2.104 \\
Parieto-Occipital Sulcus Area 2 R & Area 11l R & HC & 0.004 & 2.101 \\
Anterior IntraParietal Area L & Area 11l R & HC & 0.004 & 2.101 \\
Anterior 24 prime R & Area 11l R & HC & 0.004 & 2.099 \\
Area 43 R & Area 11l R & HC & 0.004 & 2.097 \\
Orbital Frontal Complex R & Area 11l R & HC & 0.004 & 2.091 \\
Area OP4-PV R & Area 11l R & HC & 0.004 & 2.09 \\
Area 23d L & Area 11l R & HC & 0.004 & 2.085 \\
Frontal Opercular Area 3 L & Area 11l R & HC & 0.004 & 2.083 \\
Area PGi R & Area 11l R & HC & 0.004 & 2.082 \\
Dorsal Area 24d L & Area 11l R & HC & 0.004 & 2.082 \\
Primary Sensory Cortex L & Area 11l R & HC & 0.004 & 2.08 \\
Second Visual Area R & Sixth Visual Area L & ASD & 0.004 & 2.075 \\
Area PFt L & Area 11l R & HC & 0.004 & 2.075 \\
Area 7PC L & Area 11l R & HC & 0.004 & 2.073 \\
Area anterior 32 prime R & Area 11l R & HC & 0.004 & 2.07 \\
Area 33 prime L & Area 11l R & HC & 0.004 & 2.067 \\
Area OP2-3-VS L & Area 11l R & HC & 0.004 & 2.066 \\
Area posterior 10p R & Area 11l R & HC & 0.004 & 2.064 \\
Dorsal Transitional Visual Area R & Area 11l R & HC & 0.004 & 2.064 \\
Area ventral 23 a+b L & Area 11l R & HC & 0.004 & 2.055 \\
Area 23d R & Area 11l R & HC & 0.004 & 2.054 \\
Area 7m R & Area 11l R & HC & 0.004 & 2.053 \\
Area 10v R & Area 11l R & HC & 0.004 & 2.05 \\
Lateral Area 7A L & Area 11l R & HC & 0.004 & 2.05 \\
Area IntraParietal 1 L & Area 11l R & HC & 0.004 & 2.049 \\
Area Frontal Opercular 5 R & Area 11l R & HC & 0.004 & 2.049 \\
Anterior 24 prime L & Area 11l R & HC & 0.004 & 2.049 \\
Area posterior 9-46v L & Area 11l R & HC & 0.004 & 2.046 \\
Area p32 prime R & Area 11l R & HC & 0.004 & 2.046 \\
Area ventral 23 a+b R & Area 11l R & HC & 0.004 & 2.045 \\
Area posterior 24 L & Area 11l R & HC & 0.004 & 2.045 \\
Area dorsal 32 L & Area 11l R & HC & 0.004 & 2.044 \\
PeriSylvian Language Area L & Area 11l R & HC & 0.004 & 2.044 \\
Auditory 4 Complex R & Area 11l R & HC & 0.004 & 2.044 \\
Lateral Belt Complex R & Area 11l R & HC & 0.004 & 2.043 \\
Area 8BM R & Area 11l R & HC & 0.004 & 2.043 \\
Area PGs L & Area 11l R & HC & 0.004 & 2.041 \\
Area 10d R & Area 11l R & HC & 0.004 & 2.04 \\
Fourth Visual Area L & Sixth Visual Area L & ASD & 0.005 & 2.039 \\
Rostral Area 6 R & Area 11l R & HC & 0.005 & 2.038 \\
Area 33 prime R & Area 11l R & HC & 0.005 & 2.035 \\
Insular Granular Complex L & Area 11l R & HC & 0.005 & 2.034 \\
Ventral Area 6 R & Area 11l R & HC & 0.005 & 2.034 \\
Seventh Visual Area R & Sixth Visual Area L & ASD & 0.005 & 2.032 \\
Area dorsal 23 a+b L & Area 11l R & HC & 0.005 & 2.027 \\
Orbital Frontal Complex L & Area 11l R & HC & 0.005 & 2.027 \\
Area TE2 anterior L & Area 11l R & HC & 0.005 & 2.027 \\
Medial Belt Complex R & Area 11l R & HC & 0.005 & 2.026 \\
Area 45 R & Area 11l R & HC & 0.005 & 2.025 \\
Medial Area 7P R & Area 11l R & HC & 0.005 & 2.023 \\
Ventral Area 6 L & Area 11l R & HC & 0.005 & 2.021 \\
Dorsal Area 24d R & Area 11l R & HC & 0.005 & 2.02 \\
ProStriate Area L & Area 11l R & HC & 0.005 & 2.02 \\
Area OP1-SII L & Area 11l R & HC & 0.005 & 2.02 \\
Medial IntraParietal Area L & Area 11l R & HC & 0.005 & 2.019 \\
Area anterior 47r R & Area 11l R & HC & 0.005 & 2.018 \\
Superior 6-8 Transitional Area R & Area 11l R & HC & 0.005 & 2.015 \\
Area 8Ad R & Area 11l R & HC & 0.005 & 2.012 \\
Area IFSp L & Area 11l R & HC & 0.005 & 2.012 \\
Superior Temporal Visual Area L & Area 11l R & HC & 0.005 & 2.011 \\
Ventral IntraParietal Complex L & Area 11l R & HC & 0.005 & 2.011 \\
Area p32 prime L & Area 11l R & HC & 0.005 & 2.01 \\
PreSubiculum R & Sixth Visual Area L & ASD & 0.005 & 2.007 \\
Area 8B Lateral R & Area 11l R & HC & 0.005 & 2.005 \\
Area V6A L & Area 11l R & HC & 0.005 & 2.004 \\
Area IFJp L & Area 11l R & HC & 0.005 & 2.004 \\
Area Lateral IntraParietal ventral L & Area 11l R & HC & 0.005 & 2.004 \\
Area 8Av R & Area 11l R & HC & 0.005 & 2.004 \\
Area IFJa L & Area 11l R & HC & 0.005 & 2.004 \\
Parieto-Occipital Sulcus Area 2 L & Area 11l R & HC & 0.005 & 2.003 \\
Frontal Opercular Area 4 L & Area 11l R & HC & 0.005 & 2.003 \\
Area PF Opercular L & Area 11l R & HC & 0.005 & 2.001 \\
Area posterior 24 R & Area 11l R & HC & 0.005 & 2.001 \\
Area PFm Complex R & Area 11l R & HC & 0.005 & 2.0 \\
Area 5L R & Area 11l R & HC & 0.005 & 2.0 \\
Area 10v L & Area 11l R & HC & 0.005 & 1.998 \\
Area 5m ventral L & Area 11l R & HC & 0.005 & 1.998 \\
Anterior Ventral Insular Area L & Area 11l R & HC & 0.005 & 1.996 \\
Area posterior 9-46v R & Area 11l R & HC & 0.005 & 1.994 \\
Area PGp L & Area 11l R & HC & 0.005 & 1.992 \\
Frontal Opercular Area 2 L & Area 11l R & HC & 0.005 & 1.991 \\
Area PFm Complex L & Area 11l R & HC & 0.005 & 1.99 \\
Area Lateral IntraParietal dorsal L & Area 11l R & HC & 0.005 & 1.988 \\
Parieto-Occipital Sulcus Area 1 L & Area 11l R & HC & 0.005 & 1.988 \\
Area 31p ventral L & Area 11l R & HC & 0.005 & 1.987 \\
Area V6A R & Area 11l R & HC & 0.005 & 1.987 \\

\end{longtable}
\end{landscape}

\renewcommand{\tablename}{Supplementary Table}
\begin{landscape}
\setlength{\LTcapwidth}{1.0\textwidth}
\begin{longtable}{@{}p{6cm}p{6cm}ccc@{}}
\caption{Top connectivity features in low frequency range of ABIDE ASD prediction\label{tab:connectivity_low_frequency_ABIDE_ASD}}\\
\toprule
\textbf{Region 1} & \textbf{Region 2} & \textbf{Corrected p-value} & \textbf{Disorder/HC} & \textbf{Cohen's d} \\ \midrule
\endfirsthead

\multicolumn{5}{c}%
{{\tablename\ \thetable{} -- continued from previous page}} \\
\toprule
\textbf{Region 1} & \textbf{Region 2} & \textbf{Disorder/HC}  & \textbf{Corrected p-value} & \textbf{Cohen's d} \\ \midrule
\endhead

\midrule \multicolumn{5}{r}{{Continued on next page}} \\ \midrule
\endfoot

\bottomrule
\endlastfoot

Frontal Opercular Area 3 L & Polar 10p R & HC & 0.003 & 2.128 \\
Frontal Opercular Area 2 L & Polar 10p R & HC & 0.003 & 2.123 \\
Area V3A R & Polar 10p R & HC & 0.003 & 2.117 \\
Area posterior 47r L & Polar 10p R & HC & 0.004 & 2.109 \\
IntraParietal Sulcus Area 1 L & Polar 10p R & HC & 0.004 & 2.101 \\
Area OP2-3-VS R & Polar 10p R & HC & 0.004 & 2.1 \\
Frontal Opercular Area 3 R & Polar 10p R & HC & 0.004 & 2.096 \\
Frontal Opercular Area 2 R & Polar 10p R & HC & 0.004 & 2.095 \\
Area 44 L & Polar 10p R & HC & 0.004 & 2.093 \\
Para-Insular Area R & Polar 10p R & HC & 0.004 & 2.093 \\
Eighth Visual Area R & Polar 10p R & HC & 0.004 & 2.091 \\
Ventral IntraParietal Complex R & Polar 10p R & HC & 0.004 & 2.09 \\
Area IFSp L & Polar 10p R & HC & 0.004 & 2.087 \\
Area anterior 9-46v L & Polar 10p R & HC & 0.004 & 2.083 \\
Area STGa R & Polar 10p R & HC & 0.004 & 2.082 \\
Medial IntraParietal Area L & Polar 10p R & HC & 0.004 & 2.078 \\
ParaHippocampal Area 1 R & Polar 10p R & HC & 0.004 & 2.075 \\
Rostral Area 6 R & Polar 10p R & HC & 0.004 & 2.074 \\
Area V3B L & Polar 10p R & HC & 0.004 & 2.074 \\
Pirform Cortex L & Polar 10p R & HC & 0.004 & 2.074 \\
Anterior 24 prime L & Polar 10p R & HC & 0.004 & 2.073 \\
Para-Insular Area L & Polar 10p R & HC & 0.004 & 2.072 \\
Area TF R & Polar 10p R & HC & 0.004 & 2.071 \\
Primary Auditory Cortex R & Polar 10p R & HC & 0.004 & 2.071 \\
Area PFt R & Polar 10p R & HC & 0.004 & 2.069 \\
Area V4t L & Polar 10p R & HC & 0.004 & 2.069 \\
Area OP1-SII R & Polar 10p R & HC & 0.004 & 2.069 \\
Area V3B R & Polar 10p R & HC & 0.004 & 2.069 \\
Pirform Cortex R & Polar 10p R & HC & 0.004 & 2.068 \\
Area IntraParietal 0 L & Polar 10p R & HC & 0.004 & 2.067 \\
Area Frontal Opercular 5 L & Polar 10p R & HC & 0.004 & 2.064 \\
Frontal Opercular Area 1 R & Polar 10p R & HC & 0.004 & 2.064 \\
Area 55b L & Polar 10p R & HC & 0.004 & 2.064 \\
Area OP2-3-VS L & Polar 10p R & HC & 0.004 & 2.064 \\
ProStriate Area R & Polar 10p R & HC & 0.004 & 2.063 \\
Area STSd posterior L & Polar 10p R & HC & 0.004 & 2.062 \\
Area IntraParietal 0 R & Polar 10p R & HC & 0.004 & 2.062 \\
Area 8B Lateral L & Polar 10p R & HC & 0.004 & 2.061 \\
Area STSd anterior L & Polar 10p R & HC & 0.004 & 2.061 \\
Lateral Belt Complex R & Polar 10p R & HC & 0.004 & 2.061 \\
Lateral Area 7P L & Polar 10p R & HC & 0.004 & 2.061 \\
IntraParietal Sulcus Area 1 R & Polar 10p R & HC & 0.004 & 2.06 \\
Area OP1-SII L & Polar 10p R & HC & 0.004 & 2.06 \\
Area 2 R & Polar 10p R & HC & 0.004 & 2.059 \\
Area IFJa R & Polar 10p R & HC & 0.004 & 2.059 \\
Ventral IntraParietal Complex L & Polar 10p R & HC & 0.004 & 2.058 \\
Area Posterior Insular 1 R & Polar 10p R & HC & 0.004 & 2.058 \\
Area 7PC R & Polar 10p R & HC & 0.004 & 2.058 \\
Area IFJp R & Polar 10p R & HC & 0.004 & 2.057 \\
Insular Granular Complex R & Polar 10p R & HC & 0.004 & 2.057 \\
Area IFSa L & Polar 10p R & HC & 0.004 & 2.057 \\
Dorsal Transitional Visual Area L & Polar 10p R & HC & 0.004 & 2.056 \\
Posterior Insular Area 2 R & Polar 10p R & HC & 0.004 & 2.055 \\
Area 33 prime R & Polar 10p R & HC & 0.004 & 2.055 \\
Frontal Opercular Area 4 L & Polar 10p R & HC & 0.004 & 2.054 \\
Area Lateral IntraParietal ventral R & Polar 10p R & HC & 0.004 & 2.053 \\
Area 8BM L & Polar 10p R & HC & 0.004 & 2.053 \\
Ventral Visual Complex R & Polar 10p R & HC & 0.004 & 2.053 \\
Middle Insular Area L & Polar 10p R & HC & 0.004 & 2.053 \\
Area 9 Middle R & Polar 10p R & HC & 0.004 & 2.052 \\
Insular Granular Complex L & Polar 10p R & HC & 0.004 & 2.051 \\
Sixth Visual Area R & Polar 10p R & HC & 0.004 & 2.051 \\
Lateral Area 7A R & Polar 10p R & HC & 0.004 & 2.051 \\
Ventral Visual Complex L & Polar 10p R & HC & 0.004 & 2.05 \\
Area TG dorsal R & Polar 10p R & HC & 0.004 & 2.05 \\
Frontal Opercular Area 4 R & Polar 10p R & HC & 0.004 & 2.05 \\
Anterior IntraParietal Area R & Polar 10p R & HC & 0.004 & 2.05 \\
Hippocampus R & Polar 10p R & HC & 0.004 & 2.05 \\
PreSubiculum L & Polar 10p R & HC & 0.004 & 2.05 \\
Area anterior 10p L & Polar 10p R & HC & 0.004 & 2.048 \\
Area IFJp L & Polar 10p R & HC & 0.004 & 2.047 \\
Seventh Visual Area R & Polar 10p R & HC & 0.004 & 2.047 \\
VentroMedial Visual Area 2 R & Polar 10p R & HC & 0.004 & 2.046 \\
Eighth Visual Area L & Polar 10p R & HC & 0.004 & 2.045 \\
Area TA2 R & Polar 10p R & HC & 0.004 & 2.044 \\
Fourth Visual Area R & Polar 10p R & HC & 0.004 & 2.044 \\
Area PFcm R & Polar 10p R & HC & 0.004 & 2.044 \\
Area PFm Complex R & Polar 10p R & HC & 0.004 & 2.044 \\
Area PFm Complex L & Polar 10p R & HC & 0.004 & 2.044 \\
Area TE1 anterior R & Polar 10p R & HC & 0.004 & 2.043 \\
Posterior Insular Area 2 L & Polar 10p R & HC & 0.004 & 2.043 \\
Area 8Ad L & Polar 10p R & HC & 0.004 & 2.043 \\
Area dorsal 32 R & Polar 10p R & HC & 0.004 & 2.042 \\
Area V4t R & Polar 10p R & HC & 0.004 & 2.042 \\
Frontal Eye Fields L & Polar 10p R & HC & 0.004 & 2.041 \\
Posterior InferoTemporal complex L & Polar 10p R & HC & 0.004 & 2.041 \\
Area Lateral IntraParietal dorsal R & Polar 10p R & HC & 0.004 & 2.041 \\
Area PF Complex R & Polar 10p R & HC & 0.004 & 2.04 \\
Area 9 Middle L & Polar 10p R & HC & 0.004 & 2.04 \\
Area STSd anterior R & Polar 10p R & HC & 0.004 & 2.04 \\
Area 43 L & Polar 10p R & HC & 0.004 & 2.04 \\
VentroMedial Visual Area 3 L & Polar 10p R & HC & 0.004 & 2.04 \\
Area 8Av L & Polar 10p R & HC & 0.004 & 2.039 \\
Area 6mp L & Polar 10p R & HC & 0.004 & 2.039 \\
Area TE2 anterior R & Polar 10p R & HC & 0.004 & 2.039 \\
Area 1 L & Polar 10p R & HC & 0.004 & 2.037 \\
Area V6A L & Polar 10p R & HC & 0.005 & 2.037 \\
ParaBelt Complex R & Polar 10p R & HC & 0.005 & 2.037 \\
Area 44 R & Polar 10p R & HC & 0.005 & 2.036 \\
VentroMedial Visual Area 3 R & Polar 10p R & HC & 0.005 & 2.036 \\

\end{longtable}
\end{landscape}

\renewcommand{\tablename}{Supplementary Table}
\begin{landscape}
\setlength{\LTcapwidth}{1.0\textwidth}
\begin{longtable}{@{}p{6cm}p{6cm}ccc@{}}
\caption{Top connectivity features in ultralow frequency range of ABIDE ASD prediction\label{tab:connectivity_ultralow_frequency_ABIDE_ASD}}\\
\toprule
\textbf{Region 1} & \textbf{Region 2} & \textbf{Disorder/HC} & \textbf{Corrected p-value}  & \textbf{Cohen's d} \\ \midrule
\endfirsthead

\multicolumn{5}{c}%
{{\tablename\ \thetable{} -- continued from previous page}} \\
\toprule
\textbf{Region 1} & \textbf{Region 2} & \textbf{Corrected p-value} & \textbf{Disorder/HC} & \textbf{Cohen's d} \\ \midrule
\endhead

\midrule \multicolumn{5}{r}{{Continued on next page}} \\ \midrule
\endfoot

\bottomrule
\endlastfoot

Area PHT L & Area TemporoParietoOccipital Junction 3 L & ASD & $<0.001$ & 5.637 \\
Area 9 Posterior R & Area TemporoParietoOccipital Junction 3 L & ASD & $<0.001$ & 5.334 \\
Area 8Ad R & Area TemporoParietoOccipital Junction 3 L & ASD & $<0.001$ & 3.99 \\
Area 8B Lateral L & Area TemporoParietoOccipital Junction 3 L & ASD & $<0.001$ & 3.835 \\
Orbital Frontal Complex L & Area TF R & ASD & $<0.001$ & 3.767 \\
Superior Frontal Language Area L & Area TemporoParietoOccipital Junction 3 L & ASD & $<0.001$ & 3.566 \\
Superior 6-8 Transitional Area L & Area TemporoParietoOccipital Junction 3 L & ASD & $<0.001$ & 3.566 \\
Area TE2 posterior L & Area Posterior 24 prime R & ASD & $<0.001$ & 3.549 \\
RetroInsular Cortex R & Area anterior 32 prime L & ASD & $<0.001$ & 3.54 \\
Frontal Opercular Area 2 L & Ventral IntraParietal Complex R & ASD & $<0.001$ & 3.499 \\
Fourth Visual Area L & Area anterior 32 prime L & ASD & $<0.001$ & 3.494 \\
Lateral Area 7A R & Area TemporoParietoOccipital Junction 3 L & ASD & $<0.001$ & 3.38 \\
Frontal Eye Fields L & Area TemporoParietoOccipital Junction 3 L & ASD & $<0.001$ & 3.355 \\
Area V3B R & Area 9-46d L & ASD & 0.002 & 3.339 \\
Medial Area 7P L & Area anterior 32 prime L & ASD & 0.002 & 3.328 \\
Fourth Visual Area L & Area PFm Complex L & ASD & 0.002 & 3.295 \\
Sixth Visual Area R & Area TemporoParietoOccipital Junction 3 L & ASD & $<0.001$ & 3.252 \\
Area posterior 9-46v R & Area anterior 32 prime L & ASD & $<0.001$ & 3.201 \\
Second Visual Area R & Area TemporoParietoOccipital Junction 3 L & ASD & $<0.001$ & 3.176 \\
Area 8BM R & Area TemporoParietoOccipital Junction 3 L & ASD & $<0.001$ & 3.173 \\
Area STSd posterior L & Area TemporoParietoOccipital Junction 3 L & ASD & $<0.001$ & 3.169 \\
Second Visual Area R & Area 23d R & ASD & $<0.001$ & 3.143 \\
IntraParietal Sulcus Area 1 R & Area 23d R & ASD & $<0.001$ & 3.128 \\
Lateral Area 7P L & RetroInsular Cortex L & ASD & $<0.001$ & 3.08 \\
Lateral Area 7P R & RetroInsular Cortex L & ASD & $<0.001$ & 3.076 \\
Medial Area 7P L & Primary Sensory Cortex R & ASD & $<0.001$ & 3.064 \\
Second Visual Area L & Area anterior 32 prime L & ASD & $<0.001$ & 3.024 \\
Second Visual Area L & Area PFm Complex L & ASD & $<0.001$ & 3.012 \\
Area posterior 9-46v L & RetroInsular Cortex L & ASD & $<0.001$ & 3.009 \\
Area TE2 posterior L & Area 5m R & ASD & $<0.001$ & 2.983 \\
Area 5m L & Area anterior 32 prime L & ASD & $<0.001$ & 2.966 \\
Area PH L & Area anterior 32 prime L & ASD & $<0.001$ & 2.963 \\
Area 2 R & Area anterior 32 prime L & ASD & $<0.001$ & 2.962 \\
Area V3B R & Area anterior 32 prime L & ASD & $<0.001$ & 2.95 \\
Area 9 Middle L & Area TemporoParietoOccipital Junction 3 L & ASD & $<0.001$ & 2.928 \\
Area 31a R & Area TemporoParietoOccipital Junction 3 L & ASD & $<0.001$ & 2.921 \\
Perirhinal Ectorhinal Cortex R & Area TemporoParietoOccipital Junction 3 L & ASD & $<0.001$ & 2.919 \\
Area 6m anterior R & Area TemporoParietoOccipital Junction 3 L & ASD & $<0.001$ & 2.913 \\
Area PH L & Area PFm Complex L & ASD & $<0.001$ & 2.878 \\
Ventral Area 6 R & Area TemporoParietoOccipital Junction 3 L & ASD & $<0.001$ & 2.867 \\
Parieto-Occipital Sulcus Area 1 L & Medial Superior Temporal Area R & ASD & $<0.001$ & 2.859 \\
Ventral Visual Complex R & Area anterior 32 prime L & ASD & $<0.001$ & 2.859 \\
Area anterior 32 prime R & Area 23d R & ASD & $<0.001$ & 2.844 \\
Area TemporoParietoOccipital Junction 3 R & Area anterior 32 prime L & ASD & $<0.001$ & 2.838 \\
ParaHippocampal Area 2 R & Area anterior 32 prime L & ASD & $<0.001$ & 2.821 \\
Area 9-46d R & Area TemporoParietoOccipital Junction 3 L & ASD & $<0.001$ & 2.82 \\
Area V3B R & Area Posterior 24 prime R & ASD & $<0.001$ & 2.801 \\
Area PHT L & Area dorsal 23 a+b L & ASD & $<0.001$ & 2.801 \\
Area 47m L & Area Lateral Occipital 1 L & ASD & $<0.001$ & 2.796 \\
Ventral IntraParietal Complex L & Medial Superior Temporal Area L & ASD & $<0.001$ & 2.796 \\
Area IntraParietal 2 R & Area anterior 32 prime L & ASD & $<0.001$ & 2.789 \\
Area STSv posterior L & Area 23d R & ASD & 0.007 & 2.781 \\
VentroMedial Visual Area 1 L & Primary Sensory Cortex R & ASD & $<0.001$ & 2.775 \\
Area posterior 24 L & Area TemporoParietoOccipital Junction 3 L & ASD & $<0.001$ & 2.774 \\
Medial Area 7A R & Primary Sensory Cortex R & ASD & $<0.001$ & 2.767 \\
Primary Motor Cortex R & Area anterior 32 prime L & ASD & 0.001 & 2.762 \\
Medial Area 7A L & Primary Sensory Cortex R & ASD & 0.001 & 2.753 \\
Medial Area 7P L & Area PFm Complex L & ASD & $<0.001$ & 2.751 \\
Dorsal Transitional Visual Area L & Medial Superior Temporal Area R & ASD & 0.001 & 2.744 \\
Area Lateral Occipital 2 R & RetroInsular Cortex L & ASD & 0.001 & 2.736 \\
Area Lateral Occipital 2 R & Medial Superior Temporal Area R & ASD & 0.001 & 2.734 \\
Area dorsal 32 L & Area TemporoParietoOccipital Junction 3 L & ASD & 0.001 & 2.723 \\
Area TG Ventral L & RetroInsular Cortex L & ASD & 0.001 & 2.711 \\
Area Lateral IntraParietal dorsal R & Area anterior 32 prime L & ASD & 0.001 & 2.696 \\
Primary Visual Cortex R & Area 23d R & ASD & 0.001 & 2.688 \\
Area p32 R & Medial Superior Temporal Area R & ASD & 0.005 & 2.685 \\
Area STSv posterior L & Area PFm Complex L & ASD & 0.005 & 2.683 \\
Area PFt R & Area TemporoParietoOccipital Junction 3 L & ASD & 0.001 & 2.681 \\
Area 31a R & Area PFm Complex L & ASD & 0.001 & 2.672 \\
Area TemporoParietoOccipital Junction 3 R & Area PFm Complex L & ASD & 0.001 & 2.669 \\
Dorsal Transitional Visual Area R & Area TemporoParietoOccipital Junction 3 L & ASD & 0.001 & 2.658 \\
Frontal Opercular Area 3 L & Area 8Av L & ASD & 0.001 & 2.645 \\
Anterior IntraParietal Area R & Area anterior 32 prime L & ASD & 0.001 & 2.627 \\
Dorsal Transitional Visual Area R & Area PFm Complex L & ASD & 0.001 & 2.62 \\
Posterior Insular Area 2 R & Area anterior 32 prime L & ASD & 0.001 & 2.618 \\
Primary Visual Cortex L & Area 23d R & ASD & 0.001 & 2.615 \\
Medial Area 7A L & Area anterior 32 prime L & ASD & 0.001 & 2.608 \\
Primary Visual Cortex L & Area PFm Complex L & ASD & 0.001 & 2.606 \\
Area 8Av L & Primary Sensory Cortex R & ASD & 0.001 & 2.599 \\
Area TE2 posterior L & RetroInsular Cortex L & ASD & 0.001 & 2.583 \\
Area Lateral Occipital 2 R & Area dorsal 23 a+b L & ASD & 0.001 & 2.58 \\
RetroInsular Cortex R & Area PFm Complex L & ASD & 0.001 & 2.578 \\
Area 1 R & Area TemporoParietoOccipital Junction 3 L & ASD & 0.001 & 2.576 \\
Primary Sensory Cortex R & Area TemporoParietoOccipital Junction 3 L & ASD & 0.001 & 2.572 \\
Area STSv anterior L & Area 6 anterior L & ASD & 0.001 & 2.57 \\
Area V3B R & Medial Belt Complex L & ASD & 0.001 & 2.568 \\
Seventh Visual Area L & Medial Superior Temporal Area L & ASD & 0.003 & 2.565 \\
Area V3A R & Ventral IntraParietal Complex R & ASD & 0.001 & 2.565 \\
Area STGa L & Area 23d R & ASD & 0.001 & 2.558 \\
Area 2 L & Ventral IntraParietal Complex R & ASD & 0.001 & 2.55 \\
Para-Insular Area R & Area 23d R & ASD & 0.001 & 2.548 \\
Inferior 6-8 Transitional Area L & RetroInsular Cortex L & ASD & 0.001 & 2.548 \\
Seventh Visual Area L & Area PFm Complex L & ASD & 0.001 & 2.546 \\
Dorsal area 6 R & Area 9-46d L & ASD & 0.001 & 2.541 \\
Area 8BM L & Area TemporoParietoOccipital Junction 3 L & ASD & 0.001 & 2.536 \\
Area TE2 posterior L & Area Posterior 24 prime L & ASD & 0.001 & 2.535 \\
Area 52 L & Area TemporoParietoOccipital Junction 3 L & ASD & 0.001 & 2.533 \\
Area dorsal 32 R & Area TemporoParietoOccipital Junction 3 L & ASD & 0.001 & 2.532 \\
Ventral Area 6 R & Area 23d R & ASD & 0.001 & 2.53 \\
Area 9 Posterior L & Area TemporoParietoOccipital Junction 3 L & ASD & 0.001 & 2.53 \\

\end{longtable}
\end{landscape}

\end{appendices}


\bibliography{sn-bibliography}


\begin{thebibliography}{76}
\ifx \bisbn   \undefined \def \bisbn  #1{ISBN #1}\fi
\ifx \binits  \undefined \def \binits#1{#1}\fi
\ifx \bauthor  \undefined \def \bauthor#1{#1}\fi
\ifx \batitle  \undefined \def \batitle#1{#1}\fi
\ifx \bjtitle  \undefined \def \bjtitle#1{#1}\fi
\ifx \bvolume  \undefined \def \bvolume#1{\textbf{#1}}\fi
\ifx \byear  \undefined \def \byear#1{#1}\fi
\ifx \bissue  \undefined \def \bissue#1{#1}\fi
\ifx \bfpage  \undefined \def \bfpage#1{#1}\fi
\ifx \blpage  \undefined \def \blpage #1{#1}\fi
\ifx \burl  \undefined \def \burl#1{\textsf{#1}}\fi
\ifx \doiurl  \undefined \def \doiurl#1{\url{https://doi.org/#1}}\fi
\ifx \betal  \undefined \def \betal{\textit{et al.}}\fi
\ifx \binstitute  \undefined \def \binstitute#1{#1}\fi
\ifx \binstitutionaled  \undefined \def \binstitutionaled#1{#1}\fi
\ifx \bctitle  \undefined \def \bctitle#1{#1}\fi
\ifx \beditor  \undefined \def \beditor#1{#1}\fi
\ifx \bpublisher  \undefined \def \bpublisher#1{#1}\fi
\ifx \bbtitle  \undefined \def \bbtitle#1{#1}\fi
\ifx \bedition  \undefined \def \bedition#1{#1}\fi
\ifx \bseriesno  \undefined \def \bseriesno#1{#1}\fi
\ifx \blocation  \undefined \def \blocation#1{#1}\fi
\ifx \bsertitle  \undefined \def \bsertitle#1{#1}\fi
\ifx \bsnm \undefined \def \bsnm#1{#1}\fi
\ifx \bsuffix \undefined \def \bsuffix#1{#1}\fi
\ifx \bparticle \undefined \def \bparticle#1{#1}\fi
\ifx \barticle \undefined \def \barticle#1{#1}\fi
\bibcommenthead
\ifx \bconfdate \undefined \def \bconfdate #1{#1}\fi
\ifx \botherref \undefined \def \botherref #1{#1}\fi
\ifx \url \undefined \def \url#1{\textsf{#1}}\fi
\ifx \bchapter \undefined \def \bchapter#1{#1}\fi
\ifx \bbook \undefined \def \bbook#1{#1}\fi
\ifx \bcomment \undefined \def \bcomment#1{#1}\fi
\ifx \oauthor \undefined \def \oauthor#1{#1}\fi
\ifx \citeauthoryear \undefined \def \citeauthoryear#1{#1}\fi
\ifx \endbibitem  \undefined \def \endbibitem {}\fi
\ifx \bconflocation  \undefined \def \bconflocation#1{#1}\fi
\ifx \arxivurl  \undefined \def \arxivurl#1{\textsf{#1}}\fi
\csname PreBibitemsHook\endcsname

\bibitem[\protect\citeauthoryear{Taghia et~al.}{2018}]{taghia2018uncovering}
\begin{barticle}
\bauthor{\bsnm{Taghia}, \binits{J.}},
\bauthor{\bsnm{Cai}, \binits{W.}},
\bauthor{\bsnm{Ryali}, \binits{S.}},
\bauthor{\bsnm{Kochalka}, \binits{J.}},
\bauthor{\bsnm{Nicholas}, \binits{J.}},
\bauthor{\bsnm{Chen}, \binits{T.}},
\bauthor{\bsnm{Menon}, \binits{V.}}:
\batitle{Uncovering hidden brain state dynamics that regulate performance and decision-making during cognition}.
\bjtitle{Nature communications}
\bvolume{9}(\bissue{1}),
\bfpage{2505}
(\byear{2018})
\end{barticle}
\endbibitem

\bibitem[\protect\citeauthoryear{He et~al.}{2010}]{he2010temporal}
\begin{barticle}
\bauthor{\bsnm{He}, \binits{B.J.}},
\bauthor{\bsnm{Zempel}, \binits{J.M.}},
\bauthor{\bsnm{Snyder}, \binits{A.Z.}},
\bauthor{\bsnm{Raichle}, \binits{M.E.}}:
\batitle{The temporal structures and functional significance of scale-free brain activity}.
\bjtitle{Neuron}
\bvolume{66}(\bissue{3}),
\bfpage{353}--\blpage{369}
(\byear{2010})
\end{barticle}
\endbibitem

\bibitem[\protect\citeauthoryear{Sendi et~al.}{2025}]{sendi2025brain}
\begin{botherref}
\oauthor{\bsnm{Sendi}, \binits{M.S.}},
\oauthor{\bsnm{Fu}, \binits{Z.}},
\oauthor{\bsnm{Harnett}, \binits{N.G.}},
\oauthor{\bsnm{Rooij}, \binits{S.J.}},
\oauthor{\bsnm{Vergara}, \binits{V.}},
\oauthor{\bsnm{Pizzagalli}, \binits{D.A.}},
\oauthor{\bsnm{Daskalakis}, \binits{N.P.}},
\oauthor{\bsnm{House}, \binits{S.L.}},
\oauthor{\bsnm{Beaudoin}, \binits{F.L.}},
\oauthor{\bsnm{An}, \binits{X.}}, et al.:
Brain dynamics reflecting an intra-network brain state are associated with increased post-traumatic stress symptoms in the early aftermath of trauma.
Nature Mental Health,
1--14
(2025)
\end{botherref}
\endbibitem

\bibitem[\protect\citeauthoryear{Breakspear}{2017}]{breakspear2017dynamic}
\begin{barticle}
\bauthor{\bsnm{Breakspear}, \binits{M.}}:
\batitle{Dynamic models of large-scale brain activity}.
\bjtitle{Nature neuroscience}
\bvolume{20}(\bissue{3}),
\bfpage{340}--\blpage{352}
(\byear{2017})
\end{barticle}
\endbibitem

\bibitem[\protect\citeauthoryear{Kan et~al.}{2022}]{kan2022brain}
\begin{barticle}
\bauthor{\bsnm{Kan}, \binits{X.}},
\bauthor{\bsnm{Dai}, \binits{W.}},
\bauthor{\bsnm{Cui}, \binits{H.}},
\bauthor{\bsnm{Zhang}, \binits{Z.}},
\bauthor{\bsnm{Guo}, \binits{Y.}},
\bauthor{\bsnm{Yang}, \binits{C.}}:
\batitle{Brain network transformer}.
\bjtitle{Advances in Neural Information Processing Systems}
\bvolume{35},
\bfpage{25586}--\blpage{25599}
(\byear{2022})
\end{barticle}
\endbibitem

\bibitem[\protect\citeauthoryear{Bedel et~al.}{2023}]{bedel2023bolt}
\begin{barticle}
\bauthor{\bsnm{Bedel}, \binits{H.A.}},
\bauthor{\bsnm{Sivgin}, \binits{I.}},
\bauthor{\bsnm{Dalmaz}, \binits{O.}},
\bauthor{\bsnm{Dar}, \binits{S.U.}},
\bauthor{\bsnm{{\c{C}}ukur}, \binits{T.}}:
\batitle{Bolt: Fused window transformers for fmri time series analysis}.
\bjtitle{Medical image analysis}
\bvolume{88},
\bfpage{102841}
(\byear{2023})
\end{barticle}
\endbibitem

\bibitem[\protect\citeauthoryear{Wang et~al.}{2023}]{wang2023brainbert}
\begin{botherref}
\oauthor{\bsnm{Wang}, \binits{C.}},
\oauthor{\bsnm{Subramaniam}, \binits{V.}},
\oauthor{\bsnm{Yaari}, \binits{A.U.}},
\oauthor{\bsnm{Kreiman}, \binits{G.}},
\oauthor{\bsnm{Katz}, \binits{B.}},
\oauthor{\bsnm{Cases}, \binits{I.}},
\oauthor{\bsnm{Barbu}, \binits{A.}}:
Brainbert: Self-supervised representation learning for intracranial recordings.
arXiv preprint arXiv:2302.14367
(2023)
\end{botherref}
\endbibitem

\bibitem[\protect\citeauthoryear{Ortega~Caro et~al.}{2023}]{ortega2023brainlm}
\begin{botherref}
\oauthor{\bsnm{Ortega~Caro}, \binits{J.}},
\oauthor{\bsnm{Oliveira~Fonseca}, \binits{A.H.}},
\oauthor{\bsnm{Averill}, \binits{C.}},
\oauthor{\bsnm{Rizvi}, \binits{S.A.}},
\oauthor{\bsnm{Rosati}, \binits{M.}},
\oauthor{\bsnm{Cross}, \binits{J.L.}},
\oauthor{\bsnm{Mittal}, \binits{P.}},
\oauthor{\bsnm{Zappala}, \binits{E.}},
\oauthor{\bsnm{Levine}, \binits{D.}},
\oauthor{\bsnm{Dhodapkar}, \binits{R.M.}}, et al.:
Brainlm: A foundation model for brain activity recordings.
bioRxiv,
2023--09
(2023)
\end{botherref}
\endbibitem

\bibitem[\protect\citeauthoryear{Stanley and Meakin}{1988}]{stanley1988multifractal}
\begin{barticle}
\bauthor{\bsnm{Stanley}, \binits{H.E.}},
\bauthor{\bsnm{Meakin}, \binits{P.}}:
\batitle{Multifractal phenomena in physics and chemistry}.
\bjtitle{Nature}
\bvolume{335}(\bissue{6189}),
\bfpage{405}--\blpage{409}
(\byear{1988})
\end{barticle}
\endbibitem

\bibitem[\protect\citeauthoryear{Falconer}{2013}]{falconer2013fractal}
\begin{bbook}
\bauthor{\bsnm{Falconer}, \binits{K.}}:
\bbtitle{Fractal Geometry: Mathematical Foundations and Applications},
\bedition{3rd} edn.
\bpublisher{John Wiley \& Sons},
\blocation{Chichester, UK}
(\byear{2013})
\end{bbook}
\endbibitem

\bibitem[\protect\citeauthoryear{Salat et~al.}{2017}]{salat2017multifractal}
\begin{barticle}
\bauthor{\bsnm{Salat}, \binits{H.}},
\bauthor{\bsnm{Murcio}, \binits{R.}},
\bauthor{\bsnm{Arcaute}, \binits{E.}}:
\batitle{Multifractal methodology}.
\bjtitle{Physica A: Statistical Mechanics and its Applications}
\bvolume{473},
\bfpage{467}--\blpage{487}
(\byear{2017})
\end{barticle}
\endbibitem

\bibitem[\protect\citeauthoryear{Guidolin et~al.}{2024}]{Guidolin2024}
\begin{bbook}
\bauthor{\bsnm{Guidolin}, \binits{D.}},
\bauthor{\bsnm{Tortorella}, \binits{C.}},
\bauthor{\bsnm{De~Caro}, \binits{R.}},
\bauthor{\bsnm{Agnati}, \binits{L.F.}}:
In: \beditor{\bsnm{Di~Ieva}, \binits{A.}} (ed.)
\bbtitle{A Self-Similarity Logic May Shape the Organization of the Nervous System},
pp. \bfpage{203}--\blpage{225}.
\bpublisher{Springer},
\blocation{Cham}
(\byear{2024}).
\doiurl{10.1007/978-3-031-47606-8\_10} .
\burl{https://doi.org/10.1007/978-3-031-47606-8\_10}
\end{bbook}
\endbibitem

\bibitem[\protect\citeauthoryear{Ciuciu et~al.}{2008}]{ciuciu2008log}
\begin{barticle}
\bauthor{\bsnm{Ciuciu}, \binits{P.}},
\bauthor{\bsnm{Abry}, \binits{P.}},
\bauthor{\bsnm{Rabrait}, \binits{C.}},
\bauthor{\bsnm{Wendt}, \binits{H.}}:
\batitle{Log wavelet leaders cumulant based multifractal analysis of evi fmri time series: evidence of scaling in ongoing and evoked brain activity}.
\bjtitle{IEEE Journal of Selected Topics in Signal Processing}
\bvolume{2}(\bissue{6}),
\bfpage{929}--\blpage{943}
(\byear{2008})
\end{barticle}
\endbibitem

\bibitem[\protect\citeauthoryear{Smit et~al.}{2011}]{smit2011scale}
\begin{barticle}
\bauthor{\bsnm{Smit}, \binits{D.J.}},
\bauthor{\bsnm{Geus}, \binits{E.J.}},
\bauthor{\bsnm{Nieuwenhuijzen}, \binits{M.E.}},
\bauthor{\bsnm{Beijsterveldt}, \binits{C.E.}},
\bauthor{\bsnm{Baal}, \binits{G.C.M.}},
\bauthor{\bsnm{Mansvelder}, \binits{H.D.}},
\bauthor{\bsnm{Boomsma}, \binits{D.I.}},
\bauthor{\bsnm{Linkenkaer-Hansen}, \binits{K.}}:
\batitle{Scale-free modulation of resting-state neuronal oscillations reflects prolonged brain maturation in humans}.
\bjtitle{Journal of Neuroscience}
\bvolume{31}(\bissue{37}),
\bfpage{13128}--\blpage{13136}
(\byear{2011})
\end{barticle}
\endbibitem

\bibitem[\protect\citeauthoryear{Tagliazucchi et~al.}{2013}]{tagliazucchi2013breakdown}
\begin{barticle}
\bauthor{\bsnm{Tagliazucchi}, \binits{E.}},
\bauthor{\bsnm{Wegner}, \binits{F.}},
\bauthor{\bsnm{Morzelewski}, \binits{A.}},
\bauthor{\bsnm{Brodbeck}, \binits{V.}},
\bauthor{\bsnm{Jahnke}, \binits{K.}},
\bauthor{\bsnm{Laufs}, \binits{H.}}:
\batitle{Breakdown of long-range temporal dependence in default mode and attention networks during deep sleep}.
\bjtitle{Proceedings of the National Academy of Sciences}
\bvolume{110}(\bissue{38}),
\bfpage{15419}--\blpage{15424}
(\byear{2013})
\end{barticle}
\endbibitem

\bibitem[\protect\citeauthoryear{Tolkunov et~al.}{2010}]{tolkunov2010power}
\begin{barticle}
\bauthor{\bsnm{Tolkunov}, \binits{D.}},
\bauthor{\bsnm{Rubin}, \binits{D.}},
\bauthor{\bsnm{Mujica-Parodi}, \binits{L.R.}}:
\batitle{Power spectrum scale invariance quantifies limbic dysregulation in trait anxious adults using fmri: adapting methods optimized for characterizing autonomic dysregulation to neural dynamic time series}.
\bjtitle{Neuroimage}
\bvolume{50}(\bissue{1}),
\bfpage{72}--\blpage{80}
(\byear{2010})
\end{barticle}
\endbibitem

\bibitem[\protect\citeauthoryear{Maxim et~al.}{2005}]{maxim2005fractional}
\begin{barticle}
\bauthor{\bsnm{Maxim}, \binits{V.}},
\bauthor{\bsnm{{\c{S}}endur}, \binits{L.}},
\bauthor{\bsnm{Fadili}, \binits{J.}},
\bauthor{\bsnm{Suckling}, \binits{J.}},
\bauthor{\bsnm{Gould}, \binits{R.}},
\bauthor{\bsnm{Howard}, \binits{R.}},
\bauthor{\bsnm{Bullmore}, \binits{E.}}:
\batitle{Fractional gaussian noise, functional mri and alzheimer's disease}.
\bjtitle{Neuroimage}
\bvolume{25}(\bissue{1}),
\bfpage{141}--\blpage{158}
(\byear{2005})
\end{barticle}
\endbibitem

\bibitem[\protect\citeauthoryear{Achard et~al.}{2006}]{achard2006resilient}
\begin{barticle}
\bauthor{\bsnm{Achard}, \binits{S.}},
\bauthor{\bsnm{Salvador}, \binits{R.}},
\bauthor{\bsnm{Whitcher}, \binits{B.}},
\bauthor{\bsnm{Suckling}, \binits{J.}},
\bauthor{\bsnm{Bullmore}, \binits{E.}}:
\batitle{A resilient, low-frequency, small-world human brain functional network with highly connected association cortical hubs}.
\bjtitle{Journal of Neuroscience}
\bvolume{26}(\bissue{1}),
\bfpage{63}--\blpage{72}
(\byear{2006})
\end{barticle}
\endbibitem

\bibitem[\protect\citeauthoryear{Sasai et~al.}{2021}]{sasai2021frequency}
\begin{barticle}
\bauthor{\bsnm{Sasai}, \binits{S.}},
\bauthor{\bsnm{Koike}, \binits{T.}},
\bauthor{\bsnm{Sugawara}, \binits{S.K.}},
\bauthor{\bsnm{Hamano}, \binits{Y.H.}},
\bauthor{\bsnm{Sumiya}, \binits{M.}},
\bauthor{\bsnm{Okazaki}, \binits{S.}},
\bauthor{\bsnm{Takahashi}, \binits{H.K.}},
\bauthor{\bsnm{Taga}, \binits{G.}},
\bauthor{\bsnm{Sadato}, \binits{N.}}:
\batitle{Frequency-specific task modulation of human brain functional networks: A fast fmri study}.
\bjtitle{NeuroImage}
\bvolume{224},
\bfpage{117375}
(\byear{2021})
\end{barticle}
\endbibitem

\bibitem[\protect\citeauthoryear{Hagler~Jr et~al.}{2019}]{hagler2019image}
\begin{barticle}
\bauthor{\bsnm{Hagler~Jr}, \binits{D.J.}},
\bauthor{\bsnm{Hatton}, \binits{S.}},
\bauthor{\bsnm{Cornejo}, \binits{M.D.}},
\bauthor{\bsnm{Makowski}, \binits{C.}},
\bauthor{\bsnm{Fair}, \binits{D.A.}},
\bauthor{\bsnm{Dick}, \binits{A.S.}},
\bauthor{\bsnm{Sutherland}, \binits{M.T.}},
\bauthor{\bsnm{Casey}, \binits{B.}},
\bauthor{\bsnm{Barch}, \binits{D.M.}},
\bauthor{\bsnm{Harms}, \binits{M.P.}}, \betal:
\batitle{Image processing and analysis methods for the adolescent brain cognitive development study}.
\bjtitle{Neuroimage}
\bvolume{202},
\bfpage{116091}
(\byear{2019})
\end{barticle}
\endbibitem

\bibitem[\protect\citeauthoryear{Marek et~al.}{2022}]{marek2022reproducible}
\begin{barticle}
\bauthor{\bsnm{Marek}, \binits{S.}},
\bauthor{\bsnm{Tervo-Clemmens}, \binits{B.}},
\bauthor{\bsnm{Calabro}, \binits{F.J.}},
\bauthor{\bsnm{Montez}, \binits{D.F.}},
\bauthor{\bsnm{Kay}, \binits{B.P.}},
\bauthor{\bsnm{Hatoum}, \binits{A.S.}},
\bauthor{\bsnm{Donohue}, \binits{M.R.}},
\bauthor{\bsnm{Foran}, \binits{W.}},
\bauthor{\bsnm{Miller}, \binits{R.L.}},
\bauthor{\bsnm{Hendrickson}, \binits{T.J.}}, \betal:
\batitle{Reproducible brain-wide association studies require thousands of individuals}.
\bjtitle{Nature}
\bvolume{603}(\bissue{7902}),
\bfpage{654}--\blpage{660}
(\byear{2022})
\end{barticle}
\endbibitem

\bibitem[\protect\citeauthoryear{Ingalhalikar et~al.}{2021}]{ingalhalikar2021functional}
\begin{barticle}
\bauthor{\bsnm{Ingalhalikar}, \binits{M.}},
\bauthor{\bsnm{Shinde}, \binits{S.}},
\bauthor{\bsnm{Karmarkar}, \binits{A.}},
\bauthor{\bsnm{Rajan}, \binits{A.}},
\bauthor{\bsnm{Rangaprakash}, \binits{D.}},
\bauthor{\bsnm{Deshpande}, \binits{G.}}:
\batitle{Functional connectivity-based prediction of autism on site harmonized abide dataset}.
\bjtitle{IEEE transactions on biomedical engineering}
\bvolume{68}(\bissue{12}),
\bfpage{3628}--\blpage{3637}
(\byear{2021})
\end{barticle}
\endbibitem

\bibitem[\protect\citeauthoryear{Esteban et~al.}{2019}]{esteban2019fmriprep}
\begin{barticle}
\bauthor{\bsnm{Esteban}, \binits{O.}},
\bauthor{\bsnm{Markiewicz}, \binits{C.J.}},
\bauthor{\bsnm{Blair}, \binits{R.W.}},
\bauthor{\bsnm{Moodie}, \binits{C.A.}},
\bauthor{\bsnm{Isik}, \binits{A.I.}},
\bauthor{\bsnm{Erramuzpe}, \binits{A.}},
\bauthor{\bsnm{Kent}, \binits{J.D.}},
\bauthor{\bsnm{Goncalves}, \binits{M.}},
\bauthor{\bsnm{DuPre}, \binits{E.}},
\bauthor{\bsnm{Snyder}, \binits{M.}}, \betal:
\batitle{fmriprep: a robust preprocessing pipeline for functional mri}.
\bjtitle{Nature methods}
\bvolume{16}(\bissue{1}),
\bfpage{111}--\blpage{116}
(\byear{2019})
\end{barticle}
\endbibitem

\bibitem[\protect\citeauthoryear{Behzadi et~al.}{2007}]{behzadi2007component}
\begin{barticle}
\bauthor{\bsnm{Behzadi}, \binits{Y.}},
\bauthor{\bsnm{Restom}, \binits{K.}},
\bauthor{\bsnm{Liau}, \binits{J.}},
\bauthor{\bsnm{Liu}, \binits{T.T.}}:
\batitle{A component based noise correction method (compcor) for bold and perfusion based fmri}.
\bjtitle{Neuroimage}
\bvolume{37}(\bissue{1}),
\bfpage{90}--\blpage{101}
(\byear{2007})
\end{barticle}
\endbibitem

\bibitem[\protect\citeauthoryear{Glasser et~al.}{2016}]{glasser2016multi}
\begin{barticle}
\bauthor{\bsnm{Glasser}, \binits{M.F.}},
\bauthor{\bsnm{Coalson}, \binits{T.S.}},
\bauthor{\bsnm{Robinson}, \binits{E.C.}},
\bauthor{\bsnm{Hacker}, \binits{C.D.}},
\bauthor{\bsnm{Harwell}, \binits{J.}},
\bauthor{\bsnm{Yacoub}, \binits{E.}},
\bauthor{\bsnm{Ugurbil}, \binits{K.}},
\bauthor{\bsnm{Andersson}, \binits{J.}},
\bauthor{\bsnm{Beckmann}, \binits{C.F.}},
\bauthor{\bsnm{Jenkinson}, \binits{M.}}, \betal:
\batitle{A multi-modal parcellation of human cerebral cortex}.
\bjtitle{Nature}
\bvolume{536}(\bissue{7615}),
\bfpage{171}--\blpage{178}
(\byear{2016})
\end{barticle}
\endbibitem

\bibitem[\protect\citeauthoryear{Schaefer et~al.}{2018}]{schaefer2018local}
\begin{barticle}
\bauthor{\bsnm{Schaefer}, \binits{A.}},
\bauthor{\bsnm{Kong}, \binits{R.}},
\bauthor{\bsnm{Gordon}, \binits{E.M.}},
\bauthor{\bsnm{Laumann}, \binits{T.O.}},
\bauthor{\bsnm{Zuo}, \binits{X.-N.}},
\bauthor{\bsnm{Holmes}, \binits{A.J.}},
\bauthor{\bsnm{Eickhoff}, \binits{S.B.}},
\bauthor{\bsnm{Yeo}, \binits{B.T.}}:
\batitle{Local-global parcellation of the human cerebral cortex from intrinsic functional connectivity mri}.
\bjtitle{Cerebral cortex}
\bvolume{28}(\bissue{9}),
\bfpage{3095}--\blpage{3114}
(\byear{2018})
\end{barticle}
\endbibitem

\bibitem[\protect\citeauthoryear{Cordova et~al.}{2022}]{cordova2022attention}
\begin{barticle}
\bauthor{\bsnm{Cordova}, \binits{M.M.}},
\bauthor{\bsnm{Antovich}, \binits{D.M.}},
\bauthor{\bsnm{Ryabinin}, \binits{P.}},
\bauthor{\bsnm{Neighbor}, \binits{C.}},
\bauthor{\bsnm{Mooney}, \binits{M.A.}},
\bauthor{\bsnm{Dieckmann}, \binits{N.F.}},
\bauthor{\bsnm{Miranda-Dominguez}, \binits{O.}},
\bauthor{\bsnm{Nagel}, \binits{B.J.}},
\bauthor{\bsnm{Fair}, \binits{D.A.}},
\bauthor{\bsnm{Nigg}, \binits{J.T.}}:
\batitle{Attention-deficit/hyperactivity disorder: restricted phenotypes prevalence, comorbidity, and polygenic risk sensitivity in the abcd baseline cohort}.
\bjtitle{Journal of the American Academy of Child \& Adolescent Psychiatry}
\bvolume{61}(\bissue{10}),
\bfpage{1273}--\blpage{1284}
(\byear{2022})
\end{barticle}
\endbibitem

\bibitem[\protect\citeauthoryear{Martin et~al.}{2021}]{martin2021predicting}
\begin{barticle}
\bauthor{\bsnm{Martin}, \binits{C.H.}},
\bauthor{\bsnm{Peng}, \binits{T.}},
\bauthor{\bsnm{Mahoney}, \binits{M.W.}}:
\batitle{Predicting trends in the quality of state-of-the-art neural networks without access to training or testing data}.
\bjtitle{Nature Communications}
\bvolume{12}(\bissue{1}),
\bfpage{4122}
(\byear{2021})
\end{barticle}
\endbibitem

\bibitem[\protect\citeauthoryear{Shim et~al.}{2013}]{shim2013frequency}
\begin{barticle}
\bauthor{\bsnm{Shim}, \binits{W.H.}},
\bauthor{\bsnm{Baek}, \binits{K.}},
\bauthor{\bsnm{Kim}, \binits{J.K.}},
\bauthor{\bsnm{Chae}, \binits{Y.}},
\bauthor{\bsnm{Suh}, \binits{J.-Y.}},
\bauthor{\bsnm{Rosen}, \binits{B.R.}},
\bauthor{\bsnm{Jeong}, \binits{J.}},
\bauthor{\bsnm{Kim}, \binits{Y.R.}}:
\batitle{Frequency distribution of causal connectivity in rat sensorimotor network: resting-state fmri analyses}.
\bjtitle{Journal of Neurophysiology}
\bvolume{109}(\bissue{1}),
\bfpage{238}--\blpage{248}
(\byear{2013})
\end{barticle}
\endbibitem

\bibitem[\protect\citeauthoryear{Devlin et~al.}{2018}]{devlin2018bert}
\begin{botherref}
\oauthor{\bsnm{Devlin}, \binits{J.}},
\oauthor{\bsnm{Chang}, \binits{M.-W.}},
\oauthor{\bsnm{Lee}, \binits{K.}},
\oauthor{\bsnm{Toutanova}, \binits{K.}}:
Bert: Pre-training of deep bidirectional transformers for language understanding.
arXiv preprint arXiv:1810.04805
(2018)
\end{botherref}
\endbibitem

\bibitem[\protect\citeauthoryear{Baevski et~al.}{2020}]{baevski2020wav2vec}
\begin{barticle}
\bauthor{\bsnm{Baevski}, \binits{A.}},
\bauthor{\bsnm{Zhou}, \binits{Y.}},
\bauthor{\bsnm{Mohamed}, \binits{A.}},
\bauthor{\bsnm{Auli}, \binits{M.}}:
\batitle{wav2vec 2.0: A framework for self-supervised learning of speech representations}.
\bjtitle{Advances in neural information processing systems}
\bvolume{33},
\bfpage{12449}--\blpage{12460}
(\byear{2020})
\end{barticle}
\endbibitem

\bibitem[\protect\citeauthoryear{Mohanty et~al.}{2020}]{mohanty2020rethinking}
\begin{barticle}
\bauthor{\bsnm{Mohanty}, \binits{R.}},
\bauthor{\bsnm{Sethares}, \binits{W.A.}},
\bauthor{\bsnm{Nair}, \binits{V.A.}},
\bauthor{\bsnm{Prabhakaran}, \binits{V.}}:
\batitle{Rethinking measures of functional connectivity via feature extraction}.
\bjtitle{Scientific reports}
\bvolume{10}(\bissue{1}),
\bfpage{1298}
(\byear{2020})
\end{barticle}
\endbibitem

\bibitem[\protect\citeauthoryear{Vaswani}{2017}]{vaswani2017attention}
\begin{botherref}
\oauthor{\bsnm{Vaswani}, \binits{A.}}:
Attention is all you need.
Advances in Neural Information Processing Systems
(2017)
\end{botherref}
\endbibitem

\bibitem[\protect\citeauthoryear{Selvaraju et~al.}{2017}]{selvaraju2017grad}
\begin{bchapter}
\bauthor{\bsnm{Selvaraju}, \binits{R.R.}},
\bauthor{\bsnm{Cogswell}, \binits{M.}},
\bauthor{\bsnm{Das}, \binits{A.}},
\bauthor{\bsnm{Vedantam}, \binits{R.}},
\bauthor{\bsnm{Parikh}, \binits{D.}},
\bauthor{\bsnm{Batra}, \binits{D.}}:
\bctitle{Grad-cam: Visual explanations from deep networks via gradient-based localization}.
In: \bbtitle{Proceedings of the IEEE International Conference on Computer Vision},
pp. \bfpage{618}--\blpage{626}
(\byear{2017})
\end{bchapter}
\endbibitem

\bibitem[\protect\citeauthoryear{Cohen}{2013}]{cohen2013statistical}
\begin{bbook}
\bauthor{\bsnm{Cohen}, \binits{J.}}:
\bbtitle{Statistical Power Analysis for the Behavioral Sciences},
\bedition{2nd} edn.
\bpublisher{routledge},
\blocation{New York, NY}
(\byear{2013})
\end{bbook}
\endbibitem

\bibitem[\protect\citeauthoryear{Chen et~al.}{2015}]{chen2015xgboost}
\begin{barticle}
\bauthor{\bsnm{Chen}, \binits{T.}},
\bauthor{\bsnm{He}, \binits{T.}},
\bauthor{\bsnm{Benesty}, \binits{M.}},
\bauthor{\bsnm{Khotilovich}, \binits{V.}},
\bauthor{\bsnm{Tang}, \binits{Y.}},
\bauthor{\bsnm{Cho}, \binits{H.}},
\bauthor{\bsnm{Chen}, \binits{K.}},
\bauthor{\bsnm{Mitchell}, \binits{R.}},
\bauthor{\bsnm{Cano}, \binits{I.}},
\bauthor{\bsnm{Zhou}, \binits{T.}}, \betal:
\batitle{Xgboost: extreme gradient boosting}.
\bjtitle{R package version 0.4-2}
\bvolume{1}(\bissue{4}),
\bfpage{1}--\blpage{4}
(\byear{2015})
\end{barticle}
\endbibitem

\bibitem[\protect\citeauthoryear{Kucewicz et~al.}{2024}]{kucewicz2024high}
\begin{barticle}
\bauthor{\bsnm{Kucewicz}, \binits{M.T.}},
\bauthor{\bsnm{Cimbalnik}, \binits{J.}},
\bauthor{\bsnm{Garcia-Salinas}, \binits{J.S.}},
\bauthor{\bsnm{Brazdil}, \binits{M.}},
\bauthor{\bsnm{Worrell}, \binits{G.A.}}:
\batitle{High frequency oscillations in human memory and cognition: a neurophysiological substrate of engrams?}
\bjtitle{Brain}
\bvolume{147}(\bissue{9}),
\bfpage{2966}--\blpage{2982}
(\byear{2024})
\end{barticle}
\endbibitem

\bibitem[\protect\citeauthoryear{Costa et~al.}{2024}]{costa2024distinct}
\begin{barticle}
\bauthor{\bsnm{Costa}, \binits{G.N.}},
\bauthor{\bsnm{Schaum}, \binits{M.}},
\bauthor{\bsnm{Duarte}, \binits{J.V.}},
\bauthor{\bsnm{Martins}, \binits{R.}},
\bauthor{\bsnm{Duarte}, \binits{I.C.}},
\bauthor{\bsnm{Castelhano}, \binits{J.}},
\bauthor{\bsnm{Wibral}, \binits{M.}},
\bauthor{\bsnm{Castelo-Branco}, \binits{M.}}:
\batitle{Distinct oscillatory patterns differentiate between segregation and integration processes in perceptual grouping}.
\bjtitle{Human Brain Mapping}
\bvolume{45}(\bissue{12}),
\bfpage{26779}
(\byear{2024})
\end{barticle}
\endbibitem

\bibitem[\protect\citeauthoryear{Estrada and Hatano}{2008}]{estrada2008communicability}
\begin{barticle}
\bauthor{\bsnm{Estrada}, \binits{E.}},
\bauthor{\bsnm{Hatano}, \binits{N.}}:
\batitle{Communicability in complex networks}.
\bjtitle{Physical Review E}
\bvolume{77}(\bissue{3}),
\bfpage{036111}
(\byear{2008})
\end{barticle}
\endbibitem

\bibitem[\protect\citeauthoryear{Estrada et~al.}{2012}]{estrada2012physics}
\begin{barticle}
\bauthor{\bsnm{Estrada}, \binits{E.}},
\bauthor{\bsnm{Hatano}, \binits{N.}},
\bauthor{\bsnm{Benzi}, \binits{M.}}:
\batitle{The physics of communicability in complex networks}.
\bjtitle{Physics reports}
\bvolume{514}(\bissue{3}),
\bfpage{89}--\blpage{119}
(\byear{2012})
\end{barticle}
\endbibitem

\bibitem[\protect\citeauthoryear{Bullmore et~al.}{2001}]{bullmore2001colored}
\begin{barticle}
\bauthor{\bsnm{Bullmore}, \binits{E.}},
\bauthor{\bsnm{Long}, \binits{C.}},
\bauthor{\bsnm{Suckling}, \binits{J.}},
\bauthor{\bsnm{Fadili}, \binits{J.}},
\bauthor{\bsnm{Calvert}, \binits{G.}},
\bauthor{\bsnm{Zelaya}, \binits{F.}},
\bauthor{\bsnm{Carpenter}, \binits{T.A.}},
\bauthor{\bsnm{Brammer}, \binits{M.}}:
\batitle{Colored noise and computational inference in neurophysiological (fmri) time series analysis: resampling methods in time and wavelet domains}.
\bjtitle{Human brain mapping}
\bvolume{12}(\bissue{2}),
\bfpage{61}--\blpage{78}
(\byear{2001})
\end{barticle}
\endbibitem

\bibitem[\protect\citeauthoryear{Fox and Raichle}{2007}]{fox2007spontaneous}
\begin{barticle}
\bauthor{\bsnm{Fox}, \binits{M.D.}},
\bauthor{\bsnm{Raichle}, \binits{M.E.}}:
\batitle{Spontaneous fluctuations in brain activity observed with functional magnetic resonance imaging}.
\bjtitle{Nature reviews neuroscience}
\bvolume{8}(\bissue{9}),
\bfpage{700}--\blpage{711}
(\byear{2007})
\end{barticle}
\endbibitem

\bibitem[\protect\citeauthoryear{He}{2011}]{he2011scale}
\begin{barticle}
\bauthor{\bsnm{He}, \binits{B.J.}}:
\batitle{Scale-free properties of the functional magnetic resonance imaging signal during rest and task}.
\bjtitle{Journal of Neuroscience}
\bvolume{31}(\bissue{39}),
\bfpage{13786}--\blpage{13795}
(\byear{2011})
\end{barticle}
\endbibitem

\bibitem[\protect\citeauthoryear{Mandelbrot and Van~Ness}{1968}]{mandelbrot1968fractional}
\begin{barticle}
\bauthor{\bsnm{Mandelbrot}, \binits{B.B.}},
\bauthor{\bsnm{Van~Ness}, \binits{J.W.}}:
\batitle{Fractional brownian motions, fractional noises and applications}.
\bjtitle{SIAM review}
\bvolume{10}(\bissue{4}),
\bfpage{422}--\blpage{437}
(\byear{1968})
\end{barticle}
\endbibitem

\bibitem[\protect\citeauthoryear{Eke et~al.}{2002}]{eke2002fractal}
\begin{barticle}
\bauthor{\bsnm{Eke}, \binits{A.}},
\bauthor{\bsnm{Herman}, \binits{P.}},
\bauthor{\bsnm{Kocsis}, \binits{L.}},
\bauthor{\bsnm{Kozak}, \binits{L.}}:
\batitle{Fractal characterization of complexity in temporal physiological signals}.
\bjtitle{Physiological measurement}
\bvolume{23}(\bissue{1}),
\bfpage{1}
(\byear{2002})
\end{barticle}
\endbibitem

\bibitem[\protect\citeauthoryear{Radulescu et~al.}{2012}]{radulescu2012power}
\begin{barticle}
\bauthor{\bsnm{Radulescu}, \binits{A.R.}},
\bauthor{\bsnm{Rubin}, \binits{D.}},
\bauthor{\bsnm{Strey}, \binits{H.H.}},
\bauthor{\bsnm{Mujica-Parodi}, \binits{L.R.}}:
\batitle{Power spectrum scale invariance identifies prefrontal dysregulation in paranoid schizophrenia}.
\bjtitle{Human brain mapping}
\bvolume{33}(\bissue{7}),
\bfpage{1582}--\blpage{1593}
(\byear{2012})
\end{barticle}
\endbibitem

\bibitem[\protect\citeauthoryear{Cha et~al.}{2016}]{cha2016clinically}
\begin{barticle}
\bauthor{\bsnm{Cha}, \binits{J.}},
\bauthor{\bsnm{DeDora}, \binits{D.}},
\bauthor{\bsnm{Nedic}, \binits{S.}},
\bauthor{\bsnm{Ide}, \binits{J.}},
\bauthor{\bsnm{Greenberg}, \binits{T.}},
\bauthor{\bsnm{Hajcak}, \binits{G.}},
\bauthor{\bsnm{Mujica-Parodi}, \binits{L.R.}}:
\batitle{Clinically anxious individuals show disrupted feedback between inferior frontal gyrus and prefrontal-limbic control circuit}.
\bjtitle{Journal of Neuroscience}
\bvolume{36}(\bissue{17}),
\bfpage{4708}--\blpage{4718}
(\byear{2016})
\end{barticle}
\endbibitem

\bibitem[\protect\citeauthoryear{Tu et~al.}{2023}]{tu2023rare}
\begin{botherref}
\oauthor{\bsnm{Tu}, \binits{W.}},
\oauthor{\bsnm{Liao}, \binits{Q.}},
\oauthor{\bsnm{Zhou}, \binits{S.}},
\oauthor{\bsnm{Peng}, \binits{X.}},
\oauthor{\bsnm{Ma}, \binits{C.}},
\oauthor{\bsnm{Liu}, \binits{Z.}},
\oauthor{\bsnm{Liu}, \binits{X.}},
\oauthor{\bsnm{Cai}, \binits{Z.}},
\oauthor{\bsnm{He}, \binits{K.}}:
Rare: Robust masked graph autoencoder.
IEEE Transactions on Knowledge and Data Engineering
(2023)
\end{botherref}
\endbibitem

\bibitem[\protect\citeauthoryear{Liu et~al.}{2024}]{liu2024mask}
\begin{botherref}
\oauthor{\bsnm{Liu}, \binits{C.}},
\oauthor{\bsnm{Wang}, \binits{Y.}},
\oauthor{\bsnm{Zhan}, \binits{Y.}},
\oauthor{\bsnm{Ma}, \binits{X.}},
\oauthor{\bsnm{Tao}, \binits{D.}},
\oauthor{\bsnm{Wu}, \binits{J.}},
\oauthor{\bsnm{Hu}, \binits{W.}}:
Where to mask: Structure-guided masking for graph masked autoencoders.
arXiv preprint arXiv:2404.15806
(2024)
\end{botherref}
\endbibitem

\bibitem[\protect\citeauthoryear{Harush and Barzel}{2017}]{harush2017dynamic}
\begin{barticle}
\bauthor{\bsnm{Harush}, \binits{U.}},
\bauthor{\bsnm{Barzel}, \binits{B.}}:
\batitle{Dynamic patterns of information flow in complex networks}.
\bjtitle{Nature communications}
\bvolume{8}(\bissue{1}),
\bfpage{2181}
(\byear{2017})
\end{barticle}
\endbibitem

\bibitem[\protect\citeauthoryear{Burgess and Wu}{2013}]{burgess2013rostral}
\begin{botherref}
\oauthor{\bsnm{Burgess}, \binits{P.W.}},
\oauthor{\bsnm{Wu}, \binits{H.}}:
Rostral prefrontal cortex (brodmann area 10).
Principles of frontal lobe function,
524--544
(2013)
\end{botherref}
\endbibitem

\bibitem[\protect\citeauthoryear{Roth and Saykin}{2004}]{roth2004executive}
\begin{barticle}
\bauthor{\bsnm{Roth}, \binits{R.M.}},
\bauthor{\bsnm{Saykin}, \binits{A.J.}}:
\batitle{Executive dysfunction in attention-deficit/hyperactivity disorder: cognitive and neuroimaging findings}.
\bjtitle{Psychiatric Clinics}
\bvolume{27}(\bissue{1}),
\bfpage{83}--\blpage{96}
(\byear{2004})
\end{barticle}
\endbibitem

\bibitem[\protect\citeauthoryear{Baker et~al.}{2018}]{baker2018connectomic5}
\begin{barticle}
\bauthor{\bsnm{Baker}, \binits{C.M.}},
\bauthor{\bsnm{Burks}, \binits{J.D.}},
\bauthor{\bsnm{Briggs}, \binits{R.G.}},
\bauthor{\bsnm{Conner}, \binits{A.K.}},
\bauthor{\bsnm{Glenn}, \binits{C.A.}},
\bauthor{\bsnm{Robbins}, \binits{J.M.}},
\bauthor{\bsnm{Sheets}, \binits{J.R.}},
\bauthor{\bsnm{Sali}, \binits{G.}},
\bauthor{\bsnm{McCoy}, \binits{T.M.}},
\bauthor{\bsnm{Battiste}, \binits{J.D.}}, \betal:
\batitle{A connectomic atlas of the human cerebrum—chapter 5: The insula and opercular cortex}.
\bjtitle{Operative Neurosurgery}
\bvolume{15}(\bissue{suppl\_1}),
\bfpage{175}--\blpage{244}
(\byear{2018})
\end{barticle}
\endbibitem

\bibitem[\protect\citeauthoryear{Peterson et~al.}{2023}]{peterson2023neuroanatomy}
\begin{botherref}
\oauthor{\bsnm{Peterson}, \binits{D.}},
\oauthor{\bsnm{Reddy}, \binits{V.}},
\oauthor{\bsnm{Hamel}, \binits{R.}}:
Neuroanatomy, auditory pathway. StatPearls.
StatPearls Publishing Copyright
(2023)
\end{botherref}
\endbibitem

\bibitem[\protect\citeauthoryear{Ito et~al.}{2009}]{ito2009somatosensory}
\begin{barticle}
\bauthor{\bsnm{Ito}, \binits{T.}},
\bauthor{\bsnm{Tiede}, \binits{M.}},
\bauthor{\bsnm{Ostry}, \binits{D.J.}}:
\batitle{Somatosensory function in speech perception}.
\bjtitle{Proceedings of the National Academy of Sciences}
\bvolume{106}(\bissue{4}),
\bfpage{1245}--\blpage{1248}
(\byear{2009})
\end{barticle}
\endbibitem

\bibitem[\protect\citeauthoryear{Conant et~al.}{2014}]{conant2014speech}
\begin{barticle}
\bauthor{\bsnm{Conant}, \binits{D.}},
\bauthor{\bsnm{Bouchard}, \binits{K.E.}},
\bauthor{\bsnm{Chang}, \binits{E.F.}}:
\batitle{Speech map in the human ventral sensory-motor cortex}.
\bjtitle{Current opinion in neurobiology}
\bvolume{24},
\bfpage{63}--\blpage{67}
(\byear{2014})
\end{barticle}
\endbibitem

\bibitem[\protect\citeauthoryear{LaFlamme et~al.}{2021}]{laflamme2021parahippocampal}
\begin{barticle}
\bauthor{\bsnm{LaFlamme}, \binits{E.M.}},
\bauthor{\bsnm{Waguespack}, \binits{H.F.}},
\bauthor{\bsnm{Forcelli}, \binits{P.A.}},
\bauthor{\bsnm{Malkova}, \binits{L.}}:
\batitle{The parahippocampal cortex and its functional connection with the hippocampus are critical for nonnavigational spatial memory in macaques}.
\bjtitle{Cerebral Cortex}
\bvolume{31}(\bissue{4}),
\bfpage{2251}--\blpage{2267}
(\byear{2021})
\end{barticle}
\endbibitem

\bibitem[\protect\citeauthoryear{Bohbot et~al.}{1998}]{bohbot1998spatial}
\begin{barticle}
\bauthor{\bsnm{Bohbot}, \binits{V.D.}},
\bauthor{\bsnm{Kalina}, \binits{M.}},
\bauthor{\bsnm{Stepankova}, \binits{K.}},
\bauthor{\bsnm{Spackova}, \binits{N.}},
\bauthor{\bsnm{Petrides}, \binits{M.}},
\bauthor{\bsnm{Nadel}, \binits{L.}}:
\batitle{Spatial memory deficits in patients with lesions to the right hippocampus and to the right parahippocampal cortex}.
\bjtitle{Neuropsychologia}
\bvolume{36}(\bissue{11}),
\bfpage{1217}--\blpage{1238}
(\byear{1998})
\end{barticle}
\endbibitem

\bibitem[\protect\citeauthoryear{Skodzik et~al.}{2017}]{skodzik2017long}
\begin{barticle}
\bauthor{\bsnm{Skodzik}, \binits{T.}},
\bauthor{\bsnm{Holling}, \binits{H.}},
\bauthor{\bsnm{Pedersen}, \binits{A.}}:
\batitle{Long-term memory performance in adult adhd: A meta-analysis}.
\bjtitle{Journal of attention disorders}
\bvolume{21}(\bissue{4}),
\bfpage{267}--\blpage{283}
(\byear{2017})
\end{barticle}
\endbibitem

\bibitem[\protect\citeauthoryear{Steinberg and Drabick}{2015}]{steinberg2015developmental}
\begin{barticle}
\bauthor{\bsnm{Steinberg}, \binits{E.A.}},
\bauthor{\bsnm{Drabick}, \binits{D.A.}}:
\batitle{A developmental psychopathology perspective on adhd and comorbid conditions: The role of emotion regulation}.
\bjtitle{Child Psychiatry \& Human Development}
\bvolume{46},
\bfpage{951}--\blpage{966}
(\byear{2015})
\end{barticle}
\endbibitem

\bibitem[\protect\citeauthoryear{Bunford et~al.}{2015}]{bunford2015adhd}
\begin{barticle}
\bauthor{\bsnm{Bunford}, \binits{N.}},
\bauthor{\bsnm{Evans}, \binits{S.W.}},
\bauthor{\bsnm{Wymbs}, \binits{F.}}:
\batitle{Adhd and emotion dysregulation among children and adolescents}.
\bjtitle{Clinical child and family psychology review}
\bvolume{18},
\bfpage{185}--\blpage{217}
(\byear{2015})
\end{barticle}
\endbibitem

\bibitem[\protect\citeauthoryear{Rolls et~al.}{2020}]{rolls2020orbitofrontal}
\begin{barticle}
\bauthor{\bsnm{Rolls}, \binits{E.T.}},
\bauthor{\bsnm{Cheng}, \binits{W.}},
\bauthor{\bsnm{Feng}, \binits{J.}}:
\batitle{The orbitofrontal cortex: reward, emotion and depression}.
\bjtitle{Brain communications}
\bvolume{2}(\bissue{2}),
\bfpage{196}
(\byear{2020})
\end{barticle}
\endbibitem

\bibitem[\protect\citeauthoryear{Case-Smith et~al.}{2015}]{case2015systematic}
\begin{barticle}
\bauthor{\bsnm{Case-Smith}, \binits{J.}},
\bauthor{\bsnm{Weaver}, \binits{L.L.}},
\bauthor{\bsnm{Fristad}, \binits{M.A.}}:
\batitle{A systematic review of sensory processing interventions for children with autism spectrum disorders}.
\bjtitle{Autism}
\bvolume{19}(\bissue{2}),
\bfpage{133}--\blpage{148}
(\byear{2015})
\end{barticle}
\endbibitem

\bibitem[\protect\citeauthoryear{Cibralic et~al.}{2019}]{cibralic2019systematic}
\begin{barticle}
\bauthor{\bsnm{Cibralic}, \binits{S.}},
\bauthor{\bsnm{Kohlhoff}, \binits{J.}},
\bauthor{\bsnm{Wallace}, \binits{N.}},
\bauthor{\bsnm{McMahon}, \binits{C.}},
\bauthor{\bsnm{Eapen}, \binits{V.}}:
\batitle{A systematic review of emotion regulation in children with autism spectrum disorder}.
\bjtitle{Research in Autism Spectrum Disorders}
\bvolume{68},
\bfpage{101422}
(\byear{2019})
\end{barticle}
\endbibitem

\bibitem[\protect\citeauthoryear{Velikonja et~al.}{2019}]{velikonja2019patterns}
\begin{barticle}
\bauthor{\bsnm{Velikonja}, \binits{T.}},
\bauthor{\bsnm{Fett}, \binits{A.-K.}},
\bauthor{\bsnm{Velthorst}, \binits{E.}}:
\batitle{Patterns of nonsocial and social cognitive functioning in adults with autism spectrum disorder: A systematic review and meta-analysis}.
\bjtitle{JAMA psychiatry}
\bvolume{76}(\bissue{2}),
\bfpage{135}--\blpage{151}
(\byear{2019})
\end{barticle}
\endbibitem

\bibitem[\protect\citeauthoryear{Baker et~al.}{2018}]{baker2018connectomic2}
\begin{barticle}
\bauthor{\bsnm{Baker}, \binits{C.M.}},
\bauthor{\bsnm{Burks}, \binits{J.D.}},
\bauthor{\bsnm{Briggs}, \binits{R.G.}},
\bauthor{\bsnm{Conner}, \binits{A.K.}},
\bauthor{\bsnm{Glenn}, \binits{C.A.}},
\bauthor{\bsnm{Morgan}, \binits{J.P.}},
\bauthor{\bsnm{Stafford}, \binits{J.}},
\bauthor{\bsnm{Sali}, \binits{G.}},
\bauthor{\bsnm{McCoy}, \binits{T.M.}},
\bauthor{\bsnm{Battiste}, \binits{J.D.}}, \betal:
\batitle{A connectomic atlas of the human cerebrum—chapter 2: The lateral frontal lobe}.
\bjtitle{Operative Neurosurgery}
\bvolume{15}(\bissue{suppl\_1}),
\bfpage{10}--\blpage{74}
(\byear{2018})
\end{barticle}
\endbibitem

\bibitem[\protect\citeauthoryear{Law et~al.}{2023}]{law2023frontopolar}
\begin{botherref}
\oauthor{\bsnm{Law}, \binits{C.-K.}},
\oauthor{\bsnm{Kolling}, \binits{N.}},
\oauthor{\bsnm{Chan}, \binits{C.C.}},
\oauthor{\bsnm{Chau}, \binits{B.K.}}:
Frontopolar cortex represents complex features and decision value during choice between environments.
Cell reports
\textbf{42}(6)
(2023)
\end{botherref}
\endbibitem

\bibitem[\protect\citeauthoryear{Ferrucci et~al.}{2025}]{ferrucci2025reward}
\begin{barticle}
\bauthor{\bsnm{Ferrucci}, \binits{L.}},
\bauthor{\bsnm{Ceccarelli}, \binits{F.}},
\bauthor{\bsnm{Londei}, \binits{F.}},
\bauthor{\bsnm{Arena}, \binits{G.}},
\bauthor{\bsnm{Elyasizad}, \binits{L.}},
\bauthor{\bsnm{Nougaret}, \binits{S.}},
\bauthor{\bsnm{Genovesio}, \binits{A.}}:
\batitle{Reward monitoring in the frontopolar cortex of macaques}.
\bjtitle{Scientific Reports}
\bvolume{15}(\bissue{1}),
\bfpage{1}--\blpage{14}
(\byear{2025})
\end{barticle}
\endbibitem

\bibitem[\protect\citeauthoryear{Pol{\'o}nyiov{\'a} et~al.}{2024}]{polonyiova2024roots}
\begin{botherref}
\oauthor{\bsnm{Pol{\'o}nyiov{\'a}}, \binits{K.}},
\oauthor{\bsnm{Kruyt}, \binits{J.}},
\oauthor{\bsnm{Ostatn{\'\i}kov{\'a}}, \binits{D.}}:
To the roots of theory of mind deficits in autism spectrum disorder: a narrative review.
Review Journal of Autism and Developmental Disorders,
1--5
(2024)
\end{botherref}
\endbibitem

\bibitem[\protect\citeauthoryear{De~Benedictis et~al.}{2014}]{de2014anatomo}
\begin{barticle}
\bauthor{\bsnm{De~Benedictis}, \binits{A.}},
\bauthor{\bsnm{Duffau}, \binits{H.}},
\bauthor{\bsnm{Paradiso}, \binits{B.}},
\bauthor{\bsnm{Grandi}, \binits{E.}},
\bauthor{\bsnm{Balbi}, \binits{S.}},
\bauthor{\bsnm{Granieri}, \binits{E.}},
\bauthor{\bsnm{Colarusso}, \binits{E.}},
\bauthor{\bsnm{Chioffi}, \binits{F.}},
\bauthor{\bsnm{Marras}, \binits{C.E.}},
\bauthor{\bsnm{Sarubbo}, \binits{S.}}:
\batitle{Anatomo-functional study of the temporo-parieto-occipital region: dissection, tractographic and brain mapping evidence from a neurosurgical perspective}.
\bjtitle{Journal of anatomy}
\bvolume{225}(\bissue{2}),
\bfpage{132}--\blpage{151}
(\byear{2014})
\end{barticle}
\endbibitem

\bibitem[\protect\citeauthoryear{Seghatol-Eslami et~al.}{2020}]{seghatol2020hyperconnectivity}
\begin{barticle}
\bauthor{\bsnm{Seghatol-Eslami}, \binits{V.C.}},
\bauthor{\bsnm{Maximo}, \binits{J.O.}},
\bauthor{\bsnm{Ammons}, \binits{C.J.}},
\bauthor{\bsnm{Libero}, \binits{L.E.}},
\bauthor{\bsnm{Kana}, \binits{R.K.}}:
\batitle{Hyperconnectivity of social brain networks in autism during action-intention judgment}.
\bjtitle{Neuropsychologia}
\bvolume{137},
\bfpage{107303}
(\byear{2020})
\end{barticle}
\endbibitem

\bibitem[\protect\citeauthoryear{Supekar et~al.}{2013}]{supekar2013brain}
\begin{barticle}
\bauthor{\bsnm{Supekar}, \binits{K.}},
\bauthor{\bsnm{Uddin}, \binits{L.Q.}},
\bauthor{\bsnm{Khouzam}, \binits{A.}},
\bauthor{\bsnm{Phillips}, \binits{J.}},
\bauthor{\bsnm{Gaillard}, \binits{W.D.}},
\bauthor{\bsnm{Kenworthy}, \binits{L.E.}},
\bauthor{\bsnm{Yerys}, \binits{B.E.}},
\bauthor{\bsnm{Vaidya}, \binits{C.J.}},
\bauthor{\bsnm{Menon}, \binits{V.}}:
\batitle{Brain hyperconnectivity in children with autism and its links to social deficits}.
\bjtitle{Cell reports}
\bvolume{5}(\bissue{3}),
\bfpage{738}--\blpage{747}
(\byear{2013})
\end{barticle}
\endbibitem

\bibitem[\protect\citeauthoryear{Chita-Tegmark}{2016}]{chita2016social}
\begin{barticle}
\bauthor{\bsnm{Chita-Tegmark}, \binits{M.}}:
\batitle{Social attention in asd: A review and meta-analysis of eye-tracking studies}.
\bjtitle{Research in developmental disabilities}
\bvolume{48},
\bfpage{79}--\blpage{93}
(\byear{2016})
\end{barticle}
\endbibitem

\bibitem[\protect\citeauthoryear{Coifman and Wickerhauser}{1992}]{coifman1992entropy}
\begin{barticle}
\bauthor{\bsnm{Coifman}, \binits{R.R.}},
\bauthor{\bsnm{Wickerhauser}, \binits{M.V.}}:
\batitle{Entropy-based algorithms for best basis selection}.
\bjtitle{IEEE Transactions on information theory}
\bvolume{38}(\bissue{2}),
\bfpage{713}--\blpage{718}
(\byear{1992})
\end{barticle}
\endbibitem

\bibitem[\protect\citeauthoryear{Noble et~al.}{2021}]{noble2021guide}
\begin{barticle}
\bauthor{\bsnm{Noble}, \binits{S.}},
\bauthor{\bsnm{Scheinost}, \binits{D.}},
\bauthor{\bsnm{Constable}, \binits{R.T.}}:
\batitle{A guide to the measurement and interpretation of fmri test-retest reliability}.
\bjtitle{Current opinion in behavioral sciences}
\bvolume{40},
\bfpage{27}--\blpage{32}
(\byear{2021})
\end{barticle}
\endbibitem

\bibitem[\protect\citeauthoryear{Kajimura et~al.}{2023}]{kajimura2023frequency}
\begin{barticle}
\bauthor{\bsnm{Kajimura}, \binits{S.}},
\bauthor{\bsnm{Margulies}, \binits{D.}},
\bauthor{\bsnm{Smallwood}, \binits{J.}}:
\batitle{Frequency-specific brain network architecture in resting-state fmri}.
\bjtitle{Scientific Reports}
\bvolume{13}(\bissue{1}),
\bfpage{2964}
(\byear{2023})
\end{barticle}
\endbibitem

\end{thebibliography}

\end{document}